\documentclass[floatfix,nofootinbib,twocolumn,showpacs,superscriptaddress, preprintnumbers,amssymb,letterpaper,prab]{revtex4-2}

\usepackage{lineno}
\usepackage{textgreek}

\usepackage{dcolumn}% Align table columns on decimal point
\usepackage{bm}% bold math
\usepackage{amsmath}
\usepackage{multirow}
\usepackage{graphicx}
\usepackage{float}
\usepackage[colorlinks=true,allcolors=blue]{hyperref}
\usepackage[encapsulated]{CJK}
\usepackage[english]{babel}
\usepackage{siunitx}
\usepackage{comment}

\begin{document}

 %\twocolumn
 %\baselineskip = 1.5\baselineskip

 \newcommand{\re}{\mathop{\mathrm{Re}}}
 \newcommand{\im}{\mathop{\mathrm{Im}}}
 \newcommand{\D}{\mathop{\mathrm{d}}}
 \newcommand{\E}{\mathop{\mathrm{e}}}
 \newcommand{\unite}[2]{\mbox{$#1\,{\rm #2}$}}
 \newcommand{\myvec}[1]{\mbox{$\overrightarrow{#1}$}}
 \newcommand{\mynor}[1]{\mbox{$\widehat{#1}$}}
\newcommand{\rmsemit}{\mbox{$\tilde{\varepsilon}$}}
\newcommand{\mean}[1]{\mbox{$\langle{#1}\rangle$}}
\newcommand{\warp}{{\sc Warp }}
\newcommand{\astra}{{\sc Astra }}
\newcommand{\elegant}{{\sc Elegant }}
\newcommand{\astragenerator}{{\sc AstraGenerator }}
\newcommand{\mafia}{{\sc Mafia }}

\preprint{~~~~}

\title{Formation of Temporally Shaped Electron Bunches for \\
Beam-Driven Collinear Wakefield Accelerators}
%\title{ Preparing Electron Bunches for Wakefield Accelerators
%with a Tailored Longitudinal-Phase-Space Distribution}
%via nonlinear longitudinal-phase-space manipulation}
% Force line breaks with \\
\begin{CJK*}{UTF8}{}
\author{Wei Hou Tan}
\email{wtan1@niu.edu}
\affiliation{Northern Illinois Center for Accelerator \& Detector Development and Department of Physics, Northern Illinois University, DeKalb, IL 60115, USA} 
\author{Philippe  Piot} 
\affiliation{Northern Illinois Center for Accelerator \& Detector Development and Department of Physics, Northern Illinois University, DeKalb, IL 60115, USA} 
\affiliation{Argonne National Laboratory, Lemont, IL 60439, USA}
\author{Alexander Zholents} 
\affiliation{Argonne National Laboratory, Lemont, IL 60439, USA}

\date{\today}% It is always \today, today,

\begin{abstract}
Beam-driven collinear wakefield accelerators (CWAs) that operate by using slow-wave structures or plasmas hold great promise toward reducing the size of contemporary accelerators. Sustainable acceleration of charged particles to high energies in the CWA relies on using field-generating relativistic electron bunches with a highly asymmetric peak current profile and a large energy chirp.  A new approach to obtaining such bunches has been proposed and illustrated with the accelerator design supported by particle tracking simulations. It has been shown that the required particle distribution in the longitudinal phase space can be obtained without collimators, giving CWAs an opportunity for employment in applications requiring a high repetition rate of operation.
\end{abstract}


\maketitle
\end{CJK*}
%\tableofcontents

\section{Introduction}
In a beam-based collinear wakefield accelerator (CWA), a high-charge drive bunch generates an electromagnetic field passing through a slow-wave structure (a dielectric-lined or corrugated waveguide) or plasma. This field, called the wakefield, is used to accelerate a  witness bunch propagating the structure in the same direction behind the drive bunch \cite{voss_wake_1982,briggs_electron_1974, friedman_autoacceleration_1973,perevedentsev_use_1978_2,sessler_free_1982,chen_acceleration_1985,chin_wake_1983,gai_experimental_1988}. An important figure of merit for a CWA is the transformer ratio, \(\mathcal R \equiv \left|\mathcal E_+/\mathcal E_-\right|\), where \(\mathcal E_+\) is the maximum accelerating field behind the drive bunch, and \(\mathcal E_-\) is the maximum decelerating field within the drive bunch. For symmetric drive-bunch current distribution in time $I(t)$, the transformer ratio is limited to \(\mathcal R<2\) ~\cite{bane_collinear_1985}. However, asymmetric $I(t)$ can significantly enhance the transformer ratio~\cite{bane_collinear_1985}, albeit at the expense of reduced \(\mathcal E_+\) and \(\mathcal E_-\)~\cite{baturin_upper_2017}. 

Bunch-shaping techniques investigated hitherto are photocathode-laser intensity shaping \cite{cornacchia_formation_2006,penco_experimental_2014,lemery_tailored_2015}, transverse-to-longitudinal phase-space~ exchange~\cite{jiang_formation_2012,ha_precision_2017,gao_observation_2018}, and use of multi-frequency linacs~\cite{piot_generation_2012}. Despite significant progress, they suffer either from their inability to deliver highly asymmetric bunches or from prohibitively large beam losses on collimators.  Consequently, producing drive bunches with an asymmetric peak current profile while preserving most of the bunch electrons has been an active research topic. 

Another important consideration for a drive bunch arises from its proneness to the transverse beam-break-up (BBU) instability caused by the strong transverse forces due to the transverse wakefield~\cite{panofsky_asymptotic_1968,neil_further_1979,chao_beam_1980}. A possible BBU-mitigation technique consists of imparting a large energy chirp along the drive bunch~\cite{balakin_vlepp:_1983,li_high_2014,shchegolkov_towards_2016} and creating a current profile $I(t)$ that stimulates a dynamic adjustment of this chirp concurrently with the wakefield-induced bunch deceleration in the CWA \cite{baturin_stability_2018}.

The work reported in this paper was motivated by a design of a high repetition rate CWA for use in a free-electron laser (FEL) facility described in Refs.~\cite{Zholents:IPAC2018-TUPMF010, waldschmidt_design_2018}. This facility plans to employ up to ten FELs individually driven by a dedicated CWA. A single conventional accelerator delivers $\sim\SI{1}{\giga\electronvolt}$ drive electron bunches with a highly asymmetric $I(t)$ and a large energy chirp to the ten CWAs. Since the drive-bunch charge considered in \cite{Zholents:IPAC2018-TUPMF010, waldschmidt_design_2018} is up to \SI{10}{\nano\coulomb} and the bunch repetition rate up to \SI{500}{\kilo\hertz}, the electron beam carries significant power. Therefore, using collimators to assist with the bunch shaping is prohibitive, and, consequently, preparing the drive bunches doing otherwise becomes a prime challenge.

To solve the problem, we undertook a new approach and distributed the task of obtaining the highly asymmetric $I(t)$ over the entire drive bunch accelerator beginning from the photocathode electron gun and ending by the final bunch compressor. 
To the best of our knowledge, our work demonstrates for the first time a pathway to obtaining electron bunches with a highly asymmetric $I(t)$, avoiding prohibitively large electron losses on collimators. The employed technique is rather generic and can be used for preparing the electron bunch peak current distribution with profiles different than those considered in this paper.

Although the main focus of the work was to obtain a drive bunch with the required distribution in the longitudinal phase space (LPS), an equally important additional objective, was to ensure the associated transverse emittances commensurate with the small CWA aperture.

\section{The Drive Bunch and the Wakefield\label{sec:wakeIntro}}
We define the longitudinal charge distribution in the electron bunch as $q(z)$ and consider bunches localized on the interval $0\leq{z}\leq{L}$, where $z$ is the distance behind the bunch head. Therefore, we have
\begin{align}
	\label{eq:dens}
	\int_{0}^Lq(z)\mathrm dz=Q,
\end{align}
where $Q$ is the total bunch charge.
Following \cite{baturin_upper_2017}, we use the Green's function $G(z)$ consisting only of a fundamental mode 
 $G(z)=2\kappa_{\parallel}\cos{(kz)}H(z)$
 \footnote{It has been shown in \cite{baturin_upper_2017} that a multi-mode Green's function is less effective in producing a high transformer ratio.}, 
 where $\kappa_{\parallel}$ is the loss factor of a point particle per unit length, $k=2\pi/\lambda$ is the wave vector, $\lambda$ is the wavelength, $H(z)$ is the Heaviside step function. The longitudinal electric field within the electron bunch can be written as~\cite{Zotter,Chao}
\begin{align}
\label{eq:Ein} 
\mathcal E_-(z)&=2\kappa_{\parallel}\int_{0}^{z}\cos{[k(z-z')]}q(z')\mathrm dz', ~z\leq L, 
\end{align}
which is a Volterra equation of the first kind for the function $q(z)$ with the trigonometric kernel $\cos{[k(z'-z)]}$. 
If we assume that  $\mathcal E_-(0)=0$ at the bunch head, then the solution of Eq.~\eqref{eq:Ein} is given by \cite{Polyanin},
\begin{align}
\label{eq:sol}
q(z)&=\frac{1}{2\kappa_{\parallel}}\left[\mathcal E^\prime_-(z)+k^2\int_{0}^{z}\mathcal E_{-}(x)\mathrm dx \right],
\end{align}
where $\mathcal E_-(z)$ is a known function, and its derivative is taken over $z$. Hence, $q(z)$ is defined.
\begin{figure}[tb]
   \centering
   \includegraphics[width=\columnwidth]{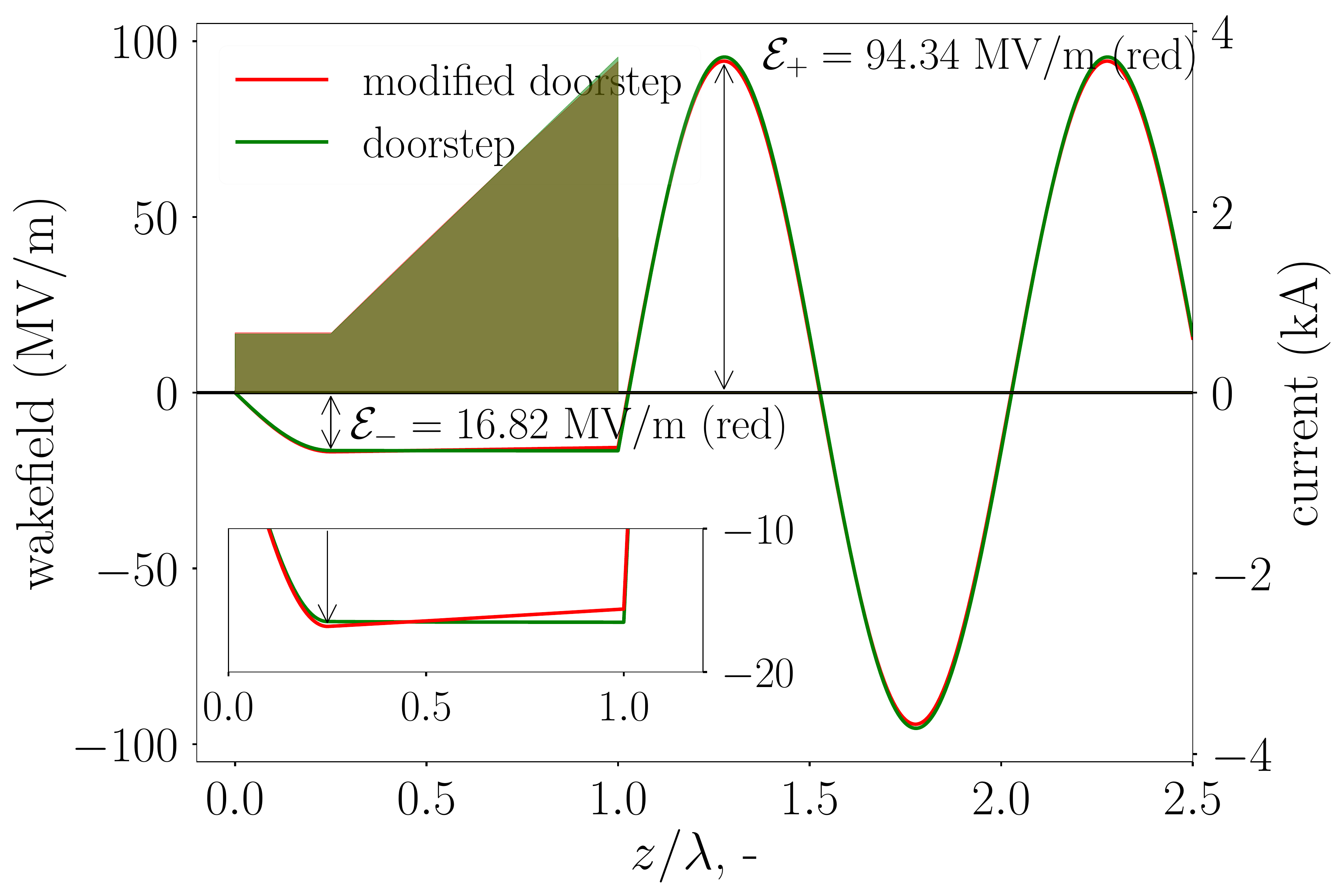}
   \caption{Nominal (green trace) and modified doorstep distributions with associated wakefields calculated using $L=\lambda$, $\chi=0$ and $\chi=-\frac{1}{10\lambda}$, respectively. The wakefields are computed for a bunch charge of \SI{10}{\nano\coulomb} and use a single-mode Green's function, where $f=\SI{180}{\giga\hertz}$ and $\kappa_{\parallel}$=\SI{14.3}{\kilo\volt/\pico\coulomb/\metre} calculated using \textsc{ECHO}\cite{zagorodnov_calculation_2015}. The transformer ratio for the modified doorstep distribution shown in the plot is $\mathcal R=\num{5.6}$.}
   \label{fig:distribution}
\end{figure}

\begin{figure}[!htb]
   \centering
   \includegraphics[width=\columnwidth]{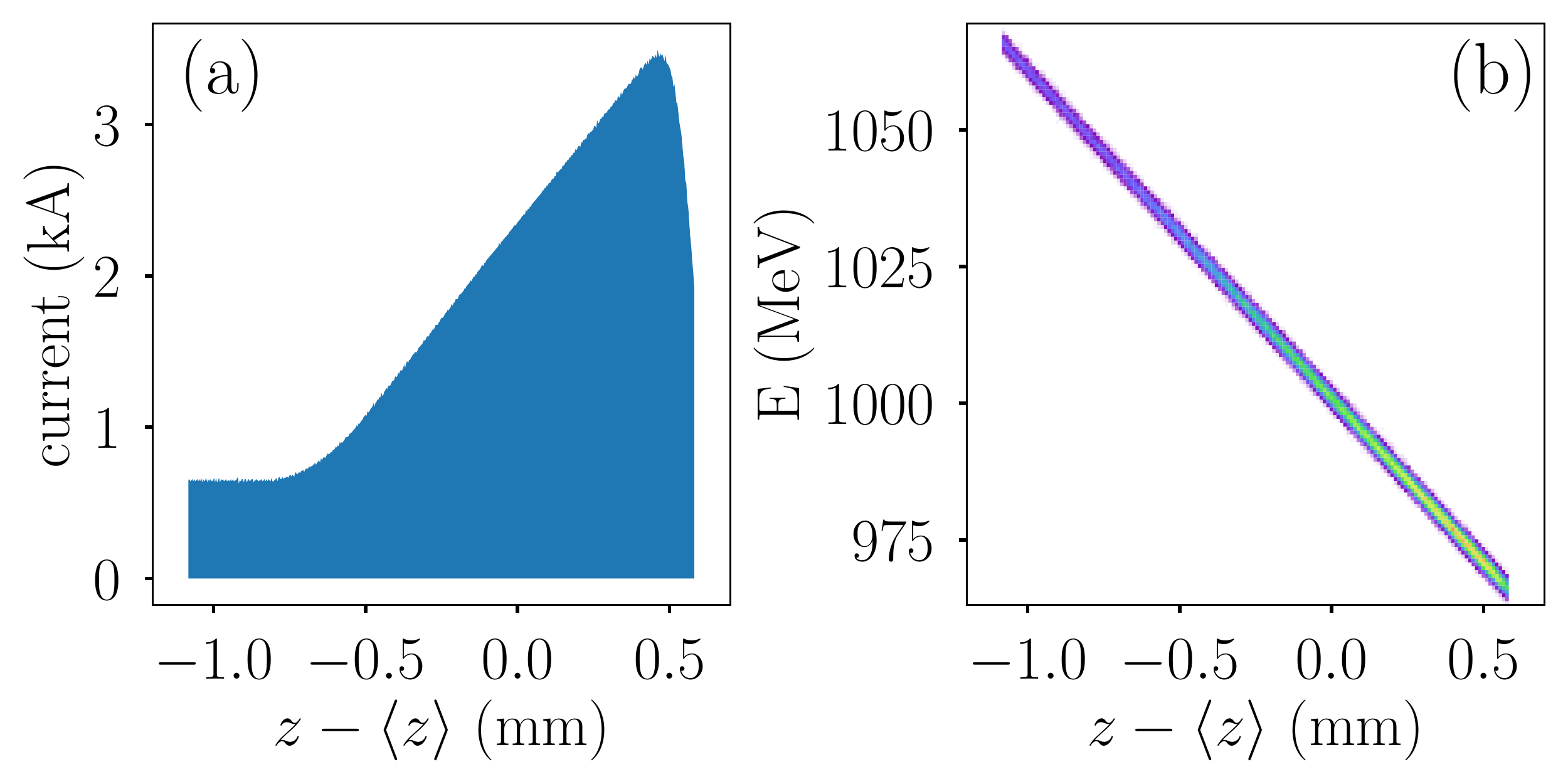}
   \caption{A target drive bunch peak current (a) and longitudinal phase space (b) distributions at the end of the drive bunch accelerator.}
   \label{fig:bunchfinalback}
\end{figure}

In order to maintain the stability of the drive bunch in the CWA throughout its deceleration, we require the bunch's relative chirp to be constant while being decelerated by the wakefield $\mathcal E_-(z)$, based on studies in \cite{baturin_stability_2018}. This requirement is achieved by having a small linear variation in energy loss within the bunch, where head particles lose more energies than tail particles such that
\begin{align}
\label{eq:chirp_req_sasha}
\chi(s) = \frac{1}{E\mathrm{_0(s)}}\frac{\partial E}{\partial z}\propto\mathcal{E'_-}(z)\equiv \text{const},
\end{align}
where $E\mathrm{_0(s)}$ is the energy of the reference particle, and $s$ is the distance propagated by the bunch in the CWA.
This is accomplished by using the electron bunch producing $\mathcal E_-$ with a linear variation in $z$. Similar to Ref.~\cite{bane_collinear_1985}, we solve Eq.~(\ref{eq:sol}) considering $q(z)$ to be constant in the range $0\leq z < \xi$ with $\xi=\frac{1}{k}\arccos(\chi/k)$, in which case the continuities of $\mathcal E_-(z)$ and $\mathcal E_-^{\prime}(z)$ are preserved over the entire bunch length
\begin{widetext}
\begin{align}
\label{eq:doormod1}
q(z) &= 
\begin{cases}
  q_0,  & 0\leq z < \xi\;, \\ 
  q_0 \left[1 - k\xi\sin(k\xi) +\frac{k^2}{2}\xi^2\cos(k\xi) + \big(k\sin(k\xi)-k^2\xi\cos(k\xi)\big)z+ \frac{k^2}{2}\cos(k\xi)z^2\right]
  , & \xi\leq z \leq L\;,\\
\end{cases}\\ 
%\label{eq:doormodnorm_sasha}
q_0 &= \frac{6Q}{6L+k^2\cos(k\xi)(L-\xi)^3+3k\sin(k\xi)(L-\xi)^2}\nonumber\;.
\end{align}
\end{widetext}
Setting $\chi=0$, simplifies $q(z)$ to one used in \cite{bane_collinear_1985}. 
Figure~\ref{fig:distribution} shows an example of a modified doorstep distribution with an associated wakefield calculated using $L=\lambda$ and $\chi=-\frac{1}{10\lambda}$. In this example we considered a corrugated waveguide with radius $a$=1 mm and fundamental mode frequency $f=\SI{180}{\giga\hertz}$, as discussed in Ref.~\cite{Asiy-napac_2019}. The current profile has sharp features that are challenging to realize. Consequently, the distribution defined by Eq.~\eqref{eq:doormod1} is used only as a starting point to construct a practically realizable distribution shown in Fig.~\ref{fig:bunchfinalback} with similar final properties listed in 
Table~\ref{tab:finaldistribution}.  
\begin{table}[!hbt]
   \centering
   \caption{Main parameters associated with the drive bunch distribution shown in Fig.~\ref{fig:bunchfinalback}.}
   \begin{tabular}{lcc}
       \toprule
       \textbf{Bunch parameter} &  \textbf{Value} &  \textbf{Unit}\\
%       Number of macroparticles & \SI{100593}{} & $--$ \\
       \hline
           Charge                  & \num 10 & \si{\nano\coulomb} \\ %[3pt]
           Reference energy        & \num 1 & \si{\GeV} \\ %[3pt]
           RMS length              & \num{419}  & \si{\micro\metre} \\ %[3pt]
           Peak current            & \num 3.5   & \si{\kilo\ampere} \\ %[3pt]
           RMS fractional energy spread       & \SI{2.51}{ } & \%\\
           RMS fractional slice energy spread & \SI{0.1}{ } & \%\\
\hline
\hline
   \end{tabular}
   \label{tab:finaldistribution}
\end{table}
%
%%%%%%%%%%%%%%%%%%%%%%%%%%%%%%%%%%%%%%%%%%%%%%%%%%%%%%%%%%
%%

%%%%%%%%%%%%%%%%%%%%%%%%%%%%%%%%%%%%%%%%%%%%%%%%%%%%%%%%%
%
\section{A preliminary design of the drive bunch accelerator}
\subsection{Basic considerations}
\begin{figure}[!hhhhhtb]
   \centering
   \includegraphics[width=\columnwidth]{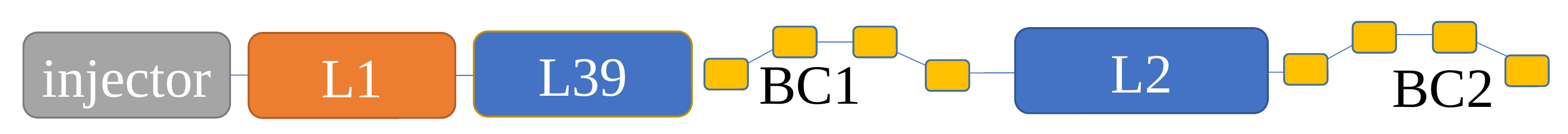}
   \caption{Block diagram of the drive bunch accelerator.}
   \label{fig:dum02}
\end{figure}

A block diagram of the drive bunch accelerator is shown in Fig.~\ref{fig:dum02}. It utilizes a commonly used configuration (see, for example, \cite{arthur_linac_2002,bosch_modeling_2008}) and includes a photocathode-gun-based injector, three linac sections, and two bunch compressors. Linac sections L1 and L2 are based on \SI{650}{\mega\hertz} superconducting (SRF) linac structures, and linac section L39  is based on \SI{3.9}{\giga\hertz} SRF structures. It is used for linearization of the electron distribution in the longitudinal phase space (LPS). Two bunch compressors are labeled as BC1 and BC2. Here we take advantage of the requirement to prepare the drive bunch with the energy chirp seen in Fig.~\ref{fig:bunchfinalback}(b) and move BC2 to the end of the linac, since we do not need to use the linac to remove the energy chirp after bunch compression. 

Using the known LPS distribution $\Phi_f(z_f, E_f)$ at the end of the accelerator,  we performed the one-dimensional (1D) backward tracking  proposed in \cite{cornacchia_formation_2006} to find the LPS distribution $\Phi_i(z_i, E_i)$ at the entrance of L1. We stopped at L1 where the beam energy is approximately \SI{50}{\MeV} considering that 1D tracking may not be reliable at lower energies where transverse and longitudinal space charge effects are stronger. The assumption is that at this point the backward tracking will produce a plausible $\Phi_i(z_i, E_i)$ that can be matched by the injector. Specifically, we constrained the peak current to $I \le\SI{300}{\ampere}$ and sought $\Phi_i(z_i, E_i)$ with minimal high-order correlations. 

A tracking program, {\sc twice}~\cite{tan_longitudinal_2018}, was developed for rapid prototyping of the longitudinal dynamics in the linac without accounting for a transverse motion. The program adopts an approach similar to that used in {\sc LiTrack} \cite{bane_litrack:_2005}. An important feature of {\sc twice} is its ability to perform backward tracking including time-reversal of the collective effect, see Appendix~\ref{app:twice}.  

The physics model implemented in {\sc twice} includes the geometric wakefields in the accelerating sections, longitudinal space charge effects (LSCs), and coherent synchrotron radiation (CSR).  The Green's functions needed for modeling of the geometric wakefield effects in the \SI{650}{\mega\hertz} and \SI{3.9}{\giga\hertz} linac sections were computed using the \textsc{echo} software and the empirical formula documented in Ref.~\cite{zagorodnov_wake_2004}. 

The backward tracking was performed to define  $\Phi_i(z_i, E_i)$ using  $\Phi_f(z_f, E_f)$, shown in Fig.~\ref{fig:bunchfinalback}.   
The following constraints for the accelerator components were observed. First, the BBU-mitigation scheme implemented in the CWA requires a drive bunch with the negative chirp $\frac{\partial E}{\partial z}<0$, which implies that the longitudinal dispersions of BC1 and BC2 should be $R_{56}^{(1)}>0$ and $R_{56}^{(2)}>0$, as we want to maintain a negative chirp throughout the entire accelerator.  
Second, a total energy gain of $\sim$\SI{950}{\mega\electronvolt} in the linac part after the injector is needed. Third, an overall compression factor of $\sim 10$ is required from two bunch compressors. 

In order to enforce all these constraints, {\sc twice} was combined with the multi-objective optimization framework {\sc deap}~\cite{deap}. The optimization was performed by analyzing the LPS distributions upstream of BC1 and L1 to extract the central energy of the beam slices at every $z$-coordinate and to fit the slice-energy dependence on $z$ with the polynomial
\begin{align}
E(z) = c_0 + c_1z+c_2z^2+c_3z^3, 
\end{align}
where $c_i$ are constants derived from the fit. The optimizer was requested to minimize the ratio $c_2/c_1$ in both locations. 

\subsection{Discussion of 1D simulation results}

\begin{figure}[!htb]
   \centering
   \includegraphics[width=\columnwidth]{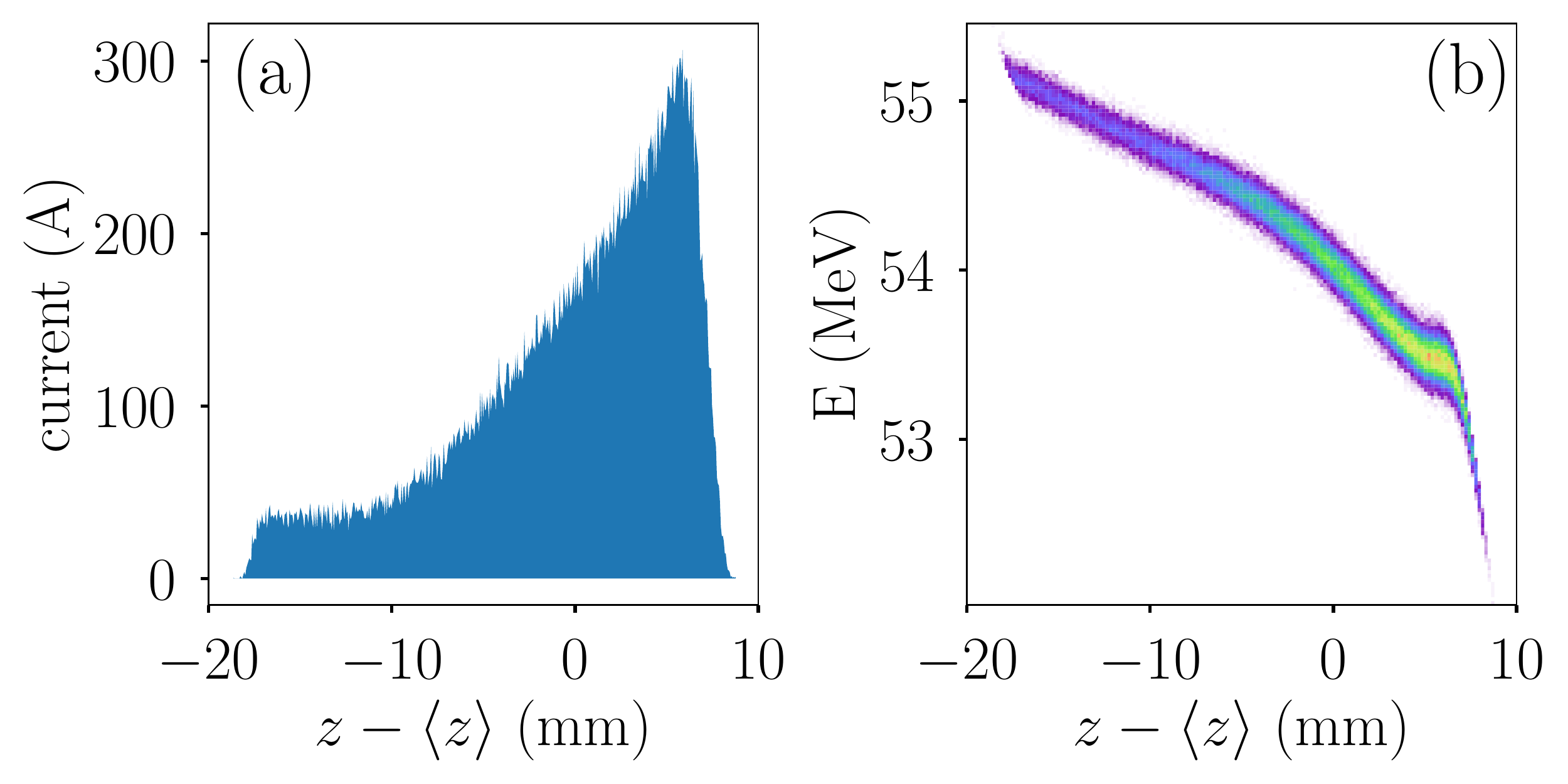}
   \includegraphics[width=\columnwidth]{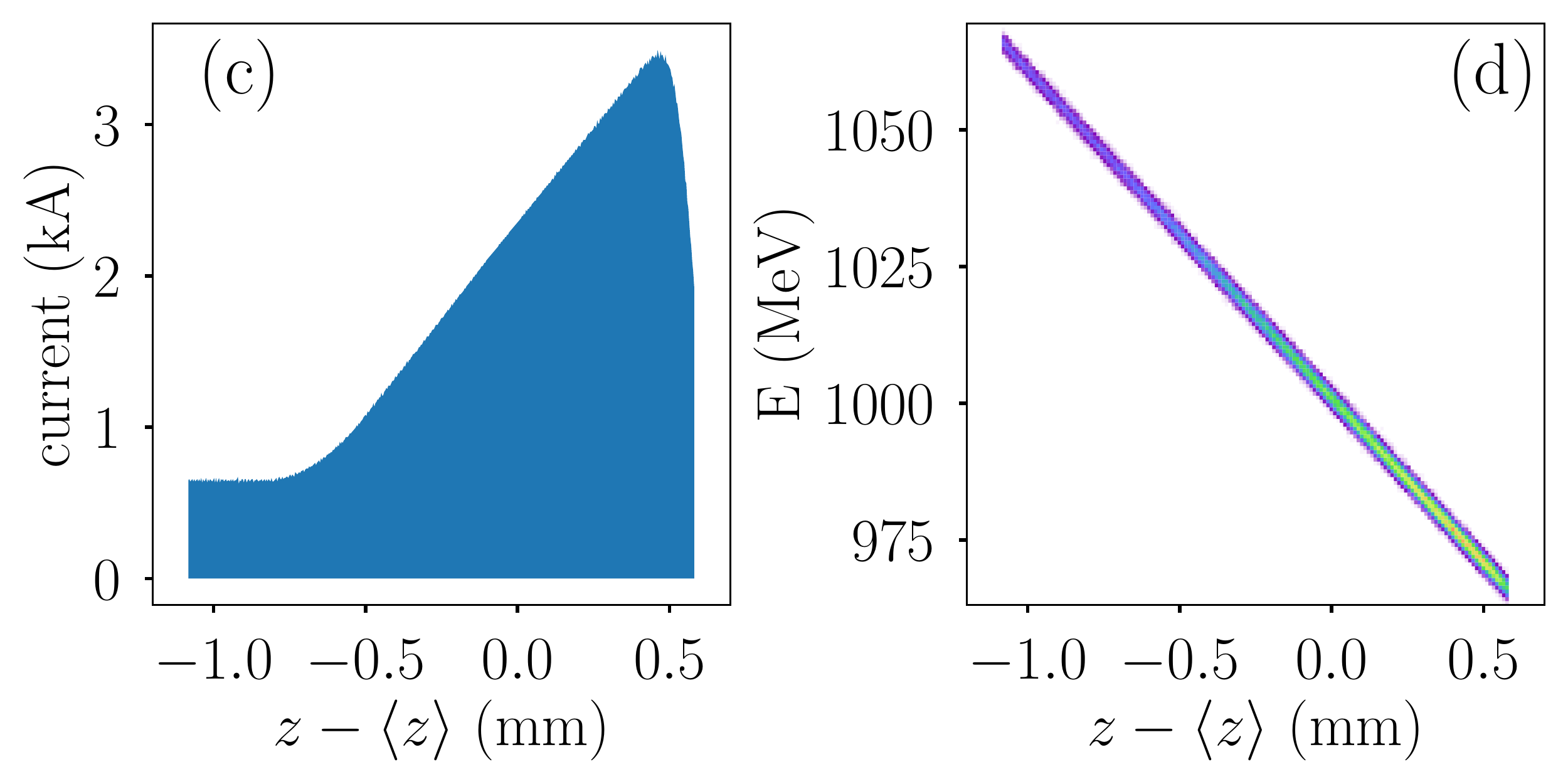}
   \caption{Current (a,c) and LPS (b,d)  distributions obtained from the backward-tracking optimization (a,b) and tracked up BC2 end (c,d) to confirm the agreement with the targeted distribution shown in Fig.~\ref{fig:bunchfinalback}.}
   \label{fig:dum03}
\end{figure}

A list of optimized accelerator settings found with {\sc twice} backward tracking is given in Table \ref{tab:02} and the resulting $\Phi_i(z_i, E_i)$ is shown in Fig.~\ref{fig:dum03}(a,b). The forward tracking using this distribution recovers $\Phi_f(z_f, E_f)$, as seen in Fig.~\ref{fig:dum03}(c,d). The excellent agreement between Fig.~\ref{fig:bunchfinalback}(a,b) and Fig.~\ref{fig:dum03}(c,d) demonstrates the ability of {\sc twice} to properly handle collective effects in both forward and backward tracking.

\begin{table}%[!hbt]
   \centering
   \caption{Optimized parameters from the one-dimensional model.}
   \begin{tabular}{lcc}
       \toprule
       \textbf{Parameter} & \textbf{Value} &   \textbf{Unit}  \\
\colrule
Accelerating voltage L1& \num{219.46} & \si{\mega\volt} \\

Phase L1 &  \num{17.81} & deg\\

Frequency L1 & 650 &  \si{\mega\hertz}\\

Accelerating voltage L39& \num{9.57} & \si{\mega\volt} \\

Phase L39 & \num{205.72} &  deg\\

Frequency L39 & \num{3.9} & \si{\giga\hertz}\\

\(R_{56}\) for bunch compressor 1 (BC1)& \num{0.1321} & \si\metre \\

\(T_{566}\) for bunch compressor 1 (BC1)& \num{-0.1581}   &  \si\metre \\

Accelerating voltage L2& \num{847.69} &  \si{\mega\volt} \\

Phase L2 & \num{28} & deg\\

Frequency L2 & \num{650} & \si{\giga\hertz}\\

\(R_{56}\) for bunch compressor 2 (BC2)& \num{0.1301} & \si{\metre} \\

\(T_{566}\) for bunch compressor 2 (BC2)& \num{0.22} & \si{\metre}\\
\colrule
   \end{tabular}
   \label{tab:02}
\end{table}

%%%

Each accelerator component serves a special role in obtaining the above-shown result. Linac section L1 provides energy gain and operates far from the crest acceleration to produce the required negative chirp. Linac section L39 corrects a second-order correlation between $E$ and $z$ imprinted on the bunch by the injector and L1 before it enters BC1. Linac section L2 operates even further off-crest to impart the necessary large chirp required for maintaining beam stability in the CWA. Both bunch compressors shorten the bunch lengths and impact the LPS distributions. The values of \(T_{566}\) selected in both bunch compressors ensure achieving $\Phi_f(z_f, E_f)$ despite the large energy chirp. The use of a negative \(T_{566}\) in BC1 and a positive \(T_{566}\) in BC2 enables the generation of a doorstep-like initial distribution without giving rise to a current spike, where $T_{566}$ has the effect of shifting the peak of current \cite{Charles:IPAC2017-MOPIK055,england_sextupole_2005}. In this paper, we adopt the convention that \(T_{566}\) with a negative (resp. positive) sign shifts the peak of current distribution to the tail (resp. head).%In this paper, we adopt the convention used in Refs.~\cite{charles_beam_2017,england_sextupole_2005}, where \(T_{566}\) with a negative (resp. positive) sign shifts the peak of current distribution to the tail (resp. head).

The result of the backward-tracking optimization provides only a starting point for obtaining a more realistic solution. For instance, the zigzag feature observed in the tail of the LPS distribution in Fig.~\ref{fig:dum03}(b) is challenging to create. In the following sections, we discuss how 1D  backward tracking results guide the design of a photocathode-gun-based injector and the downstream accelerator lattice.

\section{Injector design}\label{injSec}
Given the required initial LPS distribution obtained from the backward tracking, the next step is to explore whether such LPS distribution is achievable downstream of the injector; our approach relies on temporally shaping the photocathode laser pulse~\cite{tianzhe-AAC18}. 

The injector beamline was modeled using the particle-in-cell beam-dynamics program \textsc{astra}, which includes a quasi-static space-charge algorithm~\cite{floettmann_astra_2017}. The program was combined with the {\sc deap} multivariate optimization framework to find a possible injector configuration and the laser pulse shape that realize the desired final bunch distribution while minimizing the transverse-emittance downstream of the photoinjector. 

The injector configuration consists of a \SI{200}{\mega\hertz} quarter-wave SRF gun \cite{bisognano_progress_2011_2,bisognano_wisconsin_2013,legg_half_2008}, coupled to a \SI{650}{\mega\hertz} accelerator module composed of five 5-cell SRF cavities~\cite{tan_longitudinal-phase-space_2019}. The gun includes a high-T$_c$ superconducting solenoid~\cite{Nielsen-HTS-solenoid-2012} for emittance control. 

\begin{figure}[!htb]
   \centering
   \includegraphics*[width=\columnwidth]{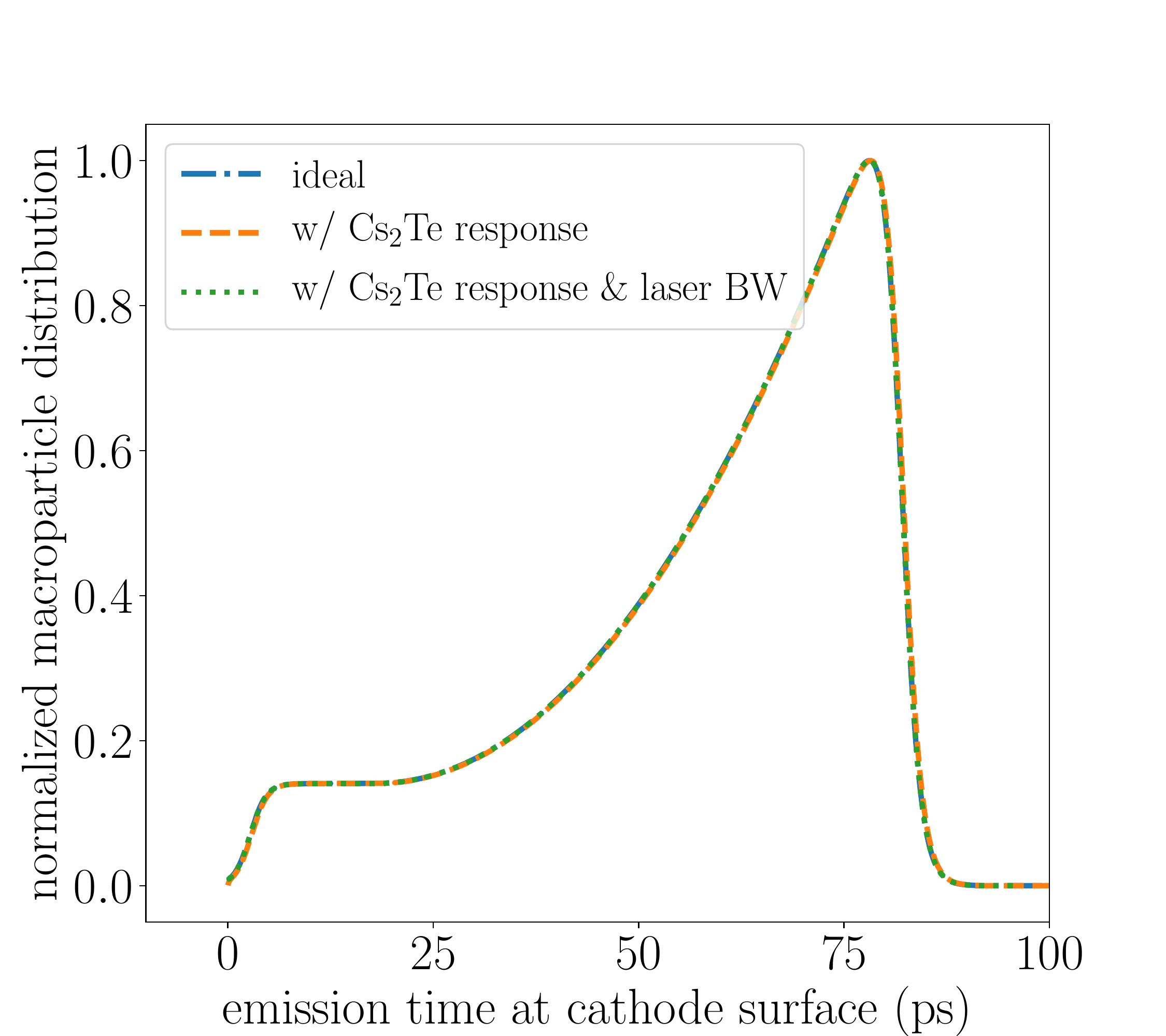}
   \caption{Programmed macroparticle distributions at the photocathode surface: for an optimized laser pulse (blue trace), taking into account the photocathode response (orange trace), and both the cathode response and finite bandwidth (BW) of the laser pulse (green trace). The laser bandwidth is taken to be $\delta f=\SI{2}{\tera\hertz}$.}
   \label{fig:respo}
\end{figure}

In the absence of collective effects, the photoemitted electron-bunch distribution mirrors the laser pulse distribution. In practice, image-charge and space-charge effects are substantial during the emission process and distort the electron bunch distribution. Consequently, devising laser-pulse distributions that compensate for the introduced deformities is critical to the generation of bunches with tailored current profiles.
The laser pulse distribution is characterized by \(I(t, r)=\Lambda(t)R(r)\), where \(\Lambda(t)\) and \(R(r)\) describe the laser temporal profile and the transverse envelope, respectively. In our simulation, we assumed the transverse distribution to be radially uniform $R(r)= H(r_c-r)$, where $H(r_c - r)$ is Heaviside step function and $r_c$ is the maximum radius. The temporal profile is parameterized as
\begin{align}
\label{eq:laser}
\Lambda(t) &= A f(t)S(a(t - f))S(-b(t-g))\text{, where}\\
f(t) &= 
\begin{cases}
  h,  & 0\leq t < c\\
  h+d(t-c)^{d-1},  & c\leq t\leq 1\\
  0, & \text{elsewhere}\nonumber
\end{cases},
\end{align}
where \(A\) is the normalization constant; and \(a\), \(b\) \(c\), \(d\), \(f\), \(g\), and \(h\) are the parameters controlling the bunch shape. The smooth edges at both ends are characterized by \(a\), \(b\), \(f\), \(g\) via the logistic function \(S(u)=1/(1+\mathrm{e}^{-u})\); \(c\) determines the length of the constant part of the laser pulse analogous to the length of the bunch head of the doorstep distribution; and \(h\) determines the relative amplitude of the constant laser pulse; see Fig.~\ref{fig:respo}. 
The overall shape resembles a smoothed version of the door-step distribution.
The laser-shape parameters introduced in Eq.~\eqref{eq:laser}, the laser spot size, the phase and accelerating voltage of all RF cavities, and the HTS solenoid peak magnetic field were taken as control parameters for the optimization algorithm. The beam kinetic energy was constrained not to exceed 60~MeV. In order to quantify the final distribution, we used the Wasserstein's distance~\cite{Xu:IPAC2019-TUPTS104} to quantify how close the shape of the simulated macroparticle distribution at the injector exit $I^{(o)}(z)$ was to the shape of the target macroparticle density distributions $I^{(t)}(z)$ obtained from backward tracking results. Specifically, the Wasserstein's distance is evaluated as
\begin{align} \label{eq:wasserstein}
{\cal D}=\sum_{i=1}{N_b} \frac{||I_i^{(t)}-I_i^{(o)}||}{N_b},
\end{align}
where $I_i^{(t,o)}$ are the corresponding histograms of the macroparticles' longitudinal positions over the interval $i$ defined as $[z_i+\delta z, z_i-\delta z]$, with $\delta z\equiv \frac{\mbox{max}(z)-\mbox{min}(z)}{N_b}$ being the longitudinal bin size and $N_b$ the number of bins used to compute the histogram. Additionally, we need to have a small beam transverse emittance. Hence, the Wasserstein's distance and the beam transverse emittance were used as our objective functions to be minimized.
\begin{table}[!hbt]
   \centering
   \caption{Optimized parameters for the injector and beam parameters at $s=\SI{11.67}{\metre}$ from the photocathode surface. The RF-cavity phases are referenced with respect to the maximum-energy phases. }
   \begin{tabular}{lcc}
       \toprule
       \textbf{Parameter} & \textbf{Value} &   \textbf{Unit}  \\
       \colrule
Laser spot radius & $2.810$ & mm \\
Laser duration & 91 & ps\\
RF gun peak electric field & 40 &  MV/m\\
RF gun phase & $1.71$ &  deg\\
Cavity C1 peak electric field & $13.25$ &  MV/m \\
Cavity C1 phase & 11.28 &  deg\\
Cavity C2 phase & -15.05&  deg\\
Cavities C2 to C5 peak electric field & $20$ &  MV/m \\
Cavities C3 to C4 phase & 0 &  deg\\
Cavity C5 phase & 20 & deg\\
Cavity C1 distance from the photocathode & 2.67 & m\\
Solenoid B-field & 0.2068 & T\\
Shape parameter $a$ & 	93.55& - \\ 
Shape parameter $b$ & 	80.70& - \\  
Shape parameter $c$ & 	0.196& - \\
Shape parameter $d$ & 	3.044& - \\
Shape parameter $f$ & 	0.030& - \\
Shape parameter $g$ & 	0.900& - \\
Shape parameter $h$ & 	0.207& - \\
\hline 
Final beam energy & $58.7$ & MeV \\
Final beam bunch length & $7.06$ & mm \\
Final beam transverse emittance & $8.36$ & \si{\micro\metre} \\
Final beam rms radius & $1.64$ & mm \\
\colrule
   \end{tabular}
   \label{tab:03}
\end{table}
%%%%%
\begin{figure}[b]
   \centering
   \includegraphics*[width=\columnwidth]{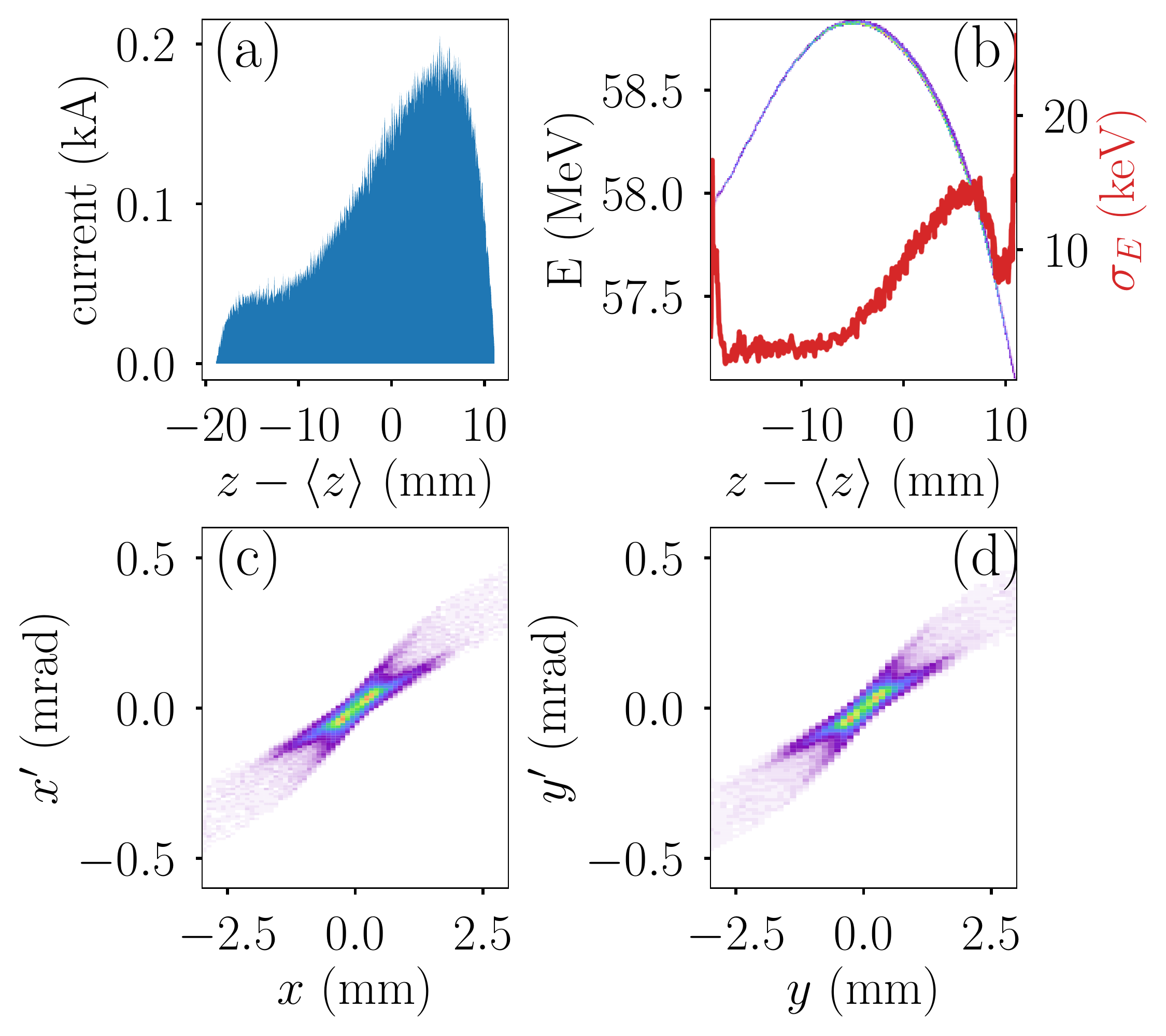}
   \caption{Current profile (a) with associated LPS (b), and horizontal (c) and vertical (d) phase-space distributions simulated with \textsc{Astra} at the end of the photoinjector ($11.67$~m from the photocathode). In plot (b), the red trace represents the slice RMS energy spread $\sigma_E$.}
   \label{fig:begin_astra}
\end{figure}
\begin{figure}[!htb]
   \centering
   \includegraphics*[width=0.5\textwidth]{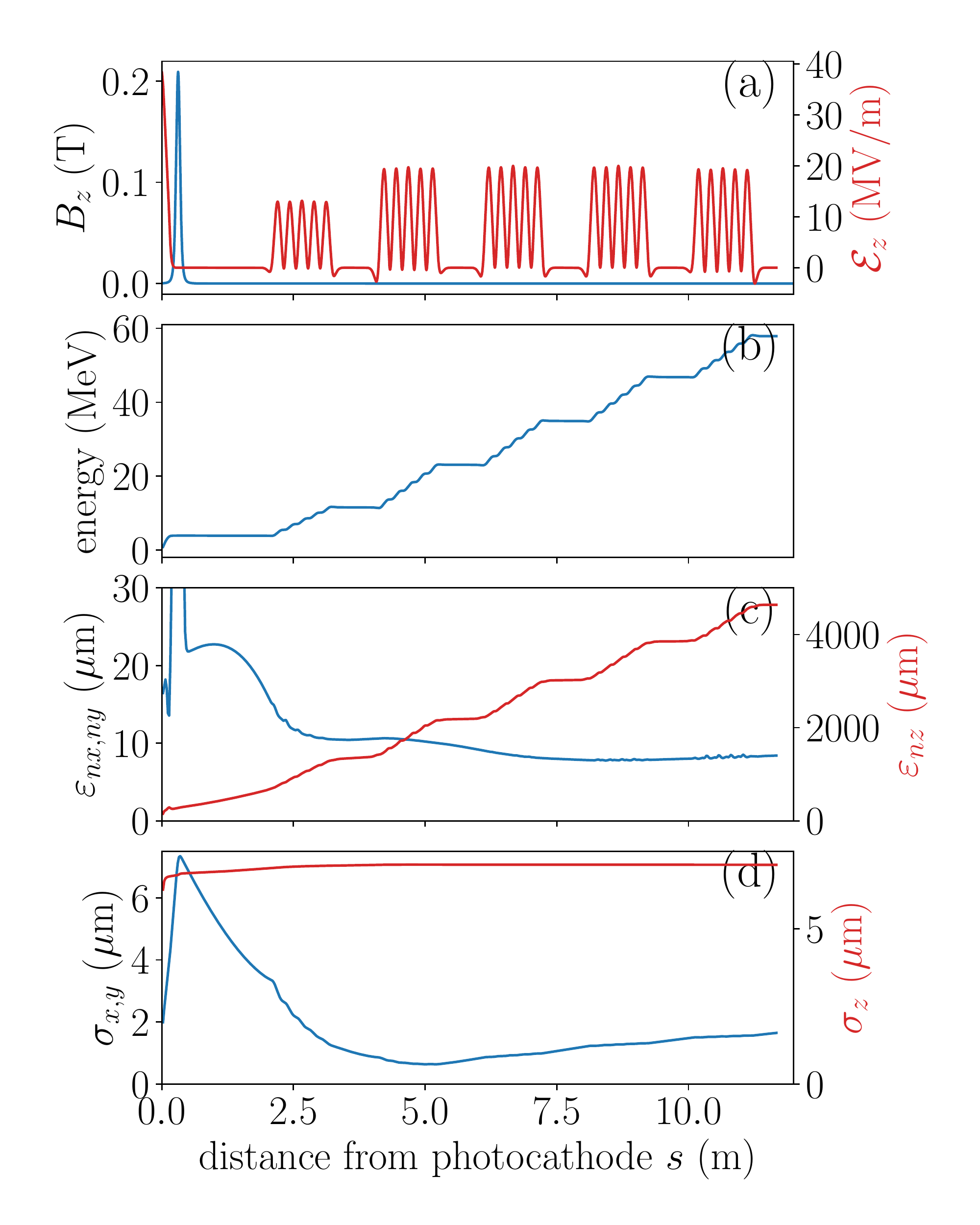}
   \caption{Axial electric  $\mathcal E_z$ (red trace) and magnetic $B_{z}$ (blue trace) fields experienced by the reference particle as it propagates along the optimized photoinjector (a) with corresponding kinetic energy (b), transverse (blue) and longitudinal (red) beam emittances (c), and sizes (d) evolving along the injector. }
   \label{fig:xemit}
\end{figure}

An example of the optimized injector settings is summarized in Table~\ref{tab:03}, and the evolution of the associated beam parameters along the beamline are presented in Figs.~\ref{fig:begin_astra} and~\ref{fig:xemit}. The final bunch distributions 11.5~m downstream of the photocathode appears in Fig.~\ref{fig:begin_astra}. The beam transverse phase space indicates some halo population. Ultimately, an alternative laser-shaping approach implementing a spatiotemporal-tailoring scheme could provide better control over the transverse emittance while producing the required shaped electron beams~\cite{tianzhe-AAC18}. We also find, as depicted in Fig.~\ref{fig:begin_astra}, that the current distribution tends to have a peak current lower than that desired from the backward tracking result shown in Fig.~\ref{fig:dum03}. Although higher currents are possible, they come at the expense of transverse emittance. Consequently, the distribution generated from the injector was considered as an input to the one-dimensional forward tracking simulations. Iterations of one-dimensional forward tracking simulation studies were done to further cross-check accelerator parameters needed for the beam-shaping process. We especially found that the desired final bunch shape at \SI{1}{\giga\electronvolt} can be recovered by altering the L39 phase and amplitude. Furthermore, the small slice rms energy spread $\sigma_{E} < 10$~keV simulated from the injector [see Fig.~\ref{fig:dum03}(b)] renders the bunch prone to microbunching instability. Consequently, a laser heater is required to increase the uncorrelated energy spread. 
\begin{figure}[!htb]
   \centering
   \includegraphics*[width=\columnwidth]{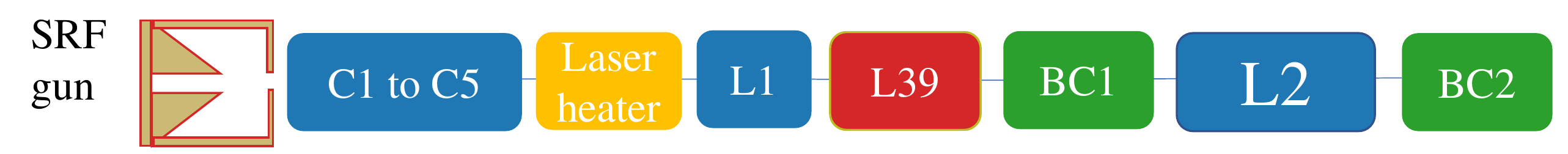}
   \caption{Updated accelerator design, with the addition of the injector beamline and a laser heater section.}
   \label{fig:dum06}
\end{figure}

The correspondingly revised diagram of the accelerator beamline shown in Fig. \ref{fig:dum06} was used as a starting point to investigate the performance of the proposed bunch-shaping process with {\sc elegant} tracking simulations taking into account the transverse beam dynamics.

Another challenge associated with the bunch formation pertains to the temporal resolution of the bunch shaping process. Ultimately, the laser pulse shape can only be controlled on a time scale $\delta t \ge 1/(2\pi \delta f_L)$ limited by the bandwidth of the photocathode laser $\delta f_L$. Contemporary laser systems are capable to $\delta t \le \SI{150}{\femto\second}$ (RMS)~\cite{Gilevich-lcls-ii_2020}. Additionally, the electron bunch shape is also affected by the time response of the photoemission process. Given the required charge of $\sim$\SI{10}{\nano\coulomb}, we consider a Cs$_2$Te photocathode with temporal response numerically investigated in Ref.~\cite{FERRINI199821, piot_formation_2013}. Recent measurements confirm that Cs$_2$Te has a photoemission response time below \SI{370}{\femto\second}~\cite{AryshevCsTe2017}. Figure~\ref{fig:respo} compares the optimized ideal laser pulse shape described by Eq.~\eqref{eq:laser} with the cases when the photocathode response time and the laser finite bandwidth are taken into account. The added effects have an insignificant impact on the final distribution due to relatively slow temporal variations in the required peak current distribution. 

%%%%%%%%%%%%%%%%%%%%%%%%%%%%%%%%%%%%%%%%%%%%%%%%%%%%%%%%%%%%%%%%%%%%%%
\section{Final accelerator design}

The strawman accelerator design developed with the help of 1D simulations provides guidance for the final design of the accelerator. 
\subsection{Accelerator components}
\paragraph{Linacs:} 
For the \SI{650}{\mega\hertz} L1 and L2 SRF linacs we adopted cryomodules proposed for the PIP-II project~\cite{jain_650_2017}. The linac L1 consists of two cryomodules, and L2 has eight cryomodules.  Each cryomodule includes six cavities containing five cells.  We assume that in CW operation each cavity provides up to \SI{20}{\mega\volt/\metre} average accelerating gradients. The quadrupole magnet doublets are located between cryomodules and produce a pseudo-periodic oscillation of the betatron functions. The two cavities used in the \SI{3.9}{\giga\hertz} L39 SRF linac are similar to the cavity described in Ref.~\cite{zagorodnov_wake_2004}. 

\paragraph{Bunch compressors:}  
We use an arc-shaped bunch compressor consisting of a series of FODO cells, where each cell contains two quadrupoles and two dipole magnets. The latter configuration nominally provides a positive \(R_{56}\) \cite{mitri_transverse_2015, akkermans_compact_2017, mitri_bunch_2018,chao_handbook_2013} 
\begin{align}
R_{56} 
\simeq\frac{\theta^2_\text{total}L_\text{total}}{4N^2_\text{cell}\sin^2{(\psi_x/2)}}\;,
\end{align}
where $\theta_\text{total}$ is the total bending angle, $L_\text{total}$ is the total path length, $N_\text{cell}$ is the total number of FODO cells, and $\psi_x$ is the horizontal phase advance per cell. 
%%%%%%%%%%%%%%%%%%%%%%%%%
%%
%%%%%%%%%%%%%%%%%%%%%%%%%
%
The dipole magnet bending angles can be used to tune the \(R_{56}\). 
The bending angle or dipole polarity from cell to cell does not need to be identical, but the number of cells should be selected to realize a  phase advance $\psi_{x,\text{total}}=2n\pi$ (with $n$ integer) over the compressor to achieve the first-order achromat.

The second-order longitudinal dispersion produced by the bunch compressor is given by~\cite{robin_quasi-isochronous_1993,williams_arclike_2020}
\begin{align}\label{eq:hod}
T_{566}= \int_0^L \left[\frac{\eta_{1,x}(s')}{\rho(s')}+\frac{\eta_{x}'^2(s')}{2} \right]\mathrm ds',
\end{align}
where $L$ is the length of the beamline, $\rho$ is the bending radius, $\eta_{1,x}(s)\equiv ({E_0}^2/2) \partial^2 x(s)/\partial E^2$ is the second-order horizontal dispersion function, and $\eta_{x}'(s)$ is the derivative of the dispersion function.
We incorporate 12 sextupole magnets to control the $T_{566}$ and 12 octupole magnets to cancel the third-order longitudinal transfer-map element $U_{5666}$ computed over BC1. If needed, a non-vanishing value of $U_{5666}$ can enable higher-order control over the LPS correlation \cite{Charles:IPAC2017-MOPIK055}. 
\begin{figure}[!!!!!!hhhhhhhhtb]
   \centering
   \includegraphics*[width=\columnwidth]{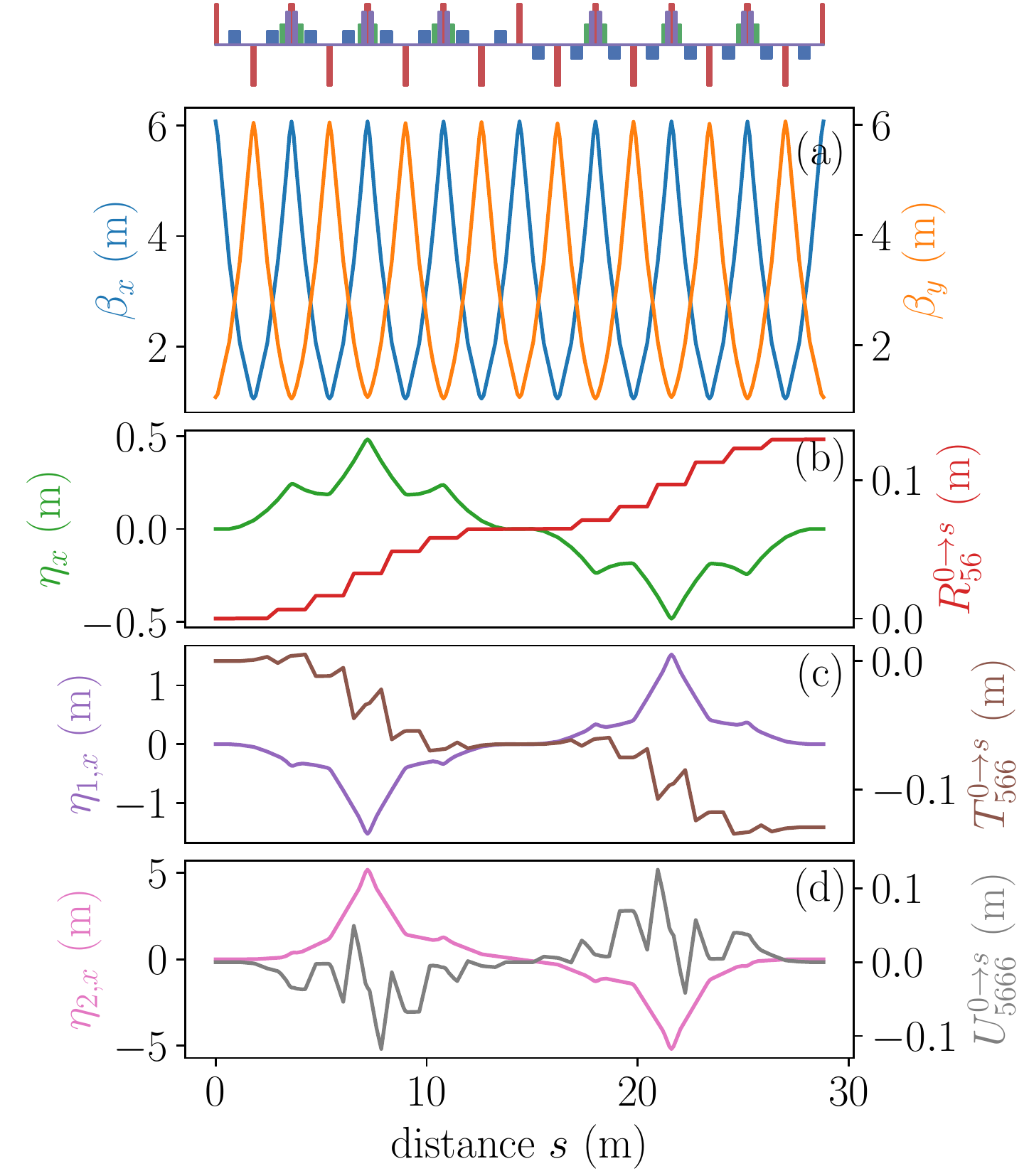}
   \caption{Layout of bunch compressor BC1 (top diagram) with evolution of associated betatron function (a) and pertinent linear (b), second-order (c), and third-order (d) transfer-map elements along the beamline (with $s=0$ corresponding to the beginning of BC1). In plots (b-d) the left and right axes refer to the horizontal chromatic functions $\eta_{i,x}$ and accumulated longitudinal transfer-map elements from $0$ to location $s$ along BC1. In the top diagram the red, blue, green, and purple rectangles correspond, respectively, to quadrupole,  dipole, sextupole, and octupole magnets.}
   \label{fig:r56bc1}
\end{figure}

The sextupole and octupole magnets are also used to zero the chromatic transfer-map elements $T_{166},T_{266}$, and $U_{1666}$, resulting in the bunch compressors being achromatic up to the third order. 

\begin{figure}[!htb]
   \centering
   \includegraphics*[width=\columnwidth]{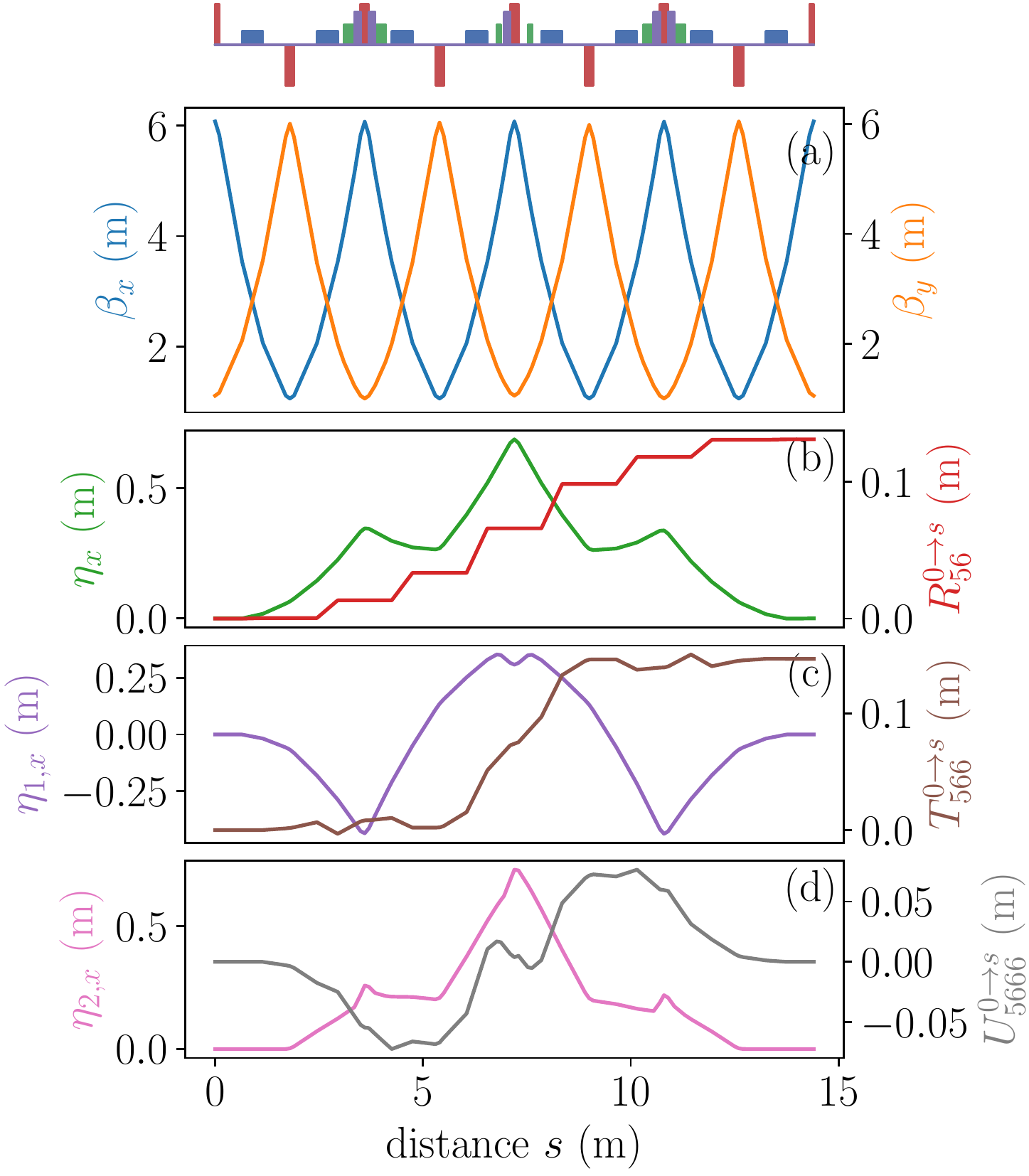}
   \caption{Layout of bunch compressor BC2 (top diagram) with evolution of associated betatron function (a) and pertinent linear (b), second-order (c), and third-order (d) transfer-map elements along the beamline (with $s=0$ corresponding to the beginning of BC2). In plots (b-d) the left and right axes refer to the horizontal chromatic functions $\eta_{i,x}$ and accumulated longitudinal transfer-map elements from $0$ to location $s$ along BC2. The top diagram follows the same conventions as in Fig.~\ref{fig:r56bc1}.}
   \label{fig:r56bc2}
\end{figure}

\begin{figure}[!htb]
   \centering
   \includegraphics*[width=0.85\columnwidth]{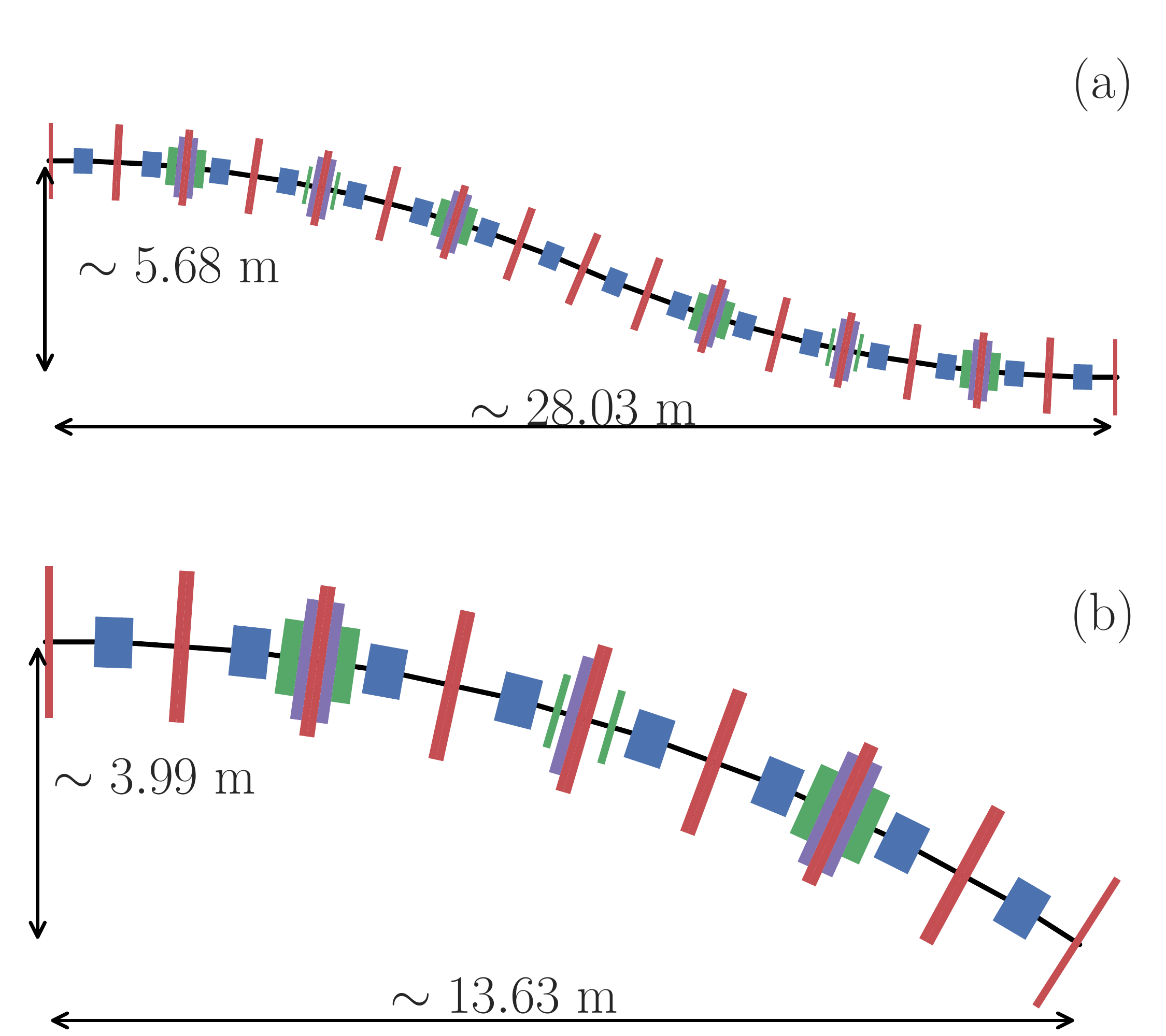}
   \caption{The geometry of the bunch compressors BC1 (a) and BC2 (b), where red, blue, green, and purple rectangles are quadrupoles, dipoles, sextupoles, and octupoles magnets, respectively.}
   \label{fig:bc1layout}
\end{figure}

Figure~\ref{fig:r56bc1} displays the BC1 configuration along with the evolution of the betatron functions and relevant horizontal chromatic ($\eta_x$, $\eta_{1,x}$, and $\eta_{2,x}$) and longitudinal accumulated transfer-map elements ($R_{56}^{0\rightarrow s}$, $T_{566}^{0\rightarrow s}$, and $U_{5666}^{0\rightarrow s}$) up to third order as a function of the beamline coordinate $s$. It has two arcs, one bending the beam trajectory by $22.92^{\circ}$ and another one bending it back. Each bending magnet has the bending angle $\theta=2.865^{\circ}$. This design eases the requirement on the sextupole-magnet strength required to provide a  $T_{566}<0$; see Table~\ref{tab:02}. The strengths of the sextupole magnets were optimized using \textsc{elegant} to achieve the required \(T_{566}\) across BC1 while obtaining a second-order achromat by constraining $T_{166}=T_{266}=0$. The three pairs of sextupole magnets in the second arc are mirror-symmetric to the first three pairs, with opposite-polarity magnet strengths. During the design process, the first pair of sextupole magnets was inserted close to the region of the first arc with the highest dispersion for tuning the desired $T_{566}$; its mirror symmetry pair was placed in the second arc and separated by \(2\pi\) phase advance. Another two pairs of sextupole magnets were subsequently inserted for tuning $T_{166}$. Similarly, their mirror symmetry pairs were separated by \(2\pi\) phase advance. Finally, six pairs of octupole magnets were inserted to zero the overall $U_{i666}$ $i=1,2,5$ transfer-map elements, where the same design process was employed.
%
%%%%%%%%%%%%%%%%%%% BC2
The BC2  compressor requires both \(R_{56}\) and \(T_{566}\) to be positive, which is naturally provided by the arc bunch compressor introduced earlier. It has a total bending angle of \SI{32.63}{\degree}, and each dipole has a bending angle of \SI{4.079}{\degree}.  Similar to BC1, we used sextupole- and octupole-magnet families to adjust both \(T_{566}\) and \(U_{5666}\) and produce the third-order achromat. The BC2 lattice appears in Fig.~\ref{fig:r56bc2} along with the evolution of the betatron functions and relevant chromatic elements. 
Finally, the layout of the two bunch compressors is presented in Fig.~\ref{fig:bc1layout}.

\paragraph{Matching sections:} All accelerator components are connected using matching sections composed of quadrupole magnets and drift spaces. 

\begin{figure*}[!!!!!hhhhhhhtb]
   \centering
   \includegraphics*[width=\linewidth]{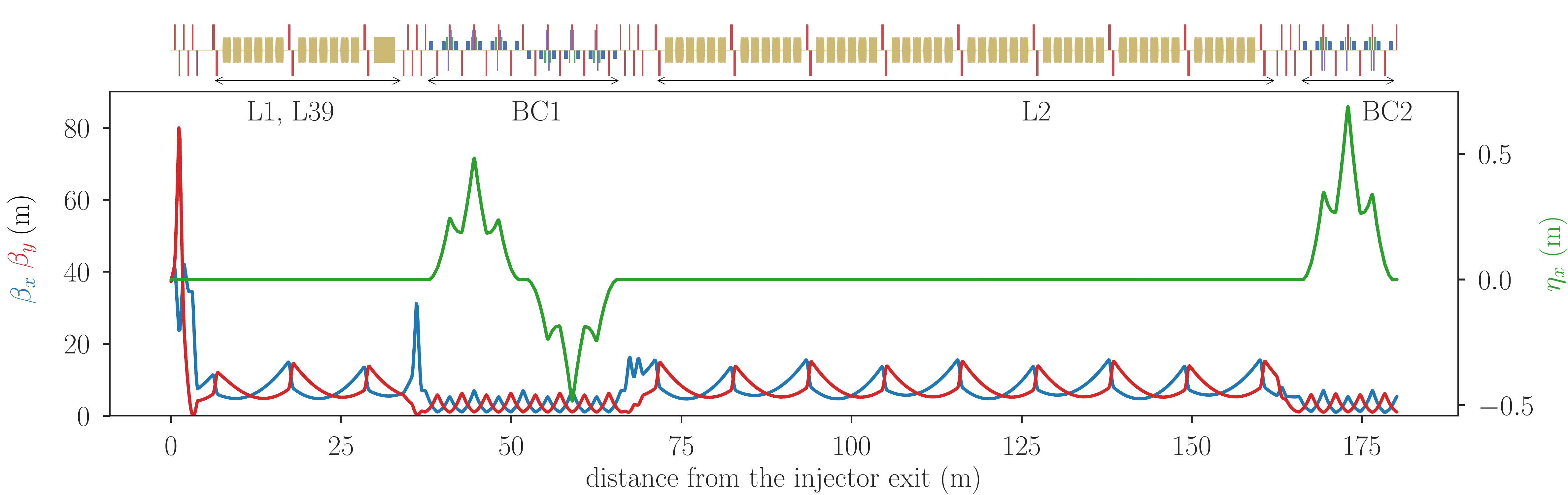}
   \caption{Evolution of the betatron (left axis) and horizontal dispersion (right axis) functions along the proposed linac. The vertical dispersion is zero throughout the linac. The magnetic-lattice color coding for the element follows Fig.~\ref{fig:r56bc1} with the accelerating cavities shown as gold rectangles.}
   \label{fig:07}
\end{figure*}

The evolution of the betatron functions from the injector exit up to the end of BC2 appears in Fig.~\ref{fig:07}. Throughout the entire accelerator, the betatron functions are maintained to values $\beta_{x,y}<\SI{30}{\metre}$. 

\subsection{Tracking and optimization} \label{sec:elegantSimu}
The beam distribution obtained at the exit of the injector was used as input to {\sc elegant} for tracking and optimization.  
\begin{figure}[!hhtb]
   \centering
   \includegraphics*[width=\columnwidth]{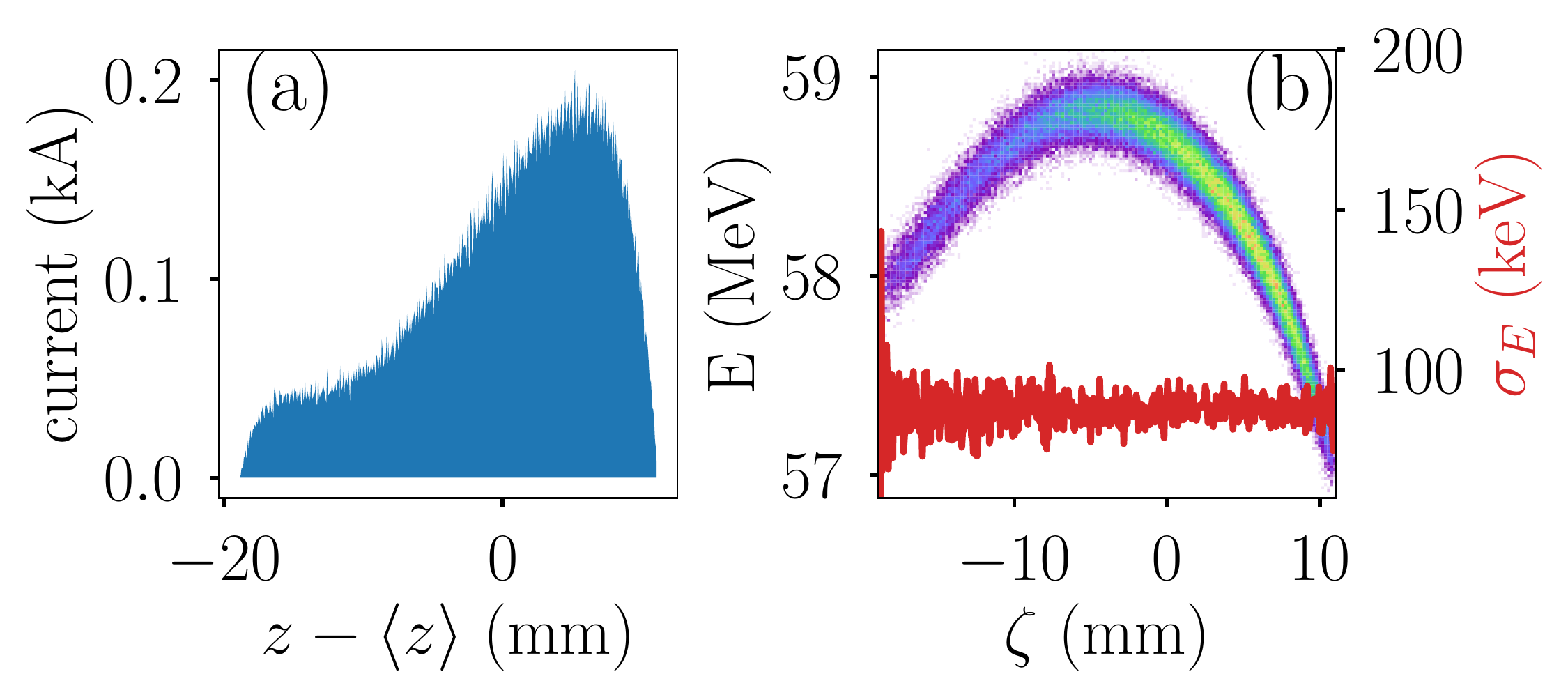}
   \caption{Current profile (a) and associated LPS (b) distributions simulated with \textsc{Astra} at the end of the photoinjector (see Fig.~\ref{fig:begin_astra}) with added uncorrelated fraction energy spread following a Gaussian distribution with RMS spread $\sigma_{E}/E=1.5\times10^{-3}$.  In plot (b) the red trace represents the slice RMS energy spread $\sigma_E$. }
   \label{fig:heat}
\end{figure}
We found that we need to increase the slice energy spread to $\sim\SI{75}{\kilo\electronvolt}$ using the laser heater to suppress the microbunching instability~\cite{SALDIN2004355,huang_suppression_2004}. 
However, in this study, we  numerically added  random noise with Gaussian distribution to the macroparticles' energy   using the {\sc scatter} element available in {\sc elegant}. Thus, Fig.~\ref{fig:heat} shows the actual LPS distribution used at the beginning of the accelerator in tracking studies. 
%%%%%%%%%%%%%%%%%%%%%%%%%%%%%%%%%%%
%

The accelerator settings obtained with {\sc twice} were used as a starting point in the accelerator optimization including transverse effects. The fine-tuning of the above-described accelerator components was accomplished using \textsc{elegant}. A multi-objective optimization was applied to determine the twelve accelerator parameters controlling the longitudinal dynamics, i.e., voltages and phases of L1, L2,   L39, and values of \(R_{56}\), \(T_{566}\) in two bunch compressors. The resulting beam distribution obtained downstream of BC2 was then used to compute the wakefield generated in a \SI{180}{\giga\hertz} corrugated waveguide considered for the role of the wakefield accelerator in ~\cite{Asiy-napac_2019}. The resulting peak accelerating field and transformer ratio were then adopted as objective functions to be maximized with the accelerator parameters as control variables. The trade-off between peak accelerating field and transformer ratio was quantified in Eq.~(30) of Ref.~\cite{baturin_upper_2017}, hence providing a good measure to verify whether our optimization reaches the optimal Pareto front.
\begin{table}[!hbt]
   \centering
   \caption{Main accelerator parameters and beam parameters at the end of BC2.}
   \begin{ruledtabular}
   \begin{tabular}{lcc}
       \textbf{Parameter} & \textbf{Value} &   \textbf{Unit}  \\
       \colrule
Accelerating voltage L1& $193.22$ & \si{\mega\volt} \\
Phase L1 & 21.64 & deg\\
Frequency L1 & 650 &  \si{\mega\hertz}\\
Accelerating voltage L39& $9.73$ &  \si{\mega\volt} \\
Phase L39 & 202.52 &  deg\\
Frequency L39 & 3.9 & \si{\giga\hertz}\\
\(R_{56}\) for bunch compressor 1 (BC1)& $0.1294$ & \si{\metre} \\
\(T_{566}\) for bunch compressor 1 (BC1)& -0.1294  &  \si{\metre}\\
\(U_{5666}\) for bunch compressor 1 (BC1)& 0  &  \si{\metre}\\
Accelerating voltage L2& $857.92$ &  \si{\mega\volt} \\
Phase L2 & 26.05 & deg\\
Frequency L2 & 650 & \si{\mega\hertz}\\
\(R_{56}\) for bunch compressor 2 (BC2)& $0.1312$ & \si{\metre}\\
\(T_{566}\) for bunch compressor 2 (BC2)& $0.1465$ & \si{\metre}\\
\(U_{5666}\) for bunch compressor 2 (BC1)& 0  &  \si{\metre}\\
\colrule
Final beam energy & $998$ & \si{\mega\electronvolt} \\
Final beam bunch length & \num{414} & \si{\micro\metre} \\
Final beam normalized emittance, \(\varepsilon_{nx}\) & 31 & \si{\micro\metre}\\
Final beam normalized emittance, \(\varepsilon_{ny}\) & 12 & \si{\micro\metre}\\
Peak accelerating wakefield $|\mathcal{E_+}|$ & \num{94.3} & \si{\mega\volt/\metre}\\
Peak decelerating wakefield $|\mathcal{E_-}|$ & \num{18.8} & \si{\mega\volt/\metre}\\
Transformer ratio $\mathcal{R}$ & \num{5.0} & -\\
   \end{tabular}
   \end{ruledtabular}
   \label{tab:04}
\end{table}
%%%%%
\begin{figure}[!htb]
   \centering
\includegraphics*[width=\columnwidth]{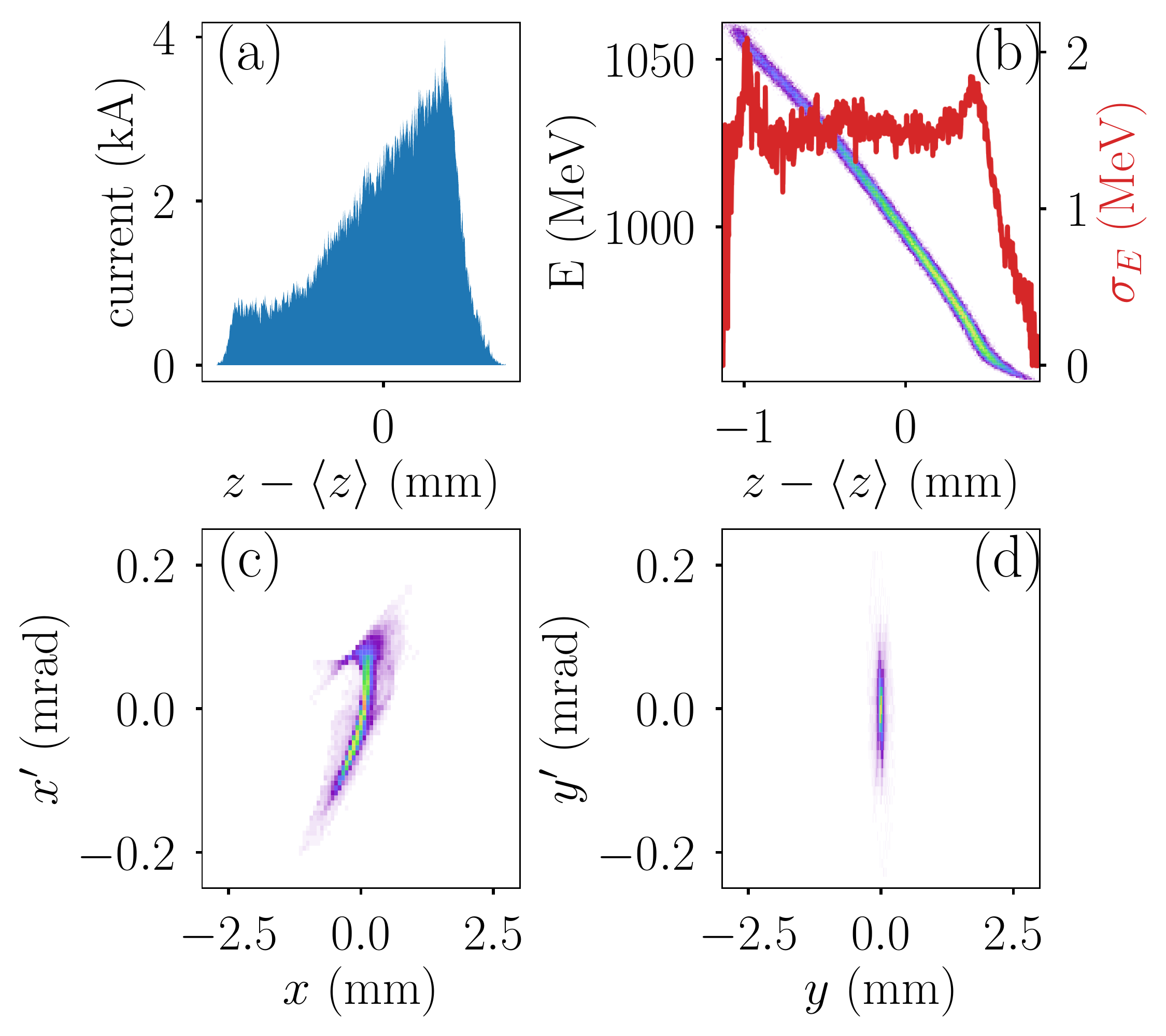}
   \caption{Current (a) with associated LPS (b), and transverse horizontal (c) and vertical (d) phase-space distributions simulated with \textsc{elegant} at the end of BC2 using the optimized linac and bunch-compressor settings summarized in Table~\ref{tab:04} and the injector distributions from Fig.~\ref{fig:heat}. In plot (b) the red trace represents the slice RMS energy spread $\sigma_E$.}
   \label{fig:dum09}
\end{figure}
%%%%
\begin{figure}[!!!hb]
\centering
\includegraphics[width=\columnwidth]{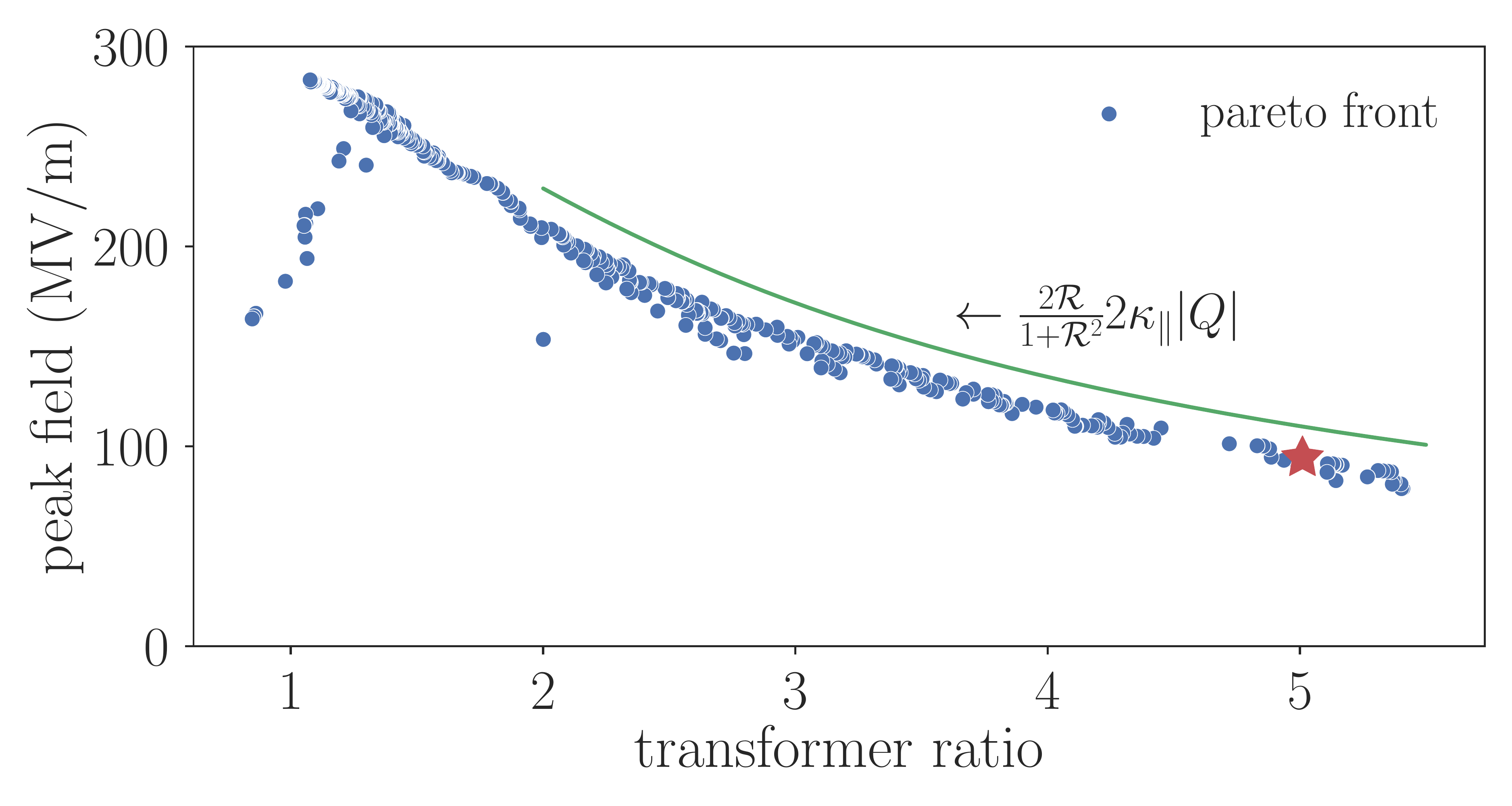}
\caption{Comparison of the Pareto front with the analytical trade-off curve between the peak field and transformer ratio described by Eq.~(30) of Ref.~\cite{baturin_upper_2017}. Each blue dot represents a numerically simulated configuration with the red star representing the configuration with parameters listed in Table~\ref{tab:04}.}
\label{fig:pareto}
\end{figure}
The optimal accelerator settings and final beam parameters are summarized in Table \ref{tab:04}. The LPS distribution at the end of the accelerator is shown in Fig.~\ref{fig:dum09}. We also calculated that the $\sim\SI{1}{\giga\electronvolt}$, \SI{10}{\nano\coulomb} electron bunch having this distribution produces a peak wakefield of \SI{94.26}{\mega\volt/\metre} with a transformer ratio of \num{5} propagating in a corrugated waveguide. Figure~\ref{fig:pareto} demonstrates that our optimization has reached the optimal set of solutions, where the Pareto front closely follows the analytically calculated tradeoff curve~\cite{baturin_upper_2017}. The obtained current profile produces a wakefield amplitude $\sim 15\%$ lower than the one expected from the ideal distribution for a transformer ratio ${\cal R} \simeq 5$. Such an agreement gives confidence in our optimization approach based on the trade-off between peak accelerating field and transformer ratio. The simulations also indicate that the horizontal transverse emittance increases to \(\varepsilon_{nx}=31\)\si{\micro\metre} due to the CSR and chromatic aberrations in the electron bunch having large correlated energy variations. Although significant, this emittance dilution is still acceptable.

\begin{figure}[!htb]
   \centering
   \includegraphics*[width=\columnwidth]{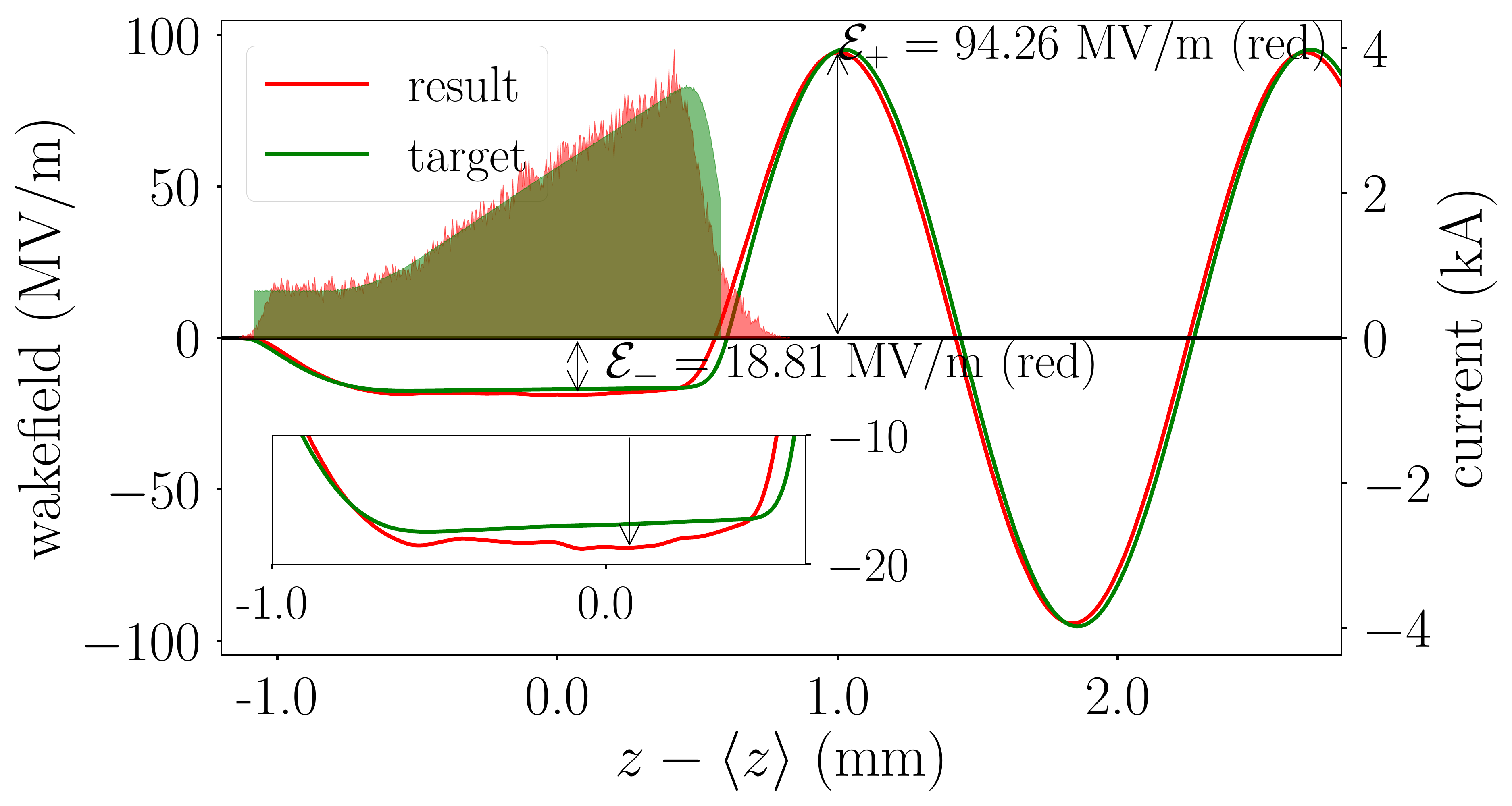}
   \caption{Target (from Fig.~\ref{fig:bunchfinalback}) and optimized final current distributions (respectively shown as green- and red-shaded curves) with associated wakefields (respectively displayed as green and red traces). The transformer ratio for the simulated distribution is ${\cal R}=5.0$.}
   \label{fig:dum10}
\end{figure}

Our main result is shown in Fig.~\ref{fig:dum10}. It compares the final distribution and wakefield with that of the target distribution and wakefield from Fig.~\ref{fig:bunchfinalback}.  A good agreement manifests that, indeed, the drive electron bunch with a highly asymmetric peak current profile can be obtained without employing the collimators. 
%%%%%%
%%

A comparison of Tables~\ref{tab:03} and \ref{tab:04} indicates that the final accelerator settings optimized by {\sc elegant} deviate less than 10$\%$ from those obtained with {\sc twice}. It justifies the strategy taken in this study to solve the difficult problem of formation of temporally shaped electron bunches for a beam-driven collinear wakefield accelerator in two steps.

The nonlinear correlation observed in the tail of the LPS distribution downstream of BC2 [see Fig.~\ref{fig:dum09}(b, blue trace)] originates from the CSR. As the beam is compressed inside the bunch compressors, its tail experiences a stronger CSR force due to its peak current being higher than the rest of the bunch. It is worth noting that \textsc{elegant} uses a 1D projected model to treat the CSR effect. The applicability of such a 1D treatment is conditioned by the Derbenev's criterion \cite{derbenev_microbunch_1995}, which suggests that projecting the bunch distribution onto a line-charge distribution may overestimate the CSR force, particularly when the bunch has a large transverse-to-longitudinal aspect ratio ${\cal A}(s) \equiv \left(\sigma_x(s)/\sigma_z(s)\right) \sqrt{\left(\sigma_x(s)/\rho(s)\right)}$. In our design, the condition ${\cal A} \ll 1$ was not rigorously followed (but rather the softer condition ${\cal A} < 1$ was achieved), suggesting that the impact of CSR may be overestimated in some regions of the bunch compressors.\\

\begin{figure}[!htb]
   \centering
   \includegraphics*[width=\columnwidth]{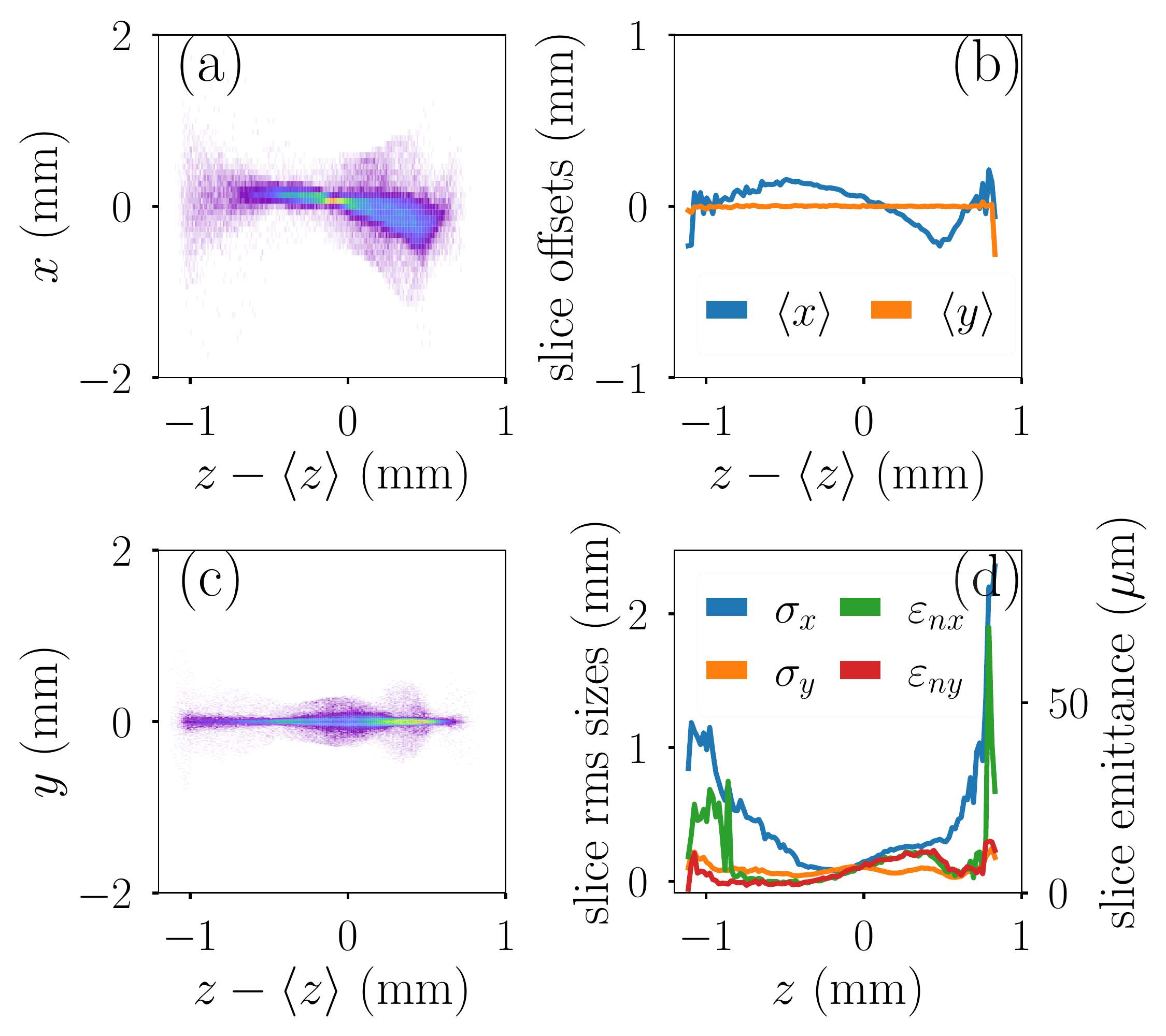}
   \caption{Final $(z,x)$ (a) and $(z,y)$ (c) beam distributions corresponding to the data shown in Fig.~\ref{fig:dum09}, and slice analysis for positions $\mean{x}$ and $\mean{y}$ (b) and RMS beam size and emittances (d).}
   \label{fig:correlationzXY}
\end{figure}

We also note that the final beam distribution exhibits significant longitudinal-horizontal ($z-x$) correlations due to CSR effects; see Fig.~\ref{fig:correlationzXY}. Although the associated projected-emittance dilution is tolerable, the electrons in the longitudinal slices with the horizontal offsets seen in Fig.~\ref{fig:correlationzXY}(c) will excite transverse wakefields in the CWA and ultimately seed the BBU instability. These offsets come from CSR-induced energy loss occurring in the BC2 that breaks the achromatic property of this beamline.  Understanding the impact of this distribution feature in the CWA linac along with finding mitigation techniques is a current research focus. 

\subsection{Impact of errors} \label{sec:erros}
\begin{figure}[!htb]
   \centering
   \includegraphics*[width=\columnwidth]{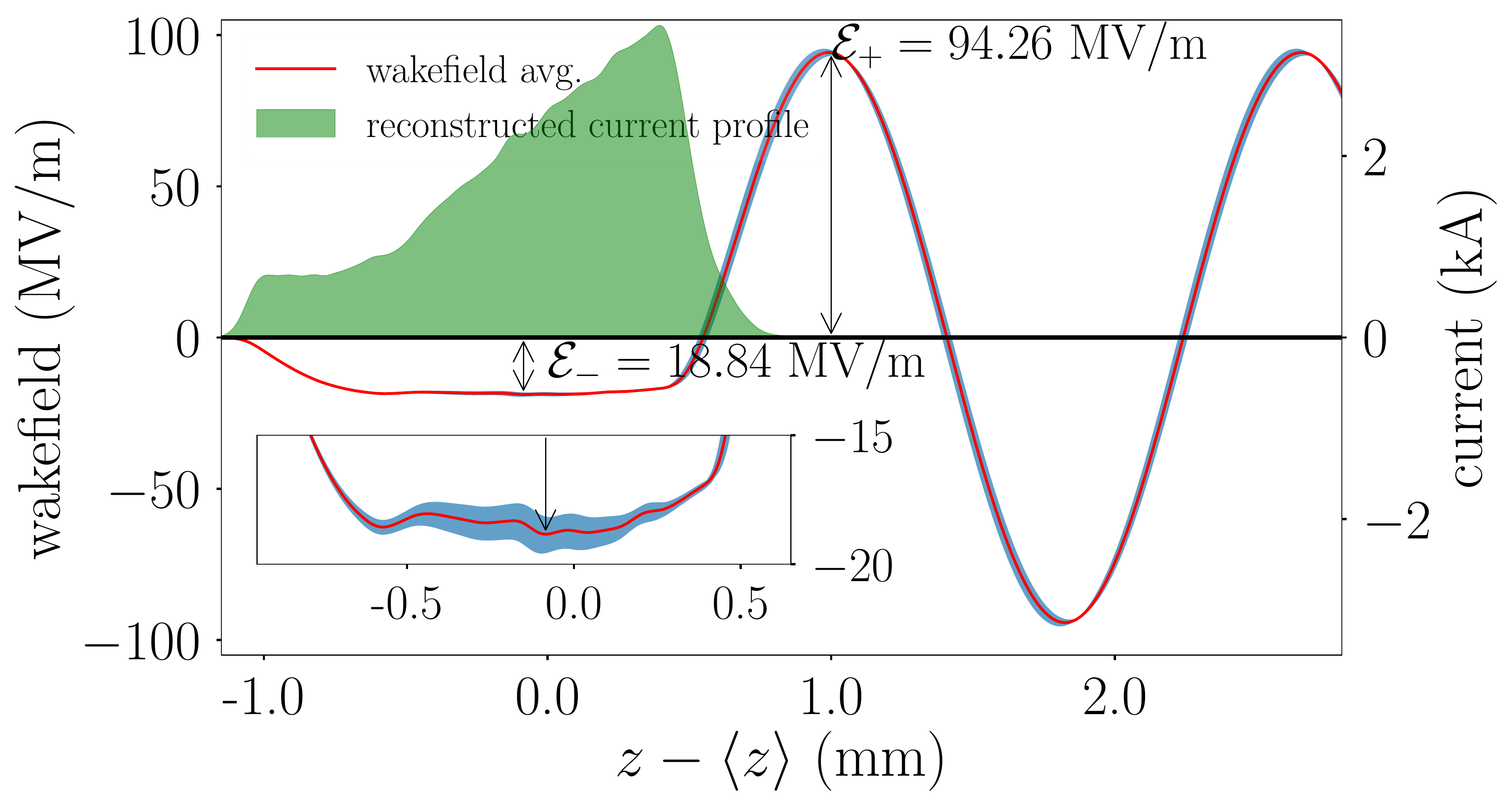}
   \caption{Wakefields obtained from 100 simulations with jitter in linacs L1, L2, and L39. All cavities are taken to have relative jitter in accelerating voltage of 0.01\% and phase jitter $0.01^{\circ}$. The red line shows the average wakefield while the blue shaded region represents the fluctuation of wakefields due to jitter over 100 random realizations of the linac settings. The average transformer ratio is \(5.00\pm 0.05\). The reconstructed current profile (green-shaded curve) is obtained numerically using Eq.~\eqref{eq:sol}.}
   \label{fig:jitter}
\end{figure}

In order to validate the robustness of the proposed design, it is instructive to investigate the sensitivity of the proposed shaping technique to shot-to-shot jitters of the amplitude and phase of the accelerating field in the linac's structures. Consistent with LCLS-II specifications~\cite{HuangLLRF2017}, we considered the relative RMS amplitude jitter of 0.01\% and the phase jitter of 0.01 degree. For simplicity, we assume that the injector produced identical bunches, as shown in Fig.~\ref{fig:begin_astra}, and performed 100 simulations of the accelerator beamline (from the injector exit to the exit of BC2) for different random realizations of the phase and amplitude for linacs L1, L2, and L39. 
The errors in linac settings were randomly generated using Gaussian probability function with standard deviations of 0.01\% and 0.01$^{\circ}$. 
Figure~\ref{fig:jitter} presents the wakefield averaged over the 100 simulations and indicates that a stable transformer ratio \(5.00\pm 0.05\) can be maintained owing to the stable beam produced in the superconducting linac.\\

\begin{figure}[!htb]
   \centering
   \includegraphics*[width=\columnwidth]{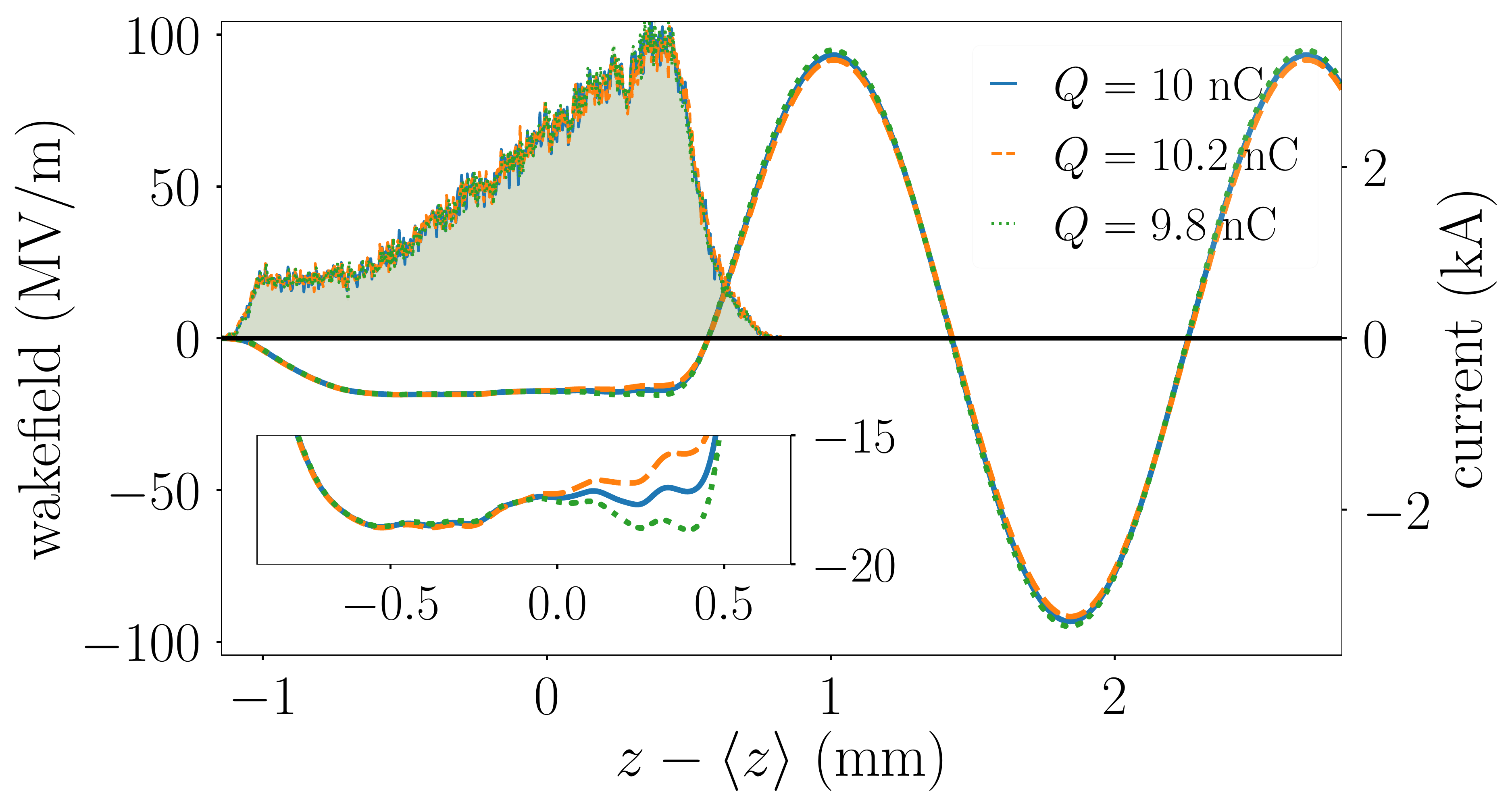}
   \caption{Current distribution (shaded curves, left axis) and associated wakefields (traces) for the nominal charge and $\pm 2$\% relative change in charge (9.8 and \SI{10.2}{\nano\coulomb}).}
   \label{fig:charge_sensitivity}
\end{figure}

Likewise, we observe the impact of charge fluctuation on the shaping to be tolerable. Cathode-to-end simulations combining {\sc astra} and {\sc elegant} indicate that a relative charge variation of +2\% (resp. -2\%) yields a relative change in the transformer ratio of -2\% (resp. +1\%) and a relative variation in peak field of -1.7\% (resp. +1.7\%); see Fig.~\ref{fig:charge_sensitivity}. 

\section{Summary}
We have presented the design of an accelerator capable of generating \SI{1}{\giga\electronvolt} electron bunches with a highly asymmetric peak current profile and a large energy chirp required for a collinear wakefield accelerator. It has been achieved without the use of collimators. Our approach is based on \textit{ab-initio} temporal shaping of the photocathode laser pulse followed by nonlinear manipulations of the electron distribution in the longitudinal phase space throughout the accelerator using collective effects and precision control of the longitudinal dispersion in two bunch compressors up to the third order. Finding the optimal design consisted of first implementing a simplified accelerator model and using it for backward tracking of the longitudinal phase space distribution of electrons through the main accelerator to provide the longitudinal phase space distribution required from the injector. The program {\sc twice} was developed to support such a capability and used to optimize the global linac parameters and time-of-flight properties of bunch compressors. Second, the simulation of the photo-injector using \textsc{astra} was performed to generate the required distribution. Third, the linac design was refined using \textsc{elegant} to account for the transverse beam dynamics. Finally, formation of longitudinally shaped drive bunches capable of producing in the collinear wakefield accelerator a transformer ratio of $\sim 5$ and a peak accelerating wakefield close to \SI{100}{\mega\volt/\metre} has been numerically demonstrated.

Although the proposed accelerator design is promising, we note that further work is required to investigate whether the same accelerator can accelerate the low-charge, low-emittance ``witness bunches" that would be accelerated to multi-\si{\giga\electronvolt} energies in the collinear wakefield accelerator and used for the generation of x-rays in the downstream free-electron laser. Discussion of this research is the subject of a forthcoming publication.

\begin{acknowledgments}
The authors are grateful to Dr. Stanislav Baturin (NIU) for useful discussions. WHT thanks Y. Park (UCLA) for several discussions on simulation studies. This work is supported by the U.S. Department of Energy, Office of Science, under award  No. DE-SC0018656 with Northern Illinois University and contract No. DE-AC02-06CH11357 with Argonne National Laboratory. 
\end{acknowledgments}
%%%%%%%%%%%%%%%%%%%%%%%%%%%%%%%%%%%%%%%%%%%%%%%%%%%%%%%%%%%%%%

\appendix
\section{One-dimensional tracking model \label{app:twice}}
A simple one-dimensional tracking program {\sc twice}~\cite{tan_longitudinal_2018} was developed for rapid assessment of the longitudinal dynamics of electrons in linear accelerators. The program adopts an approach similar to the one used in {\sc LiTrack} \cite{bane_litrack:_2005}, where only the accelerator components affecting the longitudinal beam dynamics are considered and modeled analytically. A detailed description of {\sc twice} is published in \cite{tan_longitudinal_2018}. In brief, the beam is represented by a set of $N$ macroparticles with identical charges $Q/N$ and given a set of initial LPS coordinates \((z_i, E_i)\). A transformation \((z_f, E_f) = f(z_i, E_i) \) is applied to obtain final coordinates in the LPS.
\subsubsection{Single particle dynamics}
In \textsc{twice} the transformation for a macroparticle with coordinates $(z_i, E_i) $ passing through a radiofrequency (RF) linac is given by
\begin{align}\label{eq:transRF}
\begin{pmatrix}z_f \\E_f \end{pmatrix} = \begin{pmatrix}z_i  \\ E_i(z_i) \pm eV \cos(kz_i +\varphi) \end{pmatrix},
\end{align}
where $V$, $k$, and $\varphi$ are, respectively, the accelerating voltage, wave-vector amplitude, and off-crest phase associated with the accelerating section, and $e$ is the electronic charge. In the latter and following equations the $\pm$ sign indicates the forward (+) and backward (-) tracking process detailed in Sec.~\ref{sec:backward}. 
Similarly, the transformation through a  longitudinally dispersive section, such as a bunch compressor, is given by
\begin{align}\label{eq:transZ}
 \begin{pmatrix}z_f \\E_f \end{pmatrix} = \begin{pmatrix}z_i\pm \left[R_{56}\frac{E_i-E_0}{E_0} + T_{566}\left( \frac{E_i-E_0}{E_0}  \right)^2\right]  \\ E_i \end{pmatrix} ,
\end{align}
where $E_0$ is the reference-particle energy assumed to remain constant during the transformation, and $R_{56}\equiv E_0 \frac{\partial z_f}{\partial E_i}$ and $T_{566} \equiv \frac{{E_0}^2}{2}\frac{\partial^2 z_f}{\partial E_i^2}$ are the first- and second-order longitudinal-dispersion functions introduced by the beamline. It should be noted that, given our LPS coordinate conventions, a conventional four-bend ``chicane" magnetic bunch compressor has a longitudinal dispersion $R_{56}>0$. The latter equation ignores energy loss, e.g., due to incoherent synchrotron radiation, occurring in the beamline magnets.

\subsubsection{Collective effects}
\begin{figure}[!h]
   \centering
   \includegraphics*[width=0.9\columnwidth]{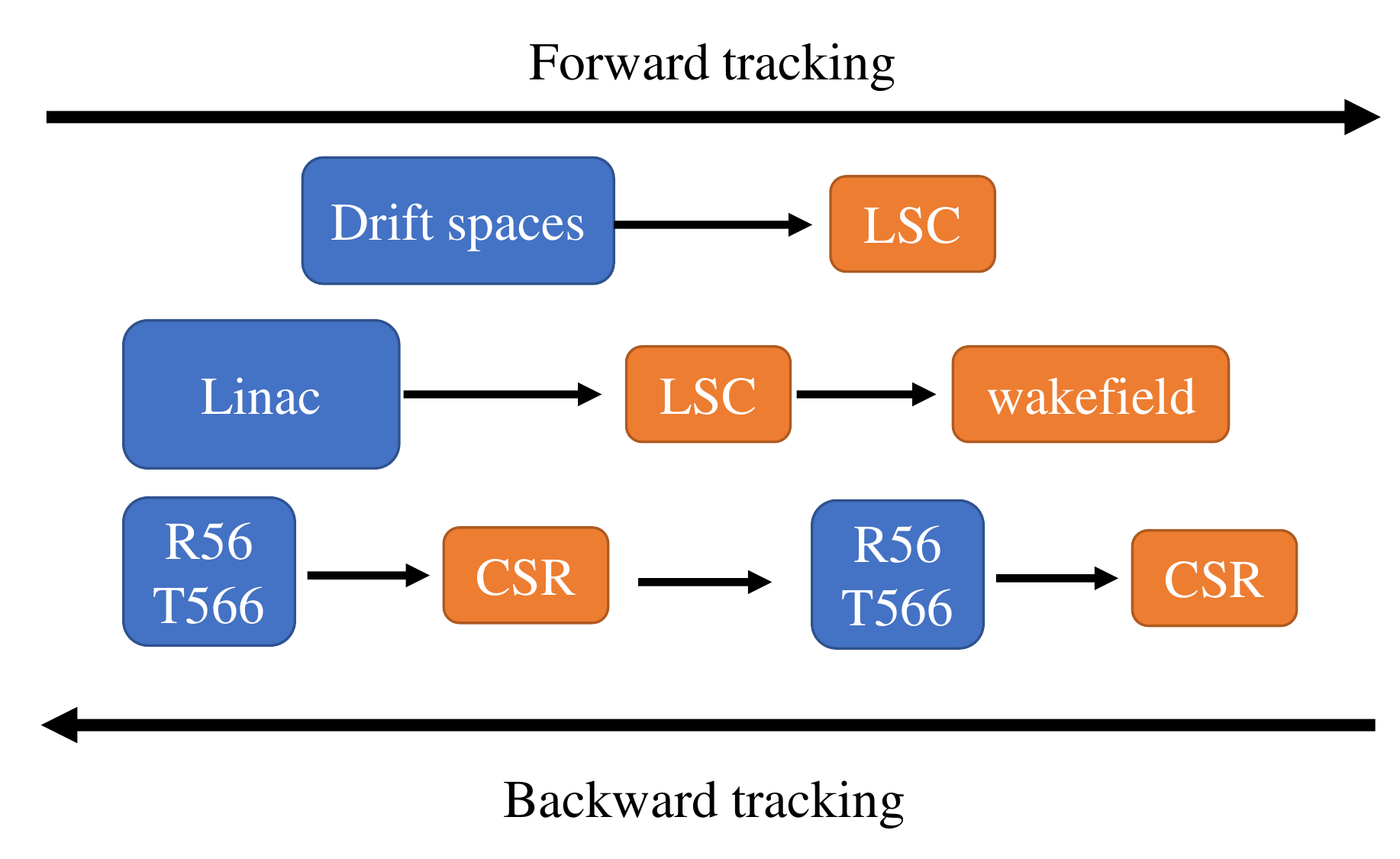}
   \caption{Treatment of collective effects as energy kicks downstream of beamline elements. In forward (resp. backward) tracking, transformations of beamline elements (resp. energy kicks) were applied, followed by energy kicks (resp. beamline elements). }
   \label{fig:collective}
\end{figure}

In \textsc{twice}, we implemented collective effects as an energy kick approximation using the transformation
\begin{align}\label{eq:transE}
%\begin{pmatrix}z_i \\E_i \end{pmatrix} \rightarrow
\begin{pmatrix}z_f \\E_f \end{pmatrix} = \begin{pmatrix}z_i  \\ E_i(z_i) \pm \Delta E(z_i) \end{pmatrix},
\end{align}
where \(\Delta E(z)\) represents the energy change associated with the considered collective effect. The treatment of collective effects is modeled as a \(z\)-dependent energy kick \(\Delta E(z)\) taken downstream of beamline elements as specified for the forward and backward tracking with the diagram shown in Fig. \ref{fig:collective}.
The implemented collective effects include wakefields modeled after a user-supplied Green's function, one-dimensional steady-state coherent synchrotron radiation (CSR), and longitudinal space charge (LSC) described via an impedance. The collective effects require the estimation of the beam's charge density, which is done in {\sc twice} either using a standard histogram binning method with noise filtering or via the kernel-density estimation technique \cite{mohayai_novel_2017}.\\

In order to model the impact of a wakefield, the charge distribution \(q(z)\) is directly used to compute the wake potential given a tabulated Green's function
\begin{align}\label{eq:wakefield}
&W(z)=\int^{z}_{0}q(z')G(z-z')\text{d}z'\;.
\end{align}
The change in energy is computed as \(\Delta E(z) = L W(z) \), where \(L\) is the effective length where the beam experiences the wakefield.\\

The LSC is implemented using a one-dimensional model detailed in \cite{qiang_high_2009}, where the impedance per unit length is
\begin{align}\label{ref:lcsimpedance}
Z(k) &= i\frac{Z_0}{\pi\gamma r_b}\frac{1- 2I_1(\xi_b)K_1(\xi_b)}{\xi_b}\;,
\end{align}
where \(\xi_b \equiv kr_b/\gamma\); \(I_1\) and \(K_1\) are modified Bessel functions of the first and second kind, respectively; and $k$, $Z_0$ and \(r_b\) are, respectively, the wave-vector amplitude, impedance of free space and a user-defined transverse beam radius, and $\gamma$ is the Lorentz factor. Given the charge density, the Fourier-transformed current density $\tilde{I}(k)$ is derived from 
\begin{align}
\tilde{I}(k)={\cal F} [c q(z)],
\end{align}
with ${\cal F}$ representing the Fourier transform. The change in energy is computed as
\begin{align}
\Delta E = -\mathcal{F}^{-1}[eZ(k)\tilde{I}(k)L],
\end{align}
where $\mathcal{F}^{-1}$ is the inverse Fourier transform, and $L$ is the effective distance along which the LSC interaction occurs. In order to account for LSC during acceleration, $\gamma$ is replaced by the geometry average $\sqrt{\gamma_i\gamma_f}$ of the Lorentz factors computed at the entrance $\gamma_i$ and exit $\gamma_f$ of the linac section. 

Finally, CSR energy kicks are applied downstream of the dispersive beamline elements. For instance, a CSR energy kick can be applied after a dispersive element with user-defined length and angle  described by \(R_{56}\) and \(T_{566}\). The effect of CSR is described using a one-dimensional model commonly implemented in other beam-dynamics program~\cite{saldin_coherent_1997-1}. To simplify the calculation, only the steady-state CSR is considered in \textsc{twice}. The energy loss associated with CSR is obtained from \cite{mitchell_fast_2013}\footnote{The variable $z$ here for CSR calculation refers to the relative position from the bunch centroid with bunch head at $z>0$.},
\begin{align}
\Delta E(z)=\rho\theta\frac{\mathrm dE}{\mathrm dct}=-\theta\frac{\gamma m_ec^2 r_e}{e}\int^z_{-\infty}\frac{\partial q(z')}{\partial z'}I_{csr}(z, z')\text{d}z'\ ,
\end{align}
with the integral kernel defined as 
\begin{align}
&I_{csr}(z, z')=\frac{4u(u^2+8)}{(u^2+4)(u^2+12)}\;,
\end{align}
where $\theta$ is the angle, $m_ec^2$ is the electron rest mass energy, $r_e$ is the classical electron radius and the variable $u$ is the solution of $\frac{\gamma^3(z-z')}{\rho}=\frac{u^3}{24}+\frac{u}{2}$. 
CSR introduces an energy loss strongly dependent on the bunch length, which varies within the dispersive sections used to compress the bunch. Consequently, the longitudinally dispersive beamlines are segmented into several elements with individual  $(R_{56}, T_{566})$ parameters. A CSR kick is applied after each of the elements. A conventional chicane-type bunch compressor is usually broken into two sections (two mirror-symmetric doglegs) but can in principle be divided into an arbitrary number of segments to improve the resolution at the expense of computational time. 

\subsubsection{Backward tracking\label{sec:backward}}
An important feature of {\sc twice} is its capability to track the beam in the forward or backward directions (indicated by the $\pm$ sign in Eqs.~\eqref{eq:transZ} and \eqref{eq:transE}) in the presence of collective effects [so far LSC, CSR, and wakefield effects are included]. The effects of LSC and wakefield are straightforward to implement as they only involve a change in energy, while handling of the CSR requires extra care since the particles' positions also change throughout the dispersive section. 
Therefore, an energy kick is applied after the beamline element in the  forward-tracking mode and before the beamline element in backward-tracking mode, as shown in Fig.~\ref{fig:collective}. Although the treatment of CSR is not exact, it nonetheless provides a good starting point to account for the effect.


\begin{thebibliography}{66}%
\makeatletter
\providecommand \@ifxundefined [1]{%
 \@ifx{#1\undefined}
}%
\providecommand \@ifnum [1]{%
 \ifnum #1\expandafter \@firstoftwo
 \else \expandafter \@secondoftwo
 \fi
}%
\providecommand \@ifx [1]{%
 \ifx #1\expandafter \@firstoftwo
 \else \expandafter \@secondoftwo
 \fi
}%
\providecommand \natexlab [1]{#1}%
\providecommand \enquote  [1]{``#1''}%
\providecommand \bibnamefont  [1]{#1}%
\providecommand \bibfnamefont [1]{#1}%
\providecommand \citenamefont [1]{#1}%
\providecommand \href@noop [0]{\@secondoftwo}%
\providecommand \href [0]{\begingroup \@sanitize@url \@href}%
\providecommand \@href[1]{\@@startlink{#1}\@@href}%
\providecommand \@@href[1]{\endgroup#1\@@endlink}%
\providecommand \@sanitize@url [0]{\catcode `\\12\catcode `\$12\catcode
  `\&12\catcode `\#12\catcode `\^12\catcode `\_12\catcode `\%12\relax}%
\providecommand \@@startlink[1]{}%
\providecommand \@@endlink[0]{}%
\providecommand \url  [0]{\begingroup\@sanitize@url \@url }%
\providecommand \@url [1]{\endgroup\@href {#1}{\urlprefix }}%
\providecommand \urlprefix  [0]{URL }%
\providecommand \Eprint [0]{\href }%
\providecommand \doibase [0]{https://doi.org/}%
\providecommand \selectlanguage [0]{\@gobble}%
\providecommand \bibinfo  [0]{\@secondoftwo}%
\providecommand \bibfield  [0]{\@secondoftwo}%
\providecommand \translation [1]{[#1]}%
\providecommand \BibitemOpen [0]{}%
\providecommand \bibitemStop [0]{}%
\providecommand \bibitemNoStop [0]{.\EOS\space}%
\providecommand \EOS [0]{\spacefactor3000\relax}%
\providecommand \BibitemShut  [1]{\csname bibitem#1\endcsname}%
\let\auto@bib@innerbib\@empty
%</preamble>
\bibitem [{\citenamefont {Voss}\ and\ \citenamefont
  {Weiland}(1982)}]{voss_wake_1982}%
  \BibitemOpen
  \bibfield  {author} {\bibinfo {author} {\bibfnamefont {G.}~\bibnamefont
  {Voss}}\ and\ \bibinfo {author} {\bibfnamefont {T.}~\bibnamefont {Weiland}},\
  }\href@noop {} {\emph {\bibinfo {title} {The wake field acceleration
  mechanism}}},\ \bibinfo {type} {Tech. Rep.}\ \bibinfo {number} {DESY-82-074}\
  (\bibinfo  {institution} {DESY},\ \bibinfo {year} {1982})\BibitemShut
  {NoStop}%
\bibitem [{\citenamefont {Briggs}\ \emph {et~al.}(1974)\citenamefont {Briggs},
  \citenamefont {Fessenden},\ and\ \citenamefont
  {Neil}}]{briggs_electron_1974}%
  \BibitemOpen
  \bibfield  {author} {\bibinfo {author} {\bibfnamefont {R.~J.}\ \bibnamefont
  {Briggs}}, \bibinfo {author} {\bibfnamefont {T.~J.}\ \bibnamefont
  {Fessenden}},\ and\ \bibinfo {author} {\bibfnamefont {V.~K.}\ \bibnamefont
  {Neil}},\ }\bibfield  {title} {\bibinfo {title} {Electron autoacceleration},\
  }in\ \href
  {https://slac.stanford.edu/pubs/slacreports/reports16/slac-r-839-b.pdf}
  {\emph {\bibinfo {booktitle} {Proceedings, 9th {International} {Conference}
  on the {High}-{Energy} {Accelerators}}}}\ (\bibinfo {year} {1974})\ p.\
  \bibinfo {pages} {278}\BibitemShut {NoStop}%
\bibitem [{\citenamefont {Friedman}(1973)}]{friedman_autoacceleration_1973}%
  \BibitemOpen
  \bibfield  {author} {\bibinfo {author} {\bibfnamefont {M.}~\bibnamefont
  {Friedman}},\ }\bibfield  {title} {\bibinfo {title} {Autoacceleration of an
  intense relativistic electron beam},\ }\href
  {https://doi.org/10.1103/PhysRevLett.31.1107} {\bibfield  {journal} {\bibinfo
   {journal} {Phys. Rev. Lett.}\ }\textbf {\bibinfo {volume} {31}},\ \bibinfo
  {pages} {1107} (\bibinfo {year} {1973})}\BibitemShut {NoStop}%
\bibitem [{\citenamefont {Perevedentsev}\ and\ \citenamefont
  {Skrinsky}(1978)}]{perevedentsev_use_1978_2}%
  \BibitemOpen
  \bibfield  {author} {\bibinfo {author} {\bibfnamefont {E.~A.}\ \bibnamefont
  {Perevedentsev}}\ and\ \bibinfo {author} {\bibfnamefont {A.~N.}\ \bibnamefont
  {Skrinsky}},\ }\bibfield  {title} {\bibinfo {title} {On the {Use} of the
  {Intense} {Beams} of {Large} {Proton} {Accelerators} to {Excite} the
  {Accelerating} {Structure} of a {Linear} {Accelerator}},\ }in\ \href
  {https://inspirehep.net/literature/1476534} {\emph {\bibinfo {booktitle}
  {Proc. 6th {All}-{Union} {Conference} {Charged} {Particle} {Accelerators},
  {Dubna} ({Institute} of {Nuclear} {Physics}, {Novosibirsk}, {USSR},
  1978)}}},\ Vol.~\bibinfo {volume} {2}\ (\bibinfo {year} {1978})\ p.\ \bibinfo
  {pages} {272},\ \bibinfo {note} {{E}nglish version is available in
  \href{http://lss.fnal.gov/conf/C791004/p61.pdf}{Proceedings of the 2nd ICFA
  Workshop on Possibilities and Limitations of Accelerators and Detectors}
  (1979) p. 61.}\BibitemShut {Stop}%
\bibitem [{\citenamefont {Sessler}(1982)}]{sessler_free_1982}%
  \BibitemOpen
  \bibfield  {author} {\bibinfo {author} {\bibfnamefont {A.~M.}\ \bibnamefont
  {Sessler}},\ }\bibfield  {title} {\bibinfo {title} {The free electron laser
  as a power source for a high‐gradient accelerating structure},\ }\href
  {https://doi.org/10.1063/1.33790} {\bibfield  {journal} {\bibinfo  {journal}
  {AIP Conference Proceedings}\ }\textbf {\bibinfo {volume} {91}},\ \bibinfo
  {pages} {154} (\bibinfo {year} {1982})}\BibitemShut {NoStop}%
\bibitem [{\citenamefont {Chen}\ \emph {et~al.}(1985)\citenamefont {Chen},
  \citenamefont {Dawson}, \citenamefont {Huff},\ and\ \citenamefont
  {Katsouleas}}]{chen_acceleration_1985}%
  \BibitemOpen
  \bibfield  {author} {\bibinfo {author} {\bibfnamefont {P.}~\bibnamefont
  {Chen}}, \bibinfo {author} {\bibfnamefont {J.~M.}\ \bibnamefont {Dawson}},
  \bibinfo {author} {\bibfnamefont {R.~W.}\ \bibnamefont {Huff}},\ and\
  \bibinfo {author} {\bibfnamefont {T.}~\bibnamefont {Katsouleas}},\ }\bibfield
   {title} {\bibinfo {title} {Acceleration of electrons by the interaction of a
  bunched electron beam with a plasma},\ }\href
  {https://doi.org/10.1103/PhysRevLett.54.693} {\bibfield  {journal} {\bibinfo
  {journal} {Phys. Rev. Lett.}\ }\textbf {\bibinfo {volume} {54}},\ \bibinfo
  {pages} {693} (\bibinfo {year} {1985})}\BibitemShut {NoStop}%
\bibitem [{\citenamefont {Chin}(1983)}]{chin_wake_1983}%
  \BibitemOpen
  \bibfield  {author} {\bibinfo {author} {\bibfnamefont {Y.}~\bibnamefont
  {Chin}},\ }\bibfield  {title} {\bibinfo {title} {The {Wake} {Field}
  {Acceleration} {Using} a {Cavity} of {Elliptical} {Cross} {Section}},\ }in\
  \href {https://accelconf.web.cern.ch/ accelconf/l84/papers/tup0026.pdf}
  {\emph {\bibinfo {booktitle} {12th international linear accelerator
  conference}}}\ (\bibinfo {year} {1983})\ pp.\ \bibinfo {pages}
  {159--161}\BibitemShut {NoStop}%
\bibitem [{\citenamefont {Gai}\ \emph {et~al.}(1988)\citenamefont {Gai},
  \citenamefont {Schoessow}, \citenamefont {Cole}, \citenamefont {Konecny},
  \citenamefont {Norem}, \citenamefont {Rosenzweig},\ and\ \citenamefont
  {Simpson}}]{gai_experimental_1988}%
  \BibitemOpen
  \bibfield  {author} {\bibinfo {author} {\bibfnamefont {W.}~\bibnamefont
  {Gai}}, \bibinfo {author} {\bibfnamefont {P.}~\bibnamefont {Schoessow}},
  \bibinfo {author} {\bibfnamefont {B.}~\bibnamefont {Cole}}, \bibinfo {author}
  {\bibfnamefont {R.}~\bibnamefont {Konecny}}, \bibinfo {author} {\bibfnamefont
  {J.}~\bibnamefont {Norem}}, \bibinfo {author} {\bibfnamefont
  {J.}~\bibnamefont {Rosenzweig}},\ and\ \bibinfo {author} {\bibfnamefont
  {J.}~\bibnamefont {Simpson}},\ }\bibfield  {title} {\bibinfo {title}
  {Experimental demonstration of wake-field effects in dielectric structures},\
  }\href {https://doi.org/10.1103/PhysRevLett.61.2756} {\bibfield  {journal}
  {\bibinfo  {journal} {Phys. Rev. Lett.}\ }\textbf {\bibinfo {volume} {61}},\
  \bibinfo {pages} {2756} (\bibinfo {year} {1988})}\BibitemShut {NoStop}%
\bibitem [{\citenamefont {Bane}\ \emph {et~al.}(1985)\citenamefont {Bane},
  \citenamefont {Chen},\ and\ \citenamefont {Wilson}}]{bane_collinear_1985}%
  \BibitemOpen
  \bibfield  {author} {\bibinfo {author} {\bibfnamefont {K.~L.}\ \bibnamefont
  {Bane}}, \bibinfo {author} {\bibfnamefont {P.}~\bibnamefont {Chen}},\ and\
  \bibinfo {author} {\bibfnamefont {P.~B.}\ \bibnamefont {Wilson}},\ }\bibfield
   {title} {\bibinfo {title} {On collinear wake field acceleration},\ }\href
  {https://doi.org/10.1109/TNS.1985.4334416} {\bibfield  {journal} {\bibinfo
  {journal} {Proceedings of the 1985 Particle Accelerator Conference (PAC1985):
  Accelerator Engineering and Technology Vancouver, BC May 13-16, 1985}\
  }\textbf {\bibinfo {volume} {32}},\ \bibinfo {pages} {3524} (\bibinfo {year}
  {1985})}\BibitemShut {NoStop}%
\bibitem [{\citenamefont {Baturin}\ and\ \citenamefont
  {Zholents}(2017)}]{baturin_upper_2017}%
  \BibitemOpen
  \bibfield  {author} {\bibinfo {author} {\bibfnamefont {S.~S.}\ \bibnamefont
  {Baturin}}\ and\ \bibinfo {author} {\bibfnamefont {A.}~\bibnamefont
  {Zholents}},\ }\bibfield  {title} {\bibinfo {title} {Upper limit for the
  accelerating gradient in the collinear wakefield accelerator as a function of
  the transformer ratio},\ }\href
  {https://doi.org/10.1103/PhysRevAccelBeams.20.061302} {\bibfield  {journal}
  {\bibinfo  {journal} {Phys. Rev. Accel. Beams}\ }\textbf {\bibinfo {volume}
  {20}},\ \bibinfo {pages} {061302} (\bibinfo {year} {2017})}\BibitemShut
  {NoStop}%
\bibitem [{\citenamefont {Cornacchia}\ \emph {et~al.}(2006)\citenamefont
  {Cornacchia}, \citenamefont {Di~Mitri}, \citenamefont {Penco},\ and\
  \citenamefont {Zholents}}]{cornacchia_formation_2006}%
  \BibitemOpen
  \bibfield  {author} {\bibinfo {author} {\bibfnamefont {M.}~\bibnamefont
  {Cornacchia}}, \bibinfo {author} {\bibfnamefont {S.}~\bibnamefont
  {Di~Mitri}}, \bibinfo {author} {\bibfnamefont {G.}~\bibnamefont {Penco}},\
  and\ \bibinfo {author} {\bibfnamefont {A.~A.}\ \bibnamefont {Zholents}},\
  }\bibfield  {title} {\bibinfo {title} {Formation of electron bunches for
  harmonic cascade x-ray free electron lasers},\ }\href
  {https://doi.org/10.1103/PhysRevSTAB.9.120701} {\bibfield  {journal}
  {\bibinfo  {journal} {Phys. Rev. ST Accel. Beams}\ }\textbf {\bibinfo
  {volume} {9}},\ \bibinfo {pages} {120701} (\bibinfo {year}
  {2006})}\BibitemShut {NoStop}%
\bibitem [{\citenamefont {Penco}\ \emph {et~al.}(2014)\citenamefont {Penco},
  \citenamefont {Danailov}, \citenamefont {Demidovich}, \citenamefont
  {Allaria}, \citenamefont {De~Ninno}, \citenamefont {Di~Mitri}, \citenamefont
  {Fawley}, \citenamefont {Ferrari}, \citenamefont {Giannessi},\ and\
  \citenamefont {Trovó}}]{penco_experimental_2014}%
  \BibitemOpen
  \bibfield  {author} {\bibinfo {author} {\bibfnamefont {G.}~\bibnamefont
  {Penco}}, \bibinfo {author} {\bibfnamefont {M.}~\bibnamefont {Danailov}},
  \bibinfo {author} {\bibfnamefont {A.}~\bibnamefont {Demidovich}}, \bibinfo
  {author} {\bibfnamefont {E.}~\bibnamefont {Allaria}}, \bibinfo {author}
  {\bibfnamefont {G.}~\bibnamefont {De~Ninno}}, \bibinfo {author}
  {\bibfnamefont {S.}~\bibnamefont {Di~Mitri}}, \bibinfo {author}
  {\bibfnamefont {W.~M.}\ \bibnamefont {Fawley}}, \bibinfo {author}
  {\bibfnamefont {E.}~\bibnamefont {Ferrari}}, \bibinfo {author} {\bibfnamefont
  {L.}~\bibnamefont {Giannessi}},\ and\ \bibinfo {author} {\bibfnamefont
  {M.}~\bibnamefont {Trovó}},\ }\bibfield  {title} {\bibinfo {title}
  {Experimental {Demonstration} of {Electron} {Longitudinal}-{Phase}-{Space}
  {Linearization} by {Shaping} the {Photoinjector} {Laser} {Pulse}},\ }\href
  {https://doi.org/10.1103/PhysRevLett.112.044801} {\bibfield  {journal}
  {\bibinfo  {journal} {Phys. Rev. Lett.}\ }\textbf {\bibinfo {volume} {112}},\
  \bibinfo {pages} {044801} (\bibinfo {year} {2014})}\BibitemShut {NoStop}%
\bibitem [{\citenamefont {Lemery}\ and\ \citenamefont
  {Piot}(2015)}]{lemery_tailored_2015}%
  \BibitemOpen
  \bibfield  {author} {\bibinfo {author} {\bibfnamefont {F.}~\bibnamefont
  {Lemery}}\ and\ \bibinfo {author} {\bibfnamefont {P.}~\bibnamefont {Piot}},\
  }\bibfield  {title} {\bibinfo {title} {Tailored electron bunches with smooth
  current profiles for enhanced transformer ratios in beam-driven
  acceleration},\ }\href {https://doi.org/10.1103/PhysRevSTAB.18.081301}
  {\bibfield  {journal} {\bibinfo  {journal} {Phys. Rev. Spec. Top. Accel.
  Beams}\ }\textbf {\bibinfo {volume} {18}},\ \bibinfo {pages} {081301}
  (\bibinfo {year} {2015})}\BibitemShut {NoStop}%
\bibitem [{\citenamefont {Jiang}\ \emph {et~al.}(2012)\citenamefont {Jiang},
  \citenamefont {Jing}, \citenamefont {Schoessow}, \citenamefont {Power},\ and\
  \citenamefont {Gai}}]{jiang_formation_2012}%
  \BibitemOpen
  \bibfield  {author} {\bibinfo {author} {\bibfnamefont {B.}~\bibnamefont
  {Jiang}}, \bibinfo {author} {\bibfnamefont {C.}~\bibnamefont {Jing}},
  \bibinfo {author} {\bibfnamefont {P.}~\bibnamefont {Schoessow}}, \bibinfo
  {author} {\bibfnamefont {J.}~\bibnamefont {Power}},\ and\ \bibinfo {author}
  {\bibfnamefont {W.}~\bibnamefont {Gai}},\ }\bibfield  {title} {\bibinfo
  {title} {Formation of a novel shaped bunch to enhance transformer ratio in
  collinear wakefield accelerators},\ }\href
  {https://doi.org/10.1103/PhysRevSTAB.15.011301} {\bibfield  {journal}
  {\bibinfo  {journal} {Phys. Rev. ST Accel. Beams}\ }\textbf {\bibinfo
  {volume} {15}},\ \bibinfo {pages} {011301} (\bibinfo {year}
  {2012})}\BibitemShut {NoStop}%
\bibitem [{\citenamefont {Ha}\ \emph {et~al.}(2017)\citenamefont {Ha},
  \citenamefont {Cho}, \citenamefont {Namkung}, \citenamefont {Power},
  \citenamefont {Doran}, \citenamefont {Wisniewski}, \citenamefont {Conde},
  \citenamefont {Gai}, \citenamefont {Liu}, \citenamefont {Whiteford},
  \citenamefont {Gao}, \citenamefont {Kim}, \citenamefont {Zholents},
  \citenamefont {Sun}, \citenamefont {Jing},\ and\ \citenamefont
  {Piot}}]{ha_precision_2017}%
  \BibitemOpen
  \bibfield  {author} {\bibinfo {author} {\bibfnamefont {G.}~\bibnamefont
  {Ha}}, \bibinfo {author} {\bibfnamefont {M.~H.}\ \bibnamefont {Cho}},
  \bibinfo {author} {\bibfnamefont {W.}~\bibnamefont {Namkung}}, \bibinfo
  {author} {\bibfnamefont {J.~G.}\ \bibnamefont {Power}}, \bibinfo {author}
  {\bibfnamefont {D.~S.}\ \bibnamefont {Doran}}, \bibinfo {author}
  {\bibfnamefont {E.~E.}\ \bibnamefont {Wisniewski}}, \bibinfo {author}
  {\bibfnamefont {M.}~\bibnamefont {Conde}}, \bibinfo {author} {\bibfnamefont
  {W.}~\bibnamefont {Gai}}, \bibinfo {author} {\bibfnamefont {W.}~\bibnamefont
  {Liu}}, \bibinfo {author} {\bibfnamefont {C.}~\bibnamefont {Whiteford}},
  \bibinfo {author} {\bibfnamefont {Q.}~\bibnamefont {Gao}}, \bibinfo {author}
  {\bibfnamefont {K.-J.}\ \bibnamefont {Kim}}, \bibinfo {author} {\bibfnamefont
  {A.}~\bibnamefont {Zholents}}, \bibinfo {author} {\bibfnamefont {Y.-E.}\
  \bibnamefont {Sun}}, \bibinfo {author} {\bibfnamefont {C.}~\bibnamefont
  {Jing}},\ and\ \bibinfo {author} {\bibfnamefont {P.}~\bibnamefont {Piot}},\
  }\bibfield  {title} {\bibinfo {title} {Precision {Control} of the {Electron}
  {Longitudinal} {Bunch} {Shape} {Using} an {Emittance}-{Exchange} {Beam}
  {Line}},\ }\href {https://doi.org/10.1103/PhysRevLett.118.104801} {\bibfield
  {journal} {\bibinfo  {journal} {Phys. Rev. Lett.}\ }\textbf {\bibinfo
  {volume} {118}},\ \bibinfo {pages} {104801} (\bibinfo {year}
  {2017})}\BibitemShut {NoStop}%
\bibitem [{\citenamefont {Gao}\ \emph {et~al.}(2018)\citenamefont {Gao},
  \citenamefont {Ha}, \citenamefont {Jing}, \citenamefont {Antipov},
  \citenamefont {Power}, \citenamefont {Conde}, \citenamefont {Gai},
  \citenamefont {Chen}, \citenamefont {Shi}, \citenamefont {Wisniewski},
  \citenamefont {Doran}, \citenamefont {Liu}, \citenamefont {Whiteford},
  \citenamefont {Zholents}, \citenamefont {Piot},\ and\ \citenamefont
  {Baturin}}]{gao_observation_2018}%
  \BibitemOpen
  \bibfield  {author} {\bibinfo {author} {\bibfnamefont {Q.}~\bibnamefont
  {Gao}}, \bibinfo {author} {\bibfnamefont {G.}~\bibnamefont {Ha}}, \bibinfo
  {author} {\bibfnamefont {C.}~\bibnamefont {Jing}}, \bibinfo {author}
  {\bibfnamefont {S.~P.}\ \bibnamefont {Antipov}}, \bibinfo {author}
  {\bibfnamefont {J.~G.}\ \bibnamefont {Power}}, \bibinfo {author}
  {\bibfnamefont {M.}~\bibnamefont {Conde}}, \bibinfo {author} {\bibfnamefont
  {W.}~\bibnamefont {Gai}}, \bibinfo {author} {\bibfnamefont {H.}~\bibnamefont
  {Chen}}, \bibinfo {author} {\bibfnamefont {J.}~\bibnamefont {Shi}}, \bibinfo
  {author} {\bibfnamefont {E.~E.}\ \bibnamefont {Wisniewski}}, \bibinfo
  {author} {\bibfnamefont {D.~S.}\ \bibnamefont {Doran}}, \bibinfo {author}
  {\bibfnamefont {W.}~\bibnamefont {Liu}}, \bibinfo {author} {\bibfnamefont
  {C.~E.}\ \bibnamefont {Whiteford}}, \bibinfo {author} {\bibfnamefont
  {A.}~\bibnamefont {Zholents}}, \bibinfo {author} {\bibfnamefont
  {P.}~\bibnamefont {Piot}},\ and\ \bibinfo {author} {\bibfnamefont {S.~S.}\
  \bibnamefont {Baturin}},\ }\bibfield  {title} {\bibinfo {title} {Observation
  of {High} {Transformer} {Ratio} of {Shaped} {Bunch} {Generated} by an
  {Emittance}-{Exchange} {Beam} {Line}},\ }\href
  {https://doi.org/10.1103/PhysRevLett.120.114801} {\bibfield  {journal}
  {\bibinfo  {journal} {Phys. Rev. Lett.}\ }\textbf {\bibinfo {volume} {120}},\
  \bibinfo {pages} {114801} (\bibinfo {year} {2018})}\BibitemShut {NoStop}%
\bibitem [{\citenamefont {Piot}\ \emph {et~al.}(2012)\citenamefont {Piot},
  \citenamefont {Behrens}, \citenamefont {Gerth}, \citenamefont {Dohlus},
  \citenamefont {Lemery}, \citenamefont {Mihalcea}, \citenamefont {Stoltz},\
  and\ \citenamefont {Vogt}}]{piot_generation_2012}%
  \BibitemOpen
  \bibfield  {author} {\bibinfo {author} {\bibfnamefont {P.}~\bibnamefont
  {Piot}}, \bibinfo {author} {\bibfnamefont {C.}~\bibnamefont {Behrens}},
  \bibinfo {author} {\bibfnamefont {C.}~\bibnamefont {Gerth}}, \bibinfo
  {author} {\bibfnamefont {M.}~\bibnamefont {Dohlus}}, \bibinfo {author}
  {\bibfnamefont {F.}~\bibnamefont {Lemery}}, \bibinfo {author} {\bibfnamefont
  {D.}~\bibnamefont {Mihalcea}}, \bibinfo {author} {\bibfnamefont
  {P.}~\bibnamefont {Stoltz}},\ and\ \bibinfo {author} {\bibfnamefont
  {M.}~\bibnamefont {Vogt}},\ }\bibfield  {title} {\bibinfo {title} {Generation
  and {Characterization} of {Electron} {Bunches} with {Ramped} {Current}
  {Profiles} in a {Dual}-{Frequency} {Superconducting} {Linear}
  {Accelerator}},\ }\href {https://doi.org/10.1103/PhysRevLett.108.034801}
  {\bibfield  {journal} {\bibinfo  {journal} {Phys. Rev. Lett.}\ }\textbf
  {\bibinfo {volume} {108}},\ \bibinfo {pages} {034801} (\bibinfo {year}
  {2012})}\BibitemShut {NoStop}%
\bibitem [{\citenamefont {Panofsky}\ and\ \citenamefont
  {Bander}(1968)}]{panofsky_asymptotic_1968}%
  \BibitemOpen
  \bibfield  {author} {\bibinfo {author} {\bibfnamefont {W.~K.~H.}\
  \bibnamefont {Panofsky}}\ and\ \bibinfo {author} {\bibfnamefont
  {M.}~\bibnamefont {Bander}},\ }\bibfield  {title} {\bibinfo {title}
  {Asymptotic {Theory} of {Beam} {Break}‐{Up} in {Linear} {Accelerators}},\
  }\href {https://doi.org/10.1063/1.1683315} {\bibfield  {journal} {\bibinfo
  {journal} {Review of Scientific Instruments}\ }\textbf {\bibinfo {volume}
  {39}},\ \bibinfo {pages} {206} (\bibinfo {year} {1968})}\BibitemShut
  {NoStop}%
\bibitem [{\citenamefont {Neil}\ \emph {et~al.}(1979)\citenamefont {Neil},
  \citenamefont {Hall},\ and\ \citenamefont {Cooper}}]{neil_further_1979}%
  \BibitemOpen
  \bibfield  {author} {\bibinfo {author} {\bibfnamefont {V.~K.}\ \bibnamefont
  {Neil}}, \bibinfo {author} {\bibfnamefont {L.~S.}\ \bibnamefont {Hall}},\
  and\ \bibinfo {author} {\bibfnamefont {R.~K.}\ \bibnamefont {Cooper}},\
  }\bibfield  {title} {\bibinfo {title} {Further {Theoretical} {Studies} {Of}
  {The} {Beam} {Breakup} {Instability}},\ }\href@noop {} {\bibfield  {journal}
  {\bibinfo  {journal} {Part. Accel.}\ }\textbf {\bibinfo {volume} {9}},\
  \bibinfo {pages} {213} (\bibinfo {year} {1979})}\BibitemShut {NoStop}%
\bibitem [{\citenamefont {Chao}\ \emph {et~al.}(1980)\citenamefont {Chao},
  \citenamefont {Richter},\ and\ \citenamefont {Yao}}]{chao_beam_1980}%
  \BibitemOpen
  \bibfield  {author} {\bibinfo {author} {\bibfnamefont {A.~W.}\ \bibnamefont
  {Chao}}, \bibinfo {author} {\bibfnamefont {B.}~\bibnamefont {Richter}},\ and\
  \bibinfo {author} {\bibfnamefont {C.-Y.}\ \bibnamefont {Yao}},\ }\bibfield
  {title} {\bibinfo {title} {Beam emittance growth caused by transverse
  deflecting fields in a linear accelerator},\ }\href
  {https://doi.org/https://doi.org/10.1016/0029-554X(80)90851-4} {\bibfield
  {journal} {\bibinfo  {journal} {Nucl. Instrum. Methods}\ }\textbf {\bibinfo
  {volume} {178}},\ \bibinfo {pages} {1 } (\bibinfo {year} {1980})}\BibitemShut
  {NoStop}%
\bibitem [{\citenamefont {Balakin}\ \emph {et~al.}(1983)\citenamefont
  {Balakin}, \citenamefont {Novokhatsky},\ and\ \citenamefont
  {Smirnov}}]{balakin_vlepp:_1983}%
  \BibitemOpen
  \bibfield  {author} {\bibinfo {author} {\bibfnamefont {V.~E.}\ \bibnamefont
  {Balakin}}, \bibinfo {author} {\bibfnamefont {A.~V.}\ \bibnamefont
  {Novokhatsky}},\ and\ \bibinfo {author} {\bibfnamefont {V.~P.}\ \bibnamefont
  {Smirnov}},\ }\bibfield  {title} {\bibinfo {title} {{VLEPP}: {TRANSVERSE}
  {BEAM} {DYNAMICS}},\ }\href@noop {} {\bibfield  {journal} {\bibinfo
  {journal} {Proceedings, 12th International Conference on High-Energy
  Accelerators, HEACC 1983: Fermilab, Batavia, August 11-16, 1983}\ }\textbf
  {\bibinfo {volume} {C830811}},\ \bibinfo {pages} {119} (\bibinfo {year}
  {1983})}\BibitemShut {NoStop}%
\bibitem [{\citenamefont {Li}\ \emph {et~al.}(2014)\citenamefont {Li},
  \citenamefont {Gai}, \citenamefont {Jing}, \citenamefont {Power},
  \citenamefont {Tang},\ and\ \citenamefont {Zholents}}]{li_high_2014}%
  \BibitemOpen
  \bibfield  {author} {\bibinfo {author} {\bibfnamefont {C.}~\bibnamefont
  {Li}}, \bibinfo {author} {\bibfnamefont {W.}~\bibnamefont {Gai}}, \bibinfo
  {author} {\bibfnamefont {C.}~\bibnamefont {Jing}}, \bibinfo {author}
  {\bibfnamefont {J.~G.}\ \bibnamefont {Power}}, \bibinfo {author}
  {\bibfnamefont {C.~X.}\ \bibnamefont {Tang}},\ and\ \bibinfo {author}
  {\bibfnamefont {A.}~\bibnamefont {Zholents}},\ }\bibfield  {title} {\bibinfo
  {title} {High gradient limits due to single bunch beam breakup in a collinear
  dielectric wakefield accelerator},\ }\href
  {https://doi.org/10.1103/PhysRevSTAB.17.091302} {\bibfield  {journal}
  {\bibinfo  {journal} {Phys. Rev. ST Accel. Beams}\ }\textbf {\bibinfo
  {volume} {17}},\ \bibinfo {pages} {091302} (\bibinfo {year}
  {2014})}\BibitemShut {NoStop}%
\bibitem [{\citenamefont {Shchegolkov}\ \emph {et~al.}(2016)\citenamefont
  {Shchegolkov}, \citenamefont {Simakov},\ and\ \citenamefont
  {Zholents}}]{shchegolkov_towards_2016}%
  \BibitemOpen
  \bibfield  {author} {\bibinfo {author} {\bibfnamefont {D.~Y.}\ \bibnamefont
  {Shchegolkov}}, \bibinfo {author} {\bibfnamefont {E.~I.}\ \bibnamefont
  {Simakov}},\ and\ \bibinfo {author} {\bibfnamefont {A.~A.}\ \bibnamefont
  {Zholents}},\ }\bibfield  {title} {\bibinfo {title} {Towards a {Practical}
  {Multi}-{Meter} {Long} {Dielectric} {Wakefield} {Accelerator}: {Problems} and
  {Solutions}},\ }\href {https://doi.org/10.1109/TNS.2015.2482820} {\bibfield
  {journal} {\bibinfo  {journal} {IEEE Transactions on Nuclear Science}\
  }\textbf {\bibinfo {volume} {63}},\ \bibinfo {pages} {804} (\bibinfo {year}
  {2016})}\BibitemShut {NoStop}%
\bibitem [{\citenamefont {Baturin}\ and\ \citenamefont
  {Zholents}(2018)}]{baturin_stability_2018}%
  \BibitemOpen
  \bibfield  {author} {\bibinfo {author} {\bibfnamefont {S.~S.}\ \bibnamefont
  {Baturin}}\ and\ \bibinfo {author} {\bibfnamefont {A.}~\bibnamefont
  {Zholents}},\ }\bibfield  {title} {\bibinfo {title} {Stability condition for
  the drive bunch in a collinear wakefield accelerator},\ }\href
  {https://doi.org/10.1103/PhysRevAccelBeams.21.031301} {\bibfield  {journal}
  {\bibinfo  {journal} {Phys. Rev. Accel. Beams}\ }\textbf {\bibinfo {volume}
  {21}},\ \bibinfo {pages} {031301} (\bibinfo {year} {2018})}\BibitemShut
  {NoStop}%
\bibitem [{\citenamefont {Zholents}\ \emph {et~al.}(2018)\citenamefont
  {Zholents} \emph {et~al.}}]{Zholents:IPAC2018-TUPMF010}%
  \BibitemOpen
  \bibfield  {author} {\bibinfo {author} {\bibfnamefont {A.}~\bibnamefont
  {Zholents}} \emph {et~al.},\ }\bibfield  {title} {{\selectlanguage
  {english}\bibinfo {title} {{A} {C}onceptual {D}esign of a {C}ompact
  {W}akefield {A}ccelerator for a {H}igh {R}epetition {R}ate {M}ulti {U}ser
  {X-}ray {F}ree{-E}lectron {L}aser {F}acility}},\ }in\ \href
  {https://doi.org/doi:10.18429/JACoW-IPAC2018-TUPMF010} {{\selectlanguage
  {english}\emph {\bibinfo {booktitle} {Proc. 9th International Particle
  Accelerator Conference (IPAC'18), Vancouver, BC, Canada, April 29-May 4,
  2018}}}},\ \bibinfo {series and number} {\bibinfo {series} {International
  Particle Accelerator Conference}\ No.~\bibinfo {number} {9}}\ (\bibinfo
  {publisher} {JACoW Publishing},\ \bibinfo {address} {Geneva, Switzerland},\
  \bibinfo {year} {2018})\ pp.\ \bibinfo {pages} {1266--1268},\ \bibinfo {note}
  {https://doi.org/10.18429/JACoW-IPAC2018-TUPMF010}\BibitemShut {NoStop}%
\bibitem [{\citenamefont {Waldschmidt}\ and\ \citenamefont
  {{others}}(2018)}]{waldschmidt_design_2018}%
  \BibitemOpen
  \bibfield  {author} {\bibinfo {author} {\bibfnamefont {G.~J.}\ \bibnamefont
  {Waldschmidt}}\ and\ \bibinfo {author} {\bibnamefont {{others}}},\ }\bibfield
   {title} {{\selectlanguage {english}\bibinfo {title} {Design and {Test}
  {Plan} for a {Prototype} {Corrugated} {Waveguide}}},\ }in\ \href
  {https://doi.org/doi:10.18429/JACoW-IPAC2018-TUPML009} {{\selectlanguage
  {english}\emph {\bibinfo {booktitle} {Proc. 9th {International} {Particle}
  {Accelerator} {Conference} ({IPAC}'18), {Vancouver}, {BC}, {Canada}, {April}
  29-{May} 4, 2018}}}},\ \bibinfo {series and number} {International {Particle}
  {Accelerator} {Conference}}\ (\bibinfo  {publisher} {JACoW Publishing},\
  \bibinfo {address} {Geneva, Switzerland},\ \bibinfo {year} {2018})\ pp.\
  \bibinfo {pages} {1550--1552}\BibitemShut {NoStop}%
\bibitem [{\citenamefont {Zotter}\ and\ \citenamefont
  {Kheifets}(1998)}]{Zotter}%
  \BibitemOpen
  \bibfield  {author} {\bibinfo {author} {\bibfnamefont {B.}~\bibnamefont
  {Zotter}}\ and\ \bibinfo {author} {\bibfnamefont {S.}~\bibnamefont
  {Kheifets}},\ }\href@noop {} {\emph {\bibinfo {title} {Impedances and Wakes
  in High Energy Particle Accelerators}}}\ (\bibinfo  {publisher} {World
  Scientific Publishing Company},\ \bibinfo {year} {1998})\BibitemShut
  {NoStop}%
\bibitem [{\citenamefont {Chao}(1993)}]{Chao}%
  \BibitemOpen
  \bibfield  {author} {\bibinfo {author} {\bibfnamefont {A.}~\bibnamefont
  {Chao}},\ }\href@noop {} {\emph {\bibinfo {title} {Physics of Collective Beam
  Instabilities in High Energy Accelerators}}}\ (\bibinfo  {publisher} {Wiley
  and Sons, New York},\ \bibinfo {year} {1993})\BibitemShut {NoStop}%
\bibitem [{\citenamefont {Polyanin}\ and\ \citenamefont
  {Manzhirov}(1998)}]{Polyanin}%
  \BibitemOpen
  \bibfield  {author} {\bibinfo {author} {\bibfnamefont {A.}~\bibnamefont
  {Polyanin}}\ and\ \bibinfo {author} {\bibfnamefont {A.}~\bibnamefont
  {Manzhirov}},\ }\href@noop {} {\emph {\bibinfo {title} {Handbook of Integral
  Equations}}}\ (\bibinfo  {publisher} {CRC Press, Boca Raton},\ \bibinfo
  {year} {1998})\BibitemShut {NoStop}%
\bibitem [{\citenamefont {Zagorodnov}\ \emph {et~al.}(2015)\citenamefont
  {Zagorodnov}, \citenamefont {Bane},\ and\ \citenamefont
  {Stupakov}}]{zagorodnov_calculation_2015}%
  \BibitemOpen
  \bibfield  {author} {\bibinfo {author} {\bibfnamefont {I.}~\bibnamefont
  {Zagorodnov}}, \bibinfo {author} {\bibfnamefont {K.~L.~F.}\ \bibnamefont
  {Bane}},\ and\ \bibinfo {author} {\bibfnamefont {G.}~\bibnamefont
  {Stupakov}},\ }\bibfield  {title} {\bibinfo {title} {Calculation of
  wakefields in {2D} rectangular structures},\ }\href
  {https://doi.org/10.1103/PhysRevSTAB.18.104401} {\bibfield  {journal}
  {\bibinfo  {journal} {Phys. Rev. ST Accel. Beams}\ }\textbf {\bibinfo
  {volume} {18}},\ \bibinfo {pages} {104401} (\bibinfo {year}
  {2015})}\BibitemShut {NoStop}%
\bibitem [{\citenamefont {Siy}\ \emph {et~al.}(2019)\citenamefont {Siy},
  \citenamefont {Waldschmidt},\ and\ \citenamefont
  {Zholents}}]{Asiy-napac_2019}%
  \BibitemOpen
  \bibfield  {author} {\bibinfo {author} {\bibfnamefont {A.~E.}\ \bibnamefont
  {Siy}}, \bibinfo {author} {\bibfnamefont {G.~J.}\ \bibnamefont
  {Waldschmidt}},\ and\ \bibinfo {author} {\bibfnamefont {A.}~\bibnamefont
  {Zholents}},\ }\bibfield  {title} {\bibinfo {title} {Design of a compact
  wakefield accelerator based on a corrugated waveguide},\ }in\ \href
  {https://doi.org/https://doi.org/10.18429/JACoW-NAPAC2019-MOPLH26} {\emph
  {\bibinfo {booktitle} {Proc. 2019 North American Particle Accelerator
  Conference (NAPAC2019), Lansing, MI, USA, September 1-6, 2019}}}\ (\bibinfo
  {publisher} {JACoW},\ \bibinfo {address} {Geneva, Switzerland},\ \bibinfo
  {year} {2019})\ pp.\ \bibinfo {pages} {232--235}\BibitemShut {NoStop}%
\bibitem [{\citenamefont {Arthur}\ and\ \citenamefont
  {{others}}(2002)}]{arthur_linac_2002}%
  \BibitemOpen
  \bibfield  {author} {\bibinfo {author} {\bibfnamefont {J.}~\bibnamefont
  {Arthur}}\ and\ \bibinfo {author} {\bibnamefont {{others}}},\ }\href
  {https://www-public.slac.stanford.edu/sciDoc/docMeta.aspx?slacPubNumber=slac-r-593}
  {\emph {\bibinfo {title} {Linac {Coherent} {Light} {Source} ({LCLS})
  conceptual design report}}},\ \bibinfo {type} {{SLAC} {Report}}\ \bibinfo
  {number} {SLAC-R-593}\ (\bibinfo  {institution} {Stanford Linear Accelerator
  Center},\ \bibinfo {year} {2002})\BibitemShut {NoStop}%
\bibitem [{\citenamefont {Bosch}\ \emph {et~al.}(2008)\citenamefont {Bosch},
  \citenamefont {Kleman},\ and\ \citenamefont {Wu}}]{bosch_modeling_2008}%
  \BibitemOpen
  \bibfield  {author} {\bibinfo {author} {\bibfnamefont {R.~A.}\ \bibnamefont
  {Bosch}}, \bibinfo {author} {\bibfnamefont {K.~J.}\ \bibnamefont {Kleman}},\
  and\ \bibinfo {author} {\bibfnamefont {J.}~\bibnamefont {Wu}},\ }\bibfield
  {title} {\bibinfo {title} {Modeling two-stage bunch compression with
  wakefields: {Macroscopic} properties and microbunching instability},\ }\href
  {https://doi.org/10.1103/PhysRevSTAB.11.090702} {\bibfield  {journal}
  {\bibinfo  {journal} {Phys. Rev. ST Accel. Beams}\ }\textbf {\bibinfo
  {volume} {11}},\ \bibinfo {pages} {090702} (\bibinfo {year}
  {2008})}\BibitemShut {NoStop}%
\bibitem [{\citenamefont {Tan}\ \emph {et~al.}(2018)\citenamefont {Tan},
  \citenamefont {Piot},\ and\ \citenamefont
  {Zholents}}]{tan_longitudinal_2018}%
  \BibitemOpen
  \bibfield  {author} {\bibinfo {author} {\bibfnamefont {W.}~\bibnamefont
  {Tan}}, \bibinfo {author} {\bibfnamefont {P.}~\bibnamefont {Piot}},\ and\
  \bibinfo {author} {\bibfnamefont {A.}~\bibnamefont {Zholents}},\ }\bibfield
  {title} {\bibinfo {title} {Longitudinal {Beam}-{Shaping} {Simulation} for
  {Enhanced} {Transformer} {Ratio} in {Beam}-{Driven} {Accelerators}},\ }in\
  \href {https://doi.org/10.1109/AAC.2018.8659429} {\emph {\bibinfo {booktitle}
  {2018 {IEEE} {Advanced} {Accelerator} {Concepts} {Workshop} ({AAC})}}}\
  (\bibinfo {year} {2018})\ pp.\ \bibinfo {pages} {190--194}\BibitemShut
  {NoStop}%
\bibitem [{\citenamefont {Bane}\ and\ \citenamefont
  {Emma}(2005)}]{bane_litrack:_2005}%
  \BibitemOpen
  \bibfield  {author} {\bibinfo {author} {\bibfnamefont {K.~L.~F.}\
  \bibnamefont {Bane}}\ and\ \bibinfo {author} {\bibfnamefont {P.}~\bibnamefont
  {Emma}},\ }\bibfield  {title} {\bibinfo {title} {Litrack: {A} {Fast}
  {Longitudinal} {Phase} {Space} {Tracking} {Code} with {Graphical} {User}
  {Interface}},\ }in\ \href {https://doi.org/10.1109/PAC.2005.1591786} {\emph
  {\bibinfo {booktitle} {Proceedings of the 2005 {Particle} {Accelerator}
  {Conference}}}}\ (\bibinfo {year} {2005})\ pp.\ \bibinfo {pages}
  {4266--4268}\BibitemShut {NoStop}%
\bibitem [{\citenamefont {Zagorodnov}\ \emph {et~al.}(2004)\citenamefont
  {Zagorodnov}, \citenamefont {Weiland},\ and\ \citenamefont
  {Dohlus}}]{zagorodnov_wake_2004}%
  \BibitemOpen
  \bibfield  {author} {\bibinfo {author} {\bibfnamefont {I.}~\bibnamefont
  {Zagorodnov}}, \bibinfo {author} {\bibfnamefont {T.}~\bibnamefont
  {Weiland}},\ and\ \bibinfo {author} {\bibfnamefont {M.}~\bibnamefont
  {Dohlus}},\ }\href@noop {} {\emph {\bibinfo {title} {Wake fields generated by
  the {LOLA}-{IV} structure and the 3rd harmonic section in {TTF}-{II}}}},\
  \bibinfo {type} {Tech. Rep.}\ \bibinfo {number} {TESLA Report 2004-01}\
  (\bibinfo  {institution} {DESY},\ \bibinfo {address} {Darmstadt, Germany},\
  \bibinfo {year} {2004})\BibitemShut {NoStop}%
\bibitem [{\citenamefont {Fortin}\ \emph {et~al.}(2012)\citenamefont {Fortin},
  \citenamefont {{De Rainville}}, \citenamefont {Gardner}, \citenamefont
  {Parizeau},\ and\ \citenamefont {Gagn\'e}}]{deap}%
  \BibitemOpen
  \bibfield  {author} {\bibinfo {author} {\bibfnamefont {F.-A.}\ \bibnamefont
  {Fortin}}, \bibinfo {author} {\bibfnamefont {F.-M.}\ \bibnamefont {{De
  Rainville}}}, \bibinfo {author} {\bibfnamefont {M.-A.}\ \bibnamefont
  {Gardner}}, \bibinfo {author} {\bibfnamefont {M.}~\bibnamefont {Parizeau}},\
  and\ \bibinfo {author} {\bibfnamefont {C.}~\bibnamefont {Gagn\'e}},\
  }\bibfield  {title} {\bibinfo {title} {{DEAP}: Evolutionary algorithms made
  easy},\ }\href@noop {} {\bibfield  {journal} {\bibinfo  {journal} {Journal of
  Machine Learning Research}\ }\textbf {\bibinfo {volume} {13}},\ \bibinfo
  {pages} {2171} (\bibinfo {year} {2012})}\BibitemShut {NoStop}%
\bibitem [{\citenamefont {Charles}\ \emph {et~al.}(2017)\citenamefont
  {Charles}, \citenamefont {Boland}, \citenamefont {Dowd},\ and\ \citenamefont
  {Paganin}}]{Charles:IPAC2017-MOPIK055}%
  \BibitemOpen
  \bibfield  {author} {\bibinfo {author} {\bibfnamefont {T.}~\bibnamefont
  {Charles}}, \bibinfo {author} {\bibfnamefont {M.}~\bibnamefont {Boland}},
  \bibinfo {author} {\bibfnamefont {R.}~\bibnamefont {Dowd}},\ and\ \bibinfo
  {author} {\bibfnamefont {D.}~\bibnamefont {Paganin}},\ }\bibfield  {title}
  {{\selectlanguage {english}\bibinfo {title} {{B}eam by {D}esign: {C}urrent
  {P}ulse {S}haping {T}hrough {L}ongitudinal {D}ispersion {C}ontrol}},\ }in\
  \href {https://doi.org/https://doi.org/10.18429/JACoW-IPAC2017-MOPIK055}
  {{\selectlanguage {english}\emph {\bibinfo {booktitle} {Proc. of
  International Particle Accelerator Conference (IPAC'17), Copenhagen, Denmark,
  14-19 May, 2017}}}},\ \bibinfo {series and number} {\bibinfo {series}
  {International Particle Accelerator Conference}\ No.~\bibinfo {number} {8}}\
  (\bibinfo  {publisher} {JACoW},\ \bibinfo {address} {Geneva, Switzerland},\
  \bibinfo {year} {2017})\ pp.\ \bibinfo {pages} {644--647},\ \bibinfo {note}
  {https://doi.org/10.18429/JACoW-IPAC2017-MOPIK055}\BibitemShut {NoStop}%
\bibitem [{\citenamefont {England}\ \emph {et~al.}(2005)\citenamefont
  {England}, \citenamefont {Rosenzweig}, \citenamefont {Andonian},
  \citenamefont {Musumeci}, \citenamefont {Travish},\ and\ \citenamefont
  {Yoder}}]{england_sextupole_2005}%
  \BibitemOpen
  \bibfield  {author} {\bibinfo {author} {\bibfnamefont {R.~J.}\ \bibnamefont
  {England}}, \bibinfo {author} {\bibfnamefont {J.~B.}\ \bibnamefont
  {Rosenzweig}}, \bibinfo {author} {\bibfnamefont {G.}~\bibnamefont
  {Andonian}}, \bibinfo {author} {\bibfnamefont {P.}~\bibnamefont {Musumeci}},
  \bibinfo {author} {\bibfnamefont {G.}~\bibnamefont {Travish}},\ and\ \bibinfo
  {author} {\bibfnamefont {R.}~\bibnamefont {Yoder}},\ }\bibfield  {title}
  {\bibinfo {title} {Sextupole correction of the longitudinal transport of
  relativistic beams in dispersionless translating sections},\ }\href
  {https://doi.org/10.1103/PhysRevSTAB.8.012801} {\bibfield  {journal}
  {\bibinfo  {journal} {Phys. Rev. ST Accel. Beams}\ }\textbf {\bibinfo
  {volume} {8}},\ \bibinfo {pages} {012801} (\bibinfo {year}
  {2005})}\BibitemShut {NoStop}%
\bibitem [{\citenamefont {{Xu}}\ \emph {et~al.}(2018)\citenamefont {{Xu}},
  \citenamefont {{Jing}}, \citenamefont {{Kanareykin}}, \citenamefont
  {{Piot}},\ and\ \citenamefont {{Power}}}]{tianzhe-AAC18}%
  \BibitemOpen
  \bibfield  {author} {\bibinfo {author} {\bibfnamefont {T.}~\bibnamefont
  {{Xu}}}, \bibinfo {author} {\bibfnamefont {C.}~\bibnamefont {{Jing}}},
  \bibinfo {author} {\bibfnamefont {A.}~\bibnamefont {{Kanareykin}}}, \bibinfo
  {author} {\bibfnamefont {P.}~\bibnamefont {{Piot}}},\ and\ \bibinfo {author}
  {\bibfnamefont {J.}~\bibnamefont {{Power}}},\ }\bibfield  {title} {\bibinfo
  {title} {Optimized electron bunch current distribution from a radiofrequency
  photo-emission source},\ }in\ \href
  {https://doi.org/10.1109/AAC.2018.8659440} {\emph {\bibinfo {booktitle}
  {Proc. 2018 IEEE Advanced Accelerator Concepts Workshop (AAC)}}}\ (\bibinfo
  {year} {2018})\ pp.\ \bibinfo {pages} {1--5}\BibitemShut {NoStop}%
\bibitem [{\citenamefont {Floettmann}(2017)}]{floettmann_astra_2017}%
  \BibitemOpen
  \bibfield  {author} {\bibinfo {author} {\bibfnamefont {K.}~\bibnamefont
  {Floettmann}},\ }\href {http://www.desy.de/ mpyflo/} {\emph {\bibinfo {title}
  {{ASTRA} – {A} {Space} {Charge} {Tracking} {Algorithm}}}}\ (\bibinfo
  {publisher} {Deutsches Elektronen-Synchrotron},\ \bibinfo {address} {Hamburg,
  Germany},\ \bibinfo {year} {2017})\BibitemShut {NoStop}%
\bibitem [{\citenamefont {Bisognano}\ \emph {et~al.}(2011)\citenamefont
  {Bisognano}, \citenamefont {Bosch}, \citenamefont {Eisert}, \citenamefont
  {Fisher}, \citenamefont {Green}, \citenamefont {Jacobs}, \citenamefont
  {Kleman}, \citenamefont {Kulpin}, \citenamefont {Rogers}, \citenamefont
  {Lawler}, \citenamefont {Yavuz},\ and\ \citenamefont
  {Legg}}]{bisognano_progress_2011_2}%
  \BibitemOpen
  \bibfield  {author} {\bibinfo {author} {\bibfnamefont {J.}~\bibnamefont
  {Bisognano}}, \bibinfo {author} {\bibfnamefont {R.}~\bibnamefont {Bosch}},
  \bibinfo {author} {\bibfnamefont {D.}~\bibnamefont {Eisert}}, \bibinfo
  {author} {\bibfnamefont {M.}~\bibnamefont {Fisher}}, \bibinfo {author}
  {\bibfnamefont {M.}~\bibnamefont {Green}}, \bibinfo {author} {\bibfnamefont
  {K.}~\bibnamefont {Jacobs}}, \bibinfo {author} {\bibfnamefont
  {K.}~\bibnamefont {Kleman}}, \bibinfo {author} {\bibfnamefont
  {J.}~\bibnamefont {Kulpin}}, \bibinfo {author} {\bibfnamefont
  {G.}~\bibnamefont {Rogers}}, \bibinfo {author} {\bibfnamefont
  {J.}~\bibnamefont {Lawler}}, \bibinfo {author} {\bibfnamefont
  {D.}~\bibnamefont {Yavuz}},\ and\ \bibinfo {author} {\bibfnamefont
  {R.}~\bibnamefont {Legg}},\ }\href
  {https://misportal.jlab.org/ul/publications/downloadFile.cfm?pub_id=10514}
  {\emph {\bibinfo {title} {Progress Toward the Wisconsin Free Electron
  Laser}}},\ \bibinfo {type} {Tech. Rep.}\ (\bibinfo  {institution}
  {{J}efferson {L}ab},\ \bibinfo {year} {2011})\ \bibinfo {note}
  {{JLAB}-{ACC}--11-1333}\BibitemShut {NoStop}%
\bibitem [{\citenamefont {Bisognano}\ \emph {et~al.}(2013)\citenamefont
  {Bisognano}, \citenamefont {Bissen}, \citenamefont {Bosch}, \citenamefont
  {Efremov}, \citenamefont {Eisert}, \citenamefont {Fisher}, \citenamefont
  {Green}, \citenamefont {Jacobs}, \citenamefont {Keil}, \citenamefont
  {Kleman}, \citenamefont {Rogers}, \citenamefont {Severson}, \citenamefont
  {D~Yavuz}, \citenamefont {Legg}, \citenamefont {Bachimanchi}, \citenamefont
  {Hovater}, \citenamefont {Plawski},\ and\ \citenamefont
  {Powers}}]{bisognano_wisconsin_2013}%
  \BibitemOpen
  \bibfield  {author} {\bibinfo {author} {\bibfnamefont {J.}~\bibnamefont
  {Bisognano}}, \bibinfo {author} {\bibfnamefont {M.}~\bibnamefont {Bissen}},
  \bibinfo {author} {\bibfnamefont {R.}~\bibnamefont {Bosch}}, \bibinfo
  {author} {\bibfnamefont {M.}~\bibnamefont {Efremov}}, \bibinfo {author}
  {\bibfnamefont {D.}~\bibnamefont {Eisert}}, \bibinfo {author} {\bibfnamefont
  {M.}~\bibnamefont {Fisher}}, \bibinfo {author} {\bibfnamefont
  {M.}~\bibnamefont {Green}}, \bibinfo {author} {\bibfnamefont
  {K.}~\bibnamefont {Jacobs}}, \bibinfo {author} {\bibfnamefont
  {R.}~\bibnamefont {Keil}}, \bibinfo {author} {\bibfnamefont {K.}~\bibnamefont
  {Kleman}}, \bibinfo {author} {\bibfnamefont {G.}~\bibnamefont {Rogers}},
  \bibinfo {author} {\bibfnamefont {M.}~\bibnamefont {Severson}}, \bibinfo
  {author} {\bibfnamefont {D.}~\bibnamefont {D~Yavuz}}, \bibinfo {author}
  {\bibfnamefont {R.}~\bibnamefont {Legg}}, \bibinfo {author} {\bibfnamefont
  {R.}~\bibnamefont {Bachimanchi}}, \bibinfo {author} {\bibfnamefont
  {C.}~\bibnamefont {Hovater}}, \bibinfo {author} {\bibfnamefont
  {T.}~\bibnamefont {Plawski}},\ and\ \bibinfo {author} {\bibfnamefont
  {T.}~\bibnamefont {Powers}},\ }\bibfield  {title} {\bibinfo {title}
  {Wisconsin {SRF} {Electron} {Gun} {Commissioning}},\ }in\ \href@noop {}
  {\emph {\bibinfo {booktitle} {Proceedings of the 25th {Particle}
  {Accelerator} {Conference}, {PAC} 2013}}}\ (\bibinfo {year} {2013})\ p.\
  \bibinfo {pages} {622}\BibitemShut {NoStop}%
\bibitem [{\citenamefont {Legg}\ \emph {et~al.}(2008)\citenamefont {Legg},
  \citenamefont {Graves}, \citenamefont {Grimm},\ and\ \citenamefont
  {Piot}}]{legg_half_2008}%
  \BibitemOpen
  \bibfield  {author} {\bibinfo {author} {\bibfnamefont {R.}~\bibnamefont
  {Legg}}, \bibinfo {author} {\bibfnamefont {W.}~\bibnamefont {Graves}},
  \bibinfo {author} {\bibfnamefont {T.}~\bibnamefont {Grimm}},\ and\ \bibinfo
  {author} {\bibfnamefont {P.}~\bibnamefont {Piot}},\ }\bibfield  {title}
  {\bibinfo {title} {Half wave injector design for {WiFEL}},\ }in\ \href@noop
  {} {\emph {\bibinfo {booktitle} {{EPAC} 2008 - {Contributions} to the
  {Proceedings}}}}\ (\bibinfo  {publisher} {European Physical Society
  Accelerator Group (EPS-AG)},\ \bibinfo {year} {2008})\ pp.\ \bibinfo {pages}
  {469--471}\BibitemShut {NoStop}%
\bibitem [{\citenamefont {Tan}\ \emph {et~al.}(2019)\citenamefont {Tan},
  \citenamefont {Piot},\ and\ \citenamefont
  {Zholents}}]{tan_longitudinal-phase-space_2019}%
  \BibitemOpen
  \bibfield  {author} {\bibinfo {author} {\bibfnamefont {W.~H.}\ \bibnamefont
  {Tan}}, \bibinfo {author} {\bibfnamefont {P.}~\bibnamefont {Piot}},\ and\
  \bibinfo {author} {\bibfnamefont {A.}~\bibnamefont {Zholents}},\ }\bibfield
  {title} {{\selectlanguage {english}\bibinfo {title}
  {Longitudinal-{Phase}-{Space} {Manipulation} for {Efficient} {Beam}-{Driven}
  {Structure} {Wakefield} {Acceleration}}},\ }in\ \href
  {https://doi.org/doi:10.18429/JACoW-IPAC2019-WEZZPLS3} {{\selectlanguage
  {english}\emph {\bibinfo {booktitle} {Proc. 10th {International} {Particle}
  {Accelerator} {Conference} ({IPAC}'19), {Melbourne}, {Australia}, 19-24 {May}
  2019}}}},\ \bibinfo {series and number} {International {Particle}
  {Accelerator} {Conference}}\ (\bibinfo  {publisher} {JACoW Publishing},\
  \bibinfo {address} {Geneva, Switzerland},\ \bibinfo {year} {2019})\ pp.\
  \bibinfo {pages} {2296--2299},\ \bibinfo {note} {issue: 10}\BibitemShut
  {NoStop}%
\bibitem [{\citenamefont {Nielsen}\ \emph {et~al.}(2020)\citenamefont
  {Nielsen}, \citenamefont {Hauge}, \citenamefont {Krauthammer},\ and\
  \citenamefont {Baurichter}}]{Nielsen-HTS-solenoid-2012}%
  \BibitemOpen
  \bibfield  {author} {\bibinfo {author} {\bibfnamefont {G.}~\bibnamefont
  {Nielsen}}, \bibinfo {author} {\bibfnamefont {N.}~\bibnamefont {Hauge}},
  \bibinfo {author} {\bibfnamefont {E.}~\bibnamefont {Krauthammer}},\ and\
  \bibinfo {author} {\bibfnamefont {A.}~\bibnamefont {Baurichter}},\ }\bibfield
   {title} {\bibinfo {title} {Compact high-{T}c 2{G} superconducting solenoid
  for superconducting {RF} electron gun},\ }in\ \href@noop {} {\emph {\bibinfo
  {booktitle} {Proc. 4th International Particle Accelerator Conference
  (IPAC2013), Shanghai, China, 12-17 May, 2013}}}\ (\bibinfo  {publisher}
  {JACoW},\ \bibinfo {address} {Geneva, Switzerland},\ \bibinfo {year} {2020})\
  pp.\ \bibinfo {pages} {3514--3516}\BibitemShut {NoStop}%
\bibitem [{\citenamefont {Xu}\ \emph {et~al.}(2019)\citenamefont {Xu},
  \citenamefont {Jing}, \citenamefont {Kanareykin}, \citenamefont {Piot},\ and\
  \citenamefont {Power}}]{Xu:IPAC2019-TUPTS104}%
  \BibitemOpen
  \bibfield  {author} {\bibinfo {author} {\bibfnamefont {T.}~\bibnamefont
  {Xu}}, \bibinfo {author} {\bibfnamefont {C.-J.}\ \bibnamefont {Jing}},
  \bibinfo {author} {\bibfnamefont {A.}~\bibnamefont {Kanareykin}}, \bibinfo
  {author} {\bibfnamefont {P.}~\bibnamefont {Piot}},\ and\ \bibinfo {author}
  {\bibfnamefont {J.}~\bibnamefont {Power}},\ }\bibfield  {title}
  {{\selectlanguage {english}\bibinfo {title} {{S}patio{-T}emporal {S}haping of
  the {P}hotocathode {L}aser {P}ulse for {L}ow{-E}mittance {S}haped {E}lectron
  {B}unches}},\ }in\ \href
  {https://doi.org/doi:10.18429/JACoW-IPAC2019-TUPTS104} {{\selectlanguage
  {english}\emph {\bibinfo {booktitle} {Proc. 10th International Particle
  Accelerator Conference (IPAC'19), Melbourne, Australia, 19-24 May 2019}}}},\
  \bibinfo {series and number} {\bibinfo {series} {International Particle
  Accelerator Conference}\ No.~\bibinfo {number} {10}}\ (\bibinfo  {publisher}
  {JACoW Publishing},\ \bibinfo {address} {Geneva, Switzerland},\ \bibinfo
  {year} {2019})\ pp.\ \bibinfo {pages} {2163--2166}\BibitemShut {NoStop}%
\bibitem [{\citenamefont {Gilevich}\ \emph {et~al.}(2020)\citenamefont
  {Gilevich}, \citenamefont {Alverson}, \citenamefont {Carbajo}, \citenamefont
  {Droste}, \citenamefont {Edstrom}, \citenamefont {Fry}, \citenamefont
  {Greenberg}, \citenamefont {Lemons}, \citenamefont {Miahnahri}, \citenamefont
  {Polzin}, \citenamefont {Vetter},\ and\ \citenamefont
  {Zhou}}]{Gilevich-lcls-ii_2020}%
  \BibitemOpen
  \bibfield  {author} {\bibinfo {author} {\bibfnamefont {S.}~\bibnamefont
  {Gilevich}}, \bibinfo {author} {\bibfnamefont {S.}~\bibnamefont {Alverson}},
  \bibinfo {author} {\bibfnamefont {S.}~\bibnamefont {Carbajo}}, \bibinfo
  {author} {\bibfnamefont {S.}~\bibnamefont {Droste}}, \bibinfo {author}
  {\bibfnamefont {S.}~\bibnamefont {Edstrom}}, \bibinfo {author} {\bibfnamefont
  {A.}~\bibnamefont {Fry}}, \bibinfo {author} {\bibfnamefont {M.}~\bibnamefont
  {Greenberg}}, \bibinfo {author} {\bibfnamefont {R.}~\bibnamefont {Lemons}},
  \bibinfo {author} {\bibfnamefont {A.}~\bibnamefont {Miahnahri}}, \bibinfo
  {author} {\bibfnamefont {W.}~\bibnamefont {Polzin}}, \bibinfo {author}
  {\bibfnamefont {S.}~\bibnamefont {Vetter}},\ and\ \bibinfo {author}
  {\bibfnamefont {F.}~\bibnamefont {Zhou}},\ }\bibfield  {title} {\bibinfo
  {title} {The {LCLS-II} photo-injector drive laser system},\ }in\ \href
  {https://doi.org/10.1364/CLEO_SI.2020.SW3E.3} {\emph {\bibinfo {booktitle}
  {Proc. Conference on Lasers and Electro-Optics}}}\ (\bibinfo  {publisher}
  {Optical Society of America},\ \bibinfo {year} {2020})\ p.\ \bibinfo {pages}
  {SW3E.3}\BibitemShut {NoStop}%
\bibitem [{\citenamefont {Ferrini}\ \emph {et~al.}(1998)\citenamefont
  {Ferrini}, \citenamefont {Michelato},\ and\ \citenamefont
  {Parmigiani}}]{FERRINI199821}%
  \BibitemOpen
  \bibfield  {author} {\bibinfo {author} {\bibfnamefont {G.}~\bibnamefont
  {Ferrini}}, \bibinfo {author} {\bibfnamefont {P.}~\bibnamefont {Michelato}},\
  and\ \bibinfo {author} {\bibfnamefont {F.}~\bibnamefont {Parmigiani}},\
  }\bibfield  {title} {\bibinfo {title} {A {Monte} {Carlo} simulation of low
  energy photoelectron scattering in {Cs$_2$Te}},\ }\href
  {https://doi.org/https://doi.org/10.1016/S0038-1098(97)10237-X} {\bibfield
  {journal} {\bibinfo  {journal} {Solid State Communications}\ }\textbf
  {\bibinfo {volume} {106}},\ \bibinfo {pages} {21 } (\bibinfo {year}
  {1998})}\BibitemShut {NoStop}%
\bibitem [{\citenamefont {Piot}\ \emph {et~al.}(2013)\citenamefont {Piot},
  \citenamefont {Sun}, \citenamefont {Maxwell}, \citenamefont {Ruan},
  \citenamefont {Secchi},\ and\ \citenamefont
  {Thangaraj}}]{piot_formation_2013}%
  \BibitemOpen
  \bibfield  {author} {\bibinfo {author} {\bibfnamefont {P.}~\bibnamefont
  {Piot}}, \bibinfo {author} {\bibfnamefont {Y.-E.}\ \bibnamefont {Sun}},
  \bibinfo {author} {\bibfnamefont {T.~J.}\ \bibnamefont {Maxwell}}, \bibinfo
  {author} {\bibfnamefont {J.}~\bibnamefont {Ruan}}, \bibinfo {author}
  {\bibfnamefont {E.}~\bibnamefont {Secchi}},\ and\ \bibinfo {author}
  {\bibfnamefont {J.~C.~T.}\ \bibnamefont {Thangaraj}},\ }\bibfield  {title}
  {\bibinfo {title} {Formation and acceleration of uniformly filled ellipsoidal
  electron bunches obtained via space-charge-driven expansion from a
  cesium-telluride photocathode},\ }\href
  {https://doi.org/10.1103/PhysRevSTAB.16.010102} {\bibfield  {journal}
  {\bibinfo  {journal} {Phys. Rev. ST Accel. Beams}\ }\textbf {\bibinfo
  {volume} {16}},\ \bibinfo {pages} {010102} (\bibinfo {year}
  {2013})}\BibitemShut {NoStop}%
\bibitem [{\citenamefont {Aryshev}\ \emph {et~al.}(2017)\citenamefont
  {Aryshev}, \citenamefont {Shevelev}, \citenamefont {Honda}, \citenamefont
  {Terunuma},\ and\ \citenamefont {Urakawa}}]{AryshevCsTe2017}%
  \BibitemOpen
  \bibfield  {author} {\bibinfo {author} {\bibfnamefont {A.}~\bibnamefont
  {Aryshev}}, \bibinfo {author} {\bibfnamefont {M.}~\bibnamefont {Shevelev}},
  \bibinfo {author} {\bibfnamefont {Y.}~\bibnamefont {Honda}}, \bibinfo
  {author} {\bibfnamefont {N.}~\bibnamefont {Terunuma}},\ and\ \bibinfo
  {author} {\bibfnamefont {J.}~\bibnamefont {Urakawa}},\ }\bibfield  {title}
  {\bibinfo {title} {Femtosecond response time measurements of a {C}s2{T}e
  photocathode},\ }\href {https://doi.org/10.1063/1.4994224} {\bibfield
  {journal} {\bibinfo  {journal} {Applied Physics Letters}\ }\textbf {\bibinfo
  {volume} {111}},\ \bibinfo {pages} {033508} (\bibinfo {year} {2017})},\
  \Eprint {https://arxiv.org/abs/https://doi.org/10.1063/1.4994224}
  {https://doi.org/10.1063/1.4994224} \BibitemShut {NoStop}%
\bibitem [{\citenamefont {Jain}\ and\ \citenamefont
  {{others}}(2017)}]{jain_650_2017}%
  \BibitemOpen
  \bibfield  {author} {\bibinfo {author} {\bibfnamefont {V.}~\bibnamefont
  {Jain}}\ and\ \bibinfo {author} {\bibnamefont {{others}}},\ }\bibfield
  {title} {\bibinfo {title} {650 {MHz} {Elliptical} {Superconducting} {RF}
  {Cavities} for {PIP}-{II} {Project}},\ }in\ \href
  {https://doi.org/10.18429/JACoW-NAPAC2016-WEB3CO03} {\emph {\bibinfo
  {booktitle} {Proceedings, 2nd {North} {American} {Particle} {Accelerator}
  {Conference} ({NAPAC2016}): {Chicago}, {Illinois}, {USA}, {October} 9-14,
  2016}}}\ (\bibinfo {year} {2017})\ p.\ \bibinfo {pages}
  {WEB3CO03}\BibitemShut {NoStop}%
\bibitem [{\citenamefont {Mitri}\ and\ \citenamefont
  {Cornacchia}(2015)}]{mitri_transverse_2015}%
  \BibitemOpen
  \bibfield  {author} {\bibinfo {author} {\bibfnamefont {S.~D.}\ \bibnamefont
  {Mitri}}\ and\ \bibinfo {author} {\bibfnamefont {M.}~\bibnamefont
  {Cornacchia}},\ }\bibfield  {title} {\bibinfo {title} {Transverse
  emittance-preserving arc compressor for high-brightness electron beam-based
  light sources and colliders},\ }\href
  {https://doi.org/10.1209/0295-5075/109/62002} {\bibfield  {journal} {\bibinfo
   {journal} {EPL (Europhysics Letters)}\ }\textbf {\bibinfo {volume} {109}},\
  \bibinfo {pages} {62002} (\bibinfo {year} {2015})}\BibitemShut {NoStop}%
\bibitem [{\citenamefont {Akkermans}\ \emph {et~al.}(2017)\citenamefont
  {Akkermans}, \citenamefont {Di~Mitri}, \citenamefont {Douglas},\ and\
  \citenamefont {Setija}}]{akkermans_compact_2017}%
  \BibitemOpen
  \bibfield  {author} {\bibinfo {author} {\bibfnamefont {J.~A.~G.}\
  \bibnamefont {Akkermans}}, \bibinfo {author} {\bibfnamefont {S.}~\bibnamefont
  {Di~Mitri}}, \bibinfo {author} {\bibfnamefont {D.}~\bibnamefont {Douglas}},\
  and\ \bibinfo {author} {\bibfnamefont {I.~D.}\ \bibnamefont {Setija}},\
  }\bibfield  {title} {\bibinfo {title} {Compact compressive arc and beam
  switchyard for energy recovery linac-driven ultraviolet free electron
  lasers},\ }\href {https://doi.org/10.1103/PhysRevAccelBeams.20.080705}
  {\bibfield  {journal} {\bibinfo  {journal} {Phys. Rev. Accel. Beams}\
  }\textbf {\bibinfo {volume} {20}},\ \bibinfo {pages} {080705} (\bibinfo
  {year} {2017})}\BibitemShut {NoStop}%
\bibitem [{\citenamefont {Mitri}(2018)}]{mitri_bunch_2018}%
  \BibitemOpen
  \bibfield  {author} {\bibinfo {author} {\bibfnamefont {S.~D.}\ \bibnamefont
  {Mitri}},\ }\bibfield  {title} {\bibinfo {title} {Bunch {Length}
  {Compressors}},\ }in\ \href {https://doi.org/10.23730/CYRSP-2018-001.363}
  {\emph {\bibinfo {booktitle} {Proceedings of the {CAS}-{CERN} {Accelerator}
  {School} on {Free} {Electron} {Lasers} and {Energy} {Recovery} {Linacs}}}},\
  Vol.~\bibinfo {volume} {1}\ (\bibinfo  {publisher} {CERN},\ \bibinfo
  {address} {Geneva, Switzerland},\ \bibinfo {year} {2018})\ p.\ \bibinfo
  {pages} {363}\BibitemShut {NoStop}%
\bibitem [{\citenamefont {Chao}\ \emph {et~al.}(2013)\citenamefont {Chao},
  \citenamefont {Mess}, \citenamefont {Tigner},\ and\ \citenamefont
  {Zimmermann}}]{chao_handbook_2013}%
  \BibitemOpen
  \bibfield  {author} {\bibinfo {author} {\bibfnamefont {A.}~\bibnamefont
  {Chao}}, \bibinfo {author} {\bibfnamefont {K.~H.}\ \bibnamefont {Mess}},
  \bibinfo {author} {\bibfnamefont {M.}~\bibnamefont {Tigner}},\ and\ \bibinfo
  {author} {\bibfnamefont {F.}~\bibnamefont {Zimmermann}},\ }\href
  {https://www.worldscientific.com/doi/abs/10.1142/8543} {\emph {\bibinfo
  {title} {Handbook of {Accelerator} {Physics} and {Engineering}}}},\ \bibinfo
  {edition} {2nd}\ ed.\ (\bibinfo  {publisher} {WORLD SCIENTIFIC},\ \bibinfo
  {year} {2013})\BibitemShut {NoStop}%
\bibitem [{\citenamefont {Robin}\ \emph {et~al.}(1993)\citenamefont {Robin},
  \citenamefont {Forest}, \citenamefont {Pellegrini},\ and\ \citenamefont
  {Amiry}}]{robin_quasi-isochronous_1993}%
  \BibitemOpen
  \bibfield  {author} {\bibinfo {author} {\bibfnamefont {D.}~\bibnamefont
  {Robin}}, \bibinfo {author} {\bibfnamefont {E.}~\bibnamefont {Forest}},
  \bibinfo {author} {\bibfnamefont {C.}~\bibnamefont {Pellegrini}},\ and\
  \bibinfo {author} {\bibfnamefont {A.}~\bibnamefont {Amiry}},\ }\bibfield
  {title} {\bibinfo {title} {Quasi-isochronous storage rings},\ }\href
  {https://doi.org/10.1103/PhysRevE.48.2149} {\bibfield  {journal} {\bibinfo
  {journal} {Phys. Rev. E}\ }\textbf {\bibinfo {volume} {48}},\ \bibinfo
  {pages} {2149} (\bibinfo {year} {1993})}\BibitemShut {NoStop}%
\bibitem [{\citenamefont {Williams}\ \emph {et~al.}(2020)\citenamefont
  {Williams}, \citenamefont {Pérez-Segurana}, \citenamefont {Bailey},
  \citenamefont {Thorin}, \citenamefont {Kyle},\ and\ \citenamefont
  {Svensson}}]{williams_arclike_2020}%
  \BibitemOpen
  \bibfield  {author} {\bibinfo {author} {\bibfnamefont {P.~H.}\ \bibnamefont
  {Williams}}, \bibinfo {author} {\bibfnamefont {G.}~\bibnamefont
  {Pérez-Segurana}}, \bibinfo {author} {\bibfnamefont {I.~R.}\ \bibnamefont
  {Bailey}}, \bibinfo {author} {\bibfnamefont {S.}~\bibnamefont {Thorin}},
  \bibinfo {author} {\bibfnamefont {B.}~\bibnamefont {Kyle}},\ and\ \bibinfo
  {author} {\bibfnamefont {J.~B.}\ \bibnamefont {Svensson}},\ }\bibfield
  {title} {\bibinfo {title} {Arclike variable bunch compressors},\ }\href
  {https://doi.org/10.1103/PhysRevAccelBeams.23.100701} {\bibfield  {journal}
  {\bibinfo  {journal} {Physical Review Accelerators and Beams}\ }\textbf
  {\bibinfo {volume} {23}},\ \bibinfo {pages} {100701} (\bibinfo {year}
  {2020})}\BibitemShut {NoStop}%
\bibitem [{\citenamefont {Saldin}\ \emph {et~al.}(2004)\citenamefont {Saldin},
  \citenamefont {Schneidmiller},\ and\ \citenamefont {Yurkov}}]{SALDIN2004355}%
  \BibitemOpen
  \bibfield  {author} {\bibinfo {author} {\bibfnamefont {E.}~\bibnamefont
  {Saldin}}, \bibinfo {author} {\bibfnamefont {E.}~\bibnamefont
  {Schneidmiller}},\ and\ \bibinfo {author} {\bibfnamefont {M.}~\bibnamefont
  {Yurkov}},\ }\bibfield  {title} {\bibinfo {title} {Longitudinal space
  charge-driven microbunching instability in the {TESLA} test facility linac},\
  }\href {https://doi.org/https://doi.org/10.1016/j.nima.2004.04.067}
  {\bibfield  {journal} {\bibinfo  {journal} {Nucl. Instrum. Methods Phys.
  Res., Sect. A}\ }\textbf {\bibinfo {volume} {528}},\ \bibinfo {pages} {355 }
  (\bibinfo {year} {2004})},\ \bibinfo {note} {also in Proc. 25th International
  Free Electron Laser Conference, and the 10th FEL Users Workshop}\BibitemShut
  {NoStop}%
\bibitem [{\citenamefont {Huang}\ \emph {et~al.}(2004)\citenamefont {Huang},
  \citenamefont {Borland}, \citenamefont {Emma}, \citenamefont {Wu},
  \citenamefont {Limborg}, \citenamefont {Stupakov},\ and\ \citenamefont
  {Welch}}]{huang_suppression_2004}%
  \BibitemOpen
  \bibfield  {author} {\bibinfo {author} {\bibfnamefont {Z.}~\bibnamefont
  {Huang}}, \bibinfo {author} {\bibfnamefont {M.}~\bibnamefont {Borland}},
  \bibinfo {author} {\bibfnamefont {P.}~\bibnamefont {Emma}}, \bibinfo {author}
  {\bibfnamefont {J.}~\bibnamefont {Wu}}, \bibinfo {author} {\bibfnamefont
  {C.}~\bibnamefont {Limborg}}, \bibinfo {author} {\bibfnamefont
  {G.}~\bibnamefont {Stupakov}},\ and\ \bibinfo {author} {\bibfnamefont
  {J.}~\bibnamefont {Welch}},\ }\bibfield  {title} {\bibinfo {title}
  {Suppression of microbunching instability in the linac coherent light
  source},\ }\href {https://doi.org/10.1103/PhysRevSTAB.7.074401} {\bibfield
  {journal} {\bibinfo  {journal} {Phys. Rev. ST Accel. Beams}\ }\textbf
  {\bibinfo {volume} {7}},\ \bibinfo {pages} {074401} (\bibinfo {year}
  {2004})}\BibitemShut {NoStop}%
\bibitem [{\citenamefont {Derbenev}\ \emph {et~al.}(1995)\citenamefont
  {Derbenev}, \citenamefont {Rossbach}, \citenamefont {Saldin},\ and\
  \citenamefont {Shiltsev}}]{derbenev_microbunch_1995}%
  \BibitemOpen
  \bibfield  {author} {\bibinfo {author} {\bibfnamefont {Y.~S.}\ \bibnamefont
  {Derbenev}}, \bibinfo {author} {\bibfnamefont {J.}~\bibnamefont {Rossbach}},
  \bibinfo {author} {\bibfnamefont {E.~L.}\ \bibnamefont {Saldin}},\ and\
  \bibinfo {author} {\bibfnamefont {V.~D.}\ \bibnamefont {Shiltsev}},\ }\href
  {https://doi.org/10.3204/PUBDB-2018-04128} {\emph {\bibinfo {title}
  {Microbunch radiative tail - head interaction}}},\ \bibinfo {type} {Tech.
  Rep.}\ \bibinfo {number} {TESLA-FEL 1995-05}\ (\bibinfo  {institution}
  {DESY},\ \bibinfo {year} {1995})\ \bibinfo {note} {series: TESLA-FEL
  Reports}\BibitemShut {NoStop}%
\bibitem [{\citenamefont {Huang}\ \emph {et~al.}(2017)\citenamefont {Huang}
  \emph {et~al.}}]{HuangLLRF2017}%
  \BibitemOpen
  \bibfield  {author} {\bibinfo {author} {\bibfnamefont {G.}~\bibnamefont
  {Huang}} \emph {et~al.},\ }\bibfield  {title} {{\selectlanguage
  {english}\bibinfo {title} {{H}igh {P}recision {RF} {C}ontrol for the
  {LCLS-II}}},\ }in\ \href
  {https://doi.org/https://doi.org/10.18429/JACoW-NAPAC2016-FRA2IO02}
  {{\selectlanguage {english}\emph {\bibinfo {booktitle} {Proc. 3rd North
  American Particle Accelerator Conference (NAPAC'16), Chicago, IL, USA,
  October 9-14, 2016}}}},\ \bibinfo {series and number} {\bibinfo {series}
  {North American Particle Accelerator Conference}\ No.~\bibinfo {number} {3}}\
  (\bibinfo  {publisher} {JACoW},\ \bibinfo {address} {Geneva, Switzerland},\
  \bibinfo {year} {2017})\ pp.\ \bibinfo {pages} {1292--1296},\ \bibinfo {note}
  {https://doi.org/10.18429/JACoW-NAPAC2016-FRA2IO02}\BibitemShut {NoStop}%
\bibitem [{\citenamefont {Mohayai}\ \emph {et~al.}(2017)\citenamefont
  {Mohayai}, \citenamefont {Snopok}, \citenamefont {Neuffer},\ and\
  \citenamefont {Rogers}}]{mohayai_novel_2017}%
  \BibitemOpen
  \bibfield  {author} {\bibinfo {author} {\bibfnamefont {T.~A.}\ \bibnamefont
  {Mohayai}}, \bibinfo {author} {\bibfnamefont {P.}~\bibnamefont {Snopok}},
  \bibinfo {author} {\bibfnamefont {D.}~\bibnamefont {Neuffer}},\ and\ \bibinfo
  {author} {\bibfnamefont {C.}~\bibnamefont {Rogers}},\ }\bibfield  {title}
  {\bibinfo {title} {Novel {Application} of {Density} {Estimation} {Techniques}
  in {Muon} {Ionization} {Cooling} {Experiment}},\ }in\ \href
  {http://lss.fnal.gov/archive/2017/conf/fermilab-conf-17-476-apc.pdf} {\emph
  {\bibinfo {booktitle} {Proceedings, {Meeting} of the {APS} {Division} of
  {Particles} and {Fields} ({DPF} 2017): {Fermilab}, {Batavia}, {Illinois},
  {USA}, {July} 31 - {August} 4, 2017}}}\ (\bibinfo {year} {2017})\BibitemShut
  {NoStop}%
\bibitem [{\citenamefont {Qiang}\ \emph {et~al.}(2009)\citenamefont {Qiang},
  \citenamefont {Ryne}, \citenamefont {Venturini}, \citenamefont {Zholents},\
  and\ \citenamefont {Pogorelov}}]{qiang_high_2009}%
  \BibitemOpen
  \bibfield  {author} {\bibinfo {author} {\bibfnamefont {J.}~\bibnamefont
  {Qiang}}, \bibinfo {author} {\bibfnamefont {R.~D.}\ \bibnamefont {Ryne}},
  \bibinfo {author} {\bibfnamefont {M.}~\bibnamefont {Venturini}}, \bibinfo
  {author} {\bibfnamefont {A.~A.}\ \bibnamefont {Zholents}},\ and\ \bibinfo
  {author} {\bibfnamefont {I.~V.}\ \bibnamefont {Pogorelov}},\ }\bibfield
  {title} {\bibinfo {title} {High resolution simulation of beam dynamics in
  electron linacs for x-ray free electron lasers},\ }\href
  {https://doi.org/10.1103/PhysRevSTAB.12.100702} {\bibfield  {journal}
  {\bibinfo  {journal} {Phys. Rev. ST Accel. Beams}\ }\textbf {\bibinfo
  {volume} {12}},\ \bibinfo {pages} {100702} (\bibinfo {year}
  {2009})}\BibitemShut {NoStop}%
\bibitem [{\citenamefont {Saldin}\ \emph {et~al.}(1997)\citenamefont {Saldin},
  \citenamefont {Schneidmiller},\ and\ \citenamefont
  {Yurkov}}]{saldin_coherent_1997-1}%
  \BibitemOpen
  \bibfield  {author} {\bibinfo {author} {\bibfnamefont {E.~L.}\ \bibnamefont
  {Saldin}}, \bibinfo {author} {\bibfnamefont {E.~A.}\ \bibnamefont
  {Schneidmiller}},\ and\ \bibinfo {author} {\bibfnamefont {M.~V.}\
  \bibnamefont {Yurkov}},\ }\bibfield  {title} {\bibinfo {title} {On the
  coherent radiation of an electron bunch moving in an arc of a circle},\
  }\href {https://doi.org/https://doi.org/10.1016/S0168-9002(97)00822-X}
  {\bibfield  {journal} {\bibinfo  {journal} {Nucl. Instrum. Methods Phys.
  Res., Sect. A}\ }\textbf {\bibinfo {volume} {398}},\ \bibinfo {pages} {373 }
  (\bibinfo {year} {1997})}\BibitemShut {NoStop}%
\bibitem [{\citenamefont {Mitchell}\ \emph {et~al.}(2013)\citenamefont
  {Mitchell}, \citenamefont {Qiang},\ and\ \citenamefont
  {Ryne}}]{mitchell_fast_2013}%
  \BibitemOpen
  \bibfield  {author} {\bibinfo {author} {\bibfnamefont {C.~E.}\ \bibnamefont
  {Mitchell}}, \bibinfo {author} {\bibfnamefont {J.}~\bibnamefont {Qiang}},\
  and\ \bibinfo {author} {\bibfnamefont {R.~D.}\ \bibnamefont {Ryne}},\
  }\bibfield  {title} {\bibinfo {title} {A fast method for computing 1-{D}
  wakefields due to coherent synchrotron radiation},\ }\href
  {https://doi.org/https://doi.org/10.1016/j.nima.2013.03.013} {\bibfield
  {journal} {\bibinfo  {journal} {Nucl. Instrum. Methods Phys. Res., Sect. A}\
  }\textbf {\bibinfo {volume} {715}},\ \bibinfo {pages} {119 } (\bibinfo {year}
  {2013})}\BibitemShut {NoStop}%
\end{thebibliography}
\end{document}